\documentclass[usenatbib,useAMS]{mnras}
\usepackage{graphicx}
\usepackage{amsmath, amssymb}
\usepackage[T1]{fontenc}
\usepackage{newtxtext,newtxmath}
\usepackage{microtype}
\usepackage{ragged2e,tabularx,booktabs}
\usepackage{soul}
\usepackage{orcidlink}
\usepackage{placeins}
\usepackage{appendix}

\makeatletter
\let\oldtheequation\theequation
\renewcommand\tagform@[1]{\maketag@@@{\ignorespaces#1\unskip\@@italiccorr}}
\renewcommand\theequation{(\oldtheequation)}
\makeatother

\usepackage{xcolor}
\definecolor{NIHAO}{HTML}{648FFF}
\definecolor{APOSTLE}{HTML}{785EF0}
\definecolor{SCALING}{HTML}{7F7F7F}

\newcommand{\SatGen}{\texttt{SatGen}}
\newcommand{\vmax}{\ensuremath{v_\text{max}}}
\newcommand{\rmax}{\ensuremath{r_\text{max}}}
\newcommand{\fobs}{\ensuremath{f_{\,i}}}
\newcommand{\fpred}{\ensuremath{f_\text{pred}}}
\newcommand{\thetab}{\ensuremath{\boldsymbol{\theta}}}
\newcommand{\xb}{\ensuremath{\mathbf{x}}}
\newcommand{\osim}{\mathord{\sim}}
\DeclareRobustCommand{\VAN}[2]{#1}
\title[The structure and orbit of MW satellites]{Probabilistic inference of the structure and orbit of Milky Way satellites with semi-analytic modelling}

\defcitealias{Rodriguez-Puebla17}{RP17}
\newcommand{\RPSMHM}{\citetalias{Rodriguez-Puebla17}}
\defcitealias{Behroozi13}{B13}
\newcommand{\BSMHM}{\citetalias{Behroozi13}}

\defcitealias{Errani18}{E18}
\newcommand{\ERV}{\citetalias{Errani18}}
\defcitealias{Kaplinghat19}{K19}
\newcommand{\KRV}{\citetalias{Kaplinghat19}}
\defcitealias{Andrade23}{A24}
\newcommand{\ARV}{\citetalias{Andrade23}}

\author[D. Folsom et al.]{
Dylan Folsom~\textsuperscript{\orcidlink{0000-0002-1544-1381}},\textsuperscript{1}\thanks{E-mail: \href{mailto:dfolsom@princeton.edu}{dfolsom@princeton.edu}}
Oren Slone~\textsuperscript{\orcidlink{0000-0002-5999-106X}},\textsuperscript{1, 2, 3}
Mariangela Lisanti~\textsuperscript{\orcidlink{0000-0002-8495-8659}},\textsuperscript{1, 4}
Fangzhou Jiang~\textsuperscript{\orcidlink{0000-0001-6115-0633}},\textsuperscript{5} and 
Manoj Kaplinghat~\textsuperscript{\orcidlink{0000-0001-8555-0164} 6} 
\\
\\
\textsuperscript{1} Department of Physics, Princeton University, Princeton, NJ 08544, USA\\
\textsuperscript{2} Center for Cosmology and Particle Physics, Department of Physics, New York University, New York, NY 10003, USA  \\
\textsuperscript{3} C.~N.~Yang Institute for Theoretical Physics, Stony Brook University, Stony Brook, NY, 11794, USA \\
\textsuperscript{4} Center for Computational Astrophysics, Flatiron Institute, New York, NY 10010, USA\\
\textsuperscript{5} Kavli Institute for Astronomy and Astrophysics, Peking University, Beijing 100871, China \\
\textsuperscript{6} Center for Cosmology, Department of Physics and Astronomy, University of California, Irvine, CA 92697, USA
}

\date{Accepted 2024 December 6. Received 2024 October 15; in original form 2024 February 21}
\volume{536}
\pubyear{2025}

\makeatletter
\def\@printed{\href{https://doi.org/10.1093/mnras/stae2736}{https://doi.org/10.1093/mnras/stae2736}}
\makeatother
\begin{document}
\pagerange{2891--2913}
\maketitle

\begin{abstract}
Semi-analytic modelling furnishes an efficient avenue for characterizing dark matter haloes associated with satellites of Milky Way-like systems, as it easily accounts for uncertainties arising from halo-to-halo variance, the orbital disruption of satellites, baryonic feedback, and the stellar-to-halo mass (SMHM) relation. We use the \SatGen{} semi-analytic satellite generator, which incorporates both empirical models of the galaxy–halo connection as well as analytic prescriptions for the orbital evolution of these satellites after accretion onto a host to create large samples of Milky Way--like systems and their satellites. By selecting satellites in the sample that match observed properties of a particular dwarf galaxy, we can infer arbitrary properties of the satellite galaxy within the cold dark matter paradigm. For the Milky Way's classical dwarfs, we provide inferred values (with associated uncertainties) for the maximum circular velocity \vmax{} and the radius \rmax{} at which it occurs, varying over two choices of baryonic feedback model and two prescriptions for the SMHM relation. While simple empirical scaling relations can recover the median inferred value for \vmax{} and \rmax{}, this approach provides realistic correlated uncertainties and aids interpretability. We also demonstrate how the internal properties of a satellite’s dark matter profile correlate with its orbit, and we show that it is difficult to reproduce observations of the Fornax dwarf without strong baryonic feedback. The technique developed in this work is flexible in its application of observational data and can leverage arbitrary information about the satellite galaxies to make inferences about their dark matter haloes and population statistics.
\end{abstract}


\section{Introduction} 
\label{sec:1}
The satellite galaxies of the Milky Way~(MW) have long been of interest as probes of cosmological structure formation and of the influence of dark matter~(DM) at small scales. The most luminous of the MW’s satellites, the `classical' satellites, are especially promising candidates for such studies. These dwarf galaxies have stellar populations large enough to provide robust measurements of galactic properties while still maintaining a low stellar-to-halo mass ratio, making them ideal for DM studies. This work proposes a novel semi-analytic formalism that enables the prediction of many properties (with associated uncertainty) for dwarf galaxies undergoing disruption in an MW-like host system, given their present-day observable properties. Using this procedure, we (i)~infer the halo structure for each of the MW’s classical satellites and compare it to existing analyses, (ii)~provide a novel means by which models of baryonic feedback may be constrained, and (iii)~demonstrate the connection between internal properties of the satellites and their orbital properties, including a study of whether or not the satellites were accreted from the field or as part of a larger system.

There is a long tradition of study surrounding the `galaxy--halo connection,' which establishes broad-strokes statistical relationships between luminous galaxies and their DM haloes -- see~\citet{Wechsler18} for a review. In the cold dark matter (CDM) paradigm, DM haloes provide the gravitational wells that capture baryonic gas and seed galaxy growth. The na\"ive expectation, which has been borne out by detailed empirical and analytical studies~\citep{Conroy06,Behroozi13,Moster13}, is that there is a direct relationship between galaxy and halo size, with more massive haloes containing more massive galaxies. This relationship, the stellar mass--halo mass (SMHM) relation, is a core ingredient in semi-analytic models and has a modest scatter, at least for systems with peak halo masses above $\osim 10^{11}~\mathrm{M}_\odot$. Below this scale, uncertainties on both the slope and scatter of the SMHM distribution can be significant -- see, e.g. \cite{Danieli22} and references therein -- and the functional form itself becomes unknown: in some simulations, the SMHM relation may be better described with an exponential suppression than a power law~\citep{Fattahi18}. Due to the unconstrained nature of the relation at these low masses, predictions for dwarf systems can vary drastically~\citep{Behroozi10, Behroozi19,Nadler19,Santos-Santos22}. This uncertain region of parameter space includes many MW satellites, and as such they provide an interesting laboratory for understanding this aspect of the galaxy--halo connection.

The galaxy--halo connection grows in complexity beyond the SMHM relation. Baryons gravitationally influence the DM haloes in which they reside, and there is a wealth of literature surrounding how these `baryonic feedback' mechanisms affect the structure of the DM halo and the uncertainties inherent in modelling this feedback~\citep[see, e.g.][]{Vogelsberger20}. The structure of an isolated halo can be described through empirical relations between, e.g. the halo mass and a concentration parameter or the maximum circular velocity~\citep{Neto07,Moline17}. However, these relations become further complicated for satellite galaxies that undergo complex tidal interactions with their hosts~\citep{Penarrubia09}. These interactions can lead to deviations from predictions for isolated haloes and potentially even to total disruption of the satellite. The tidal evolution depends on the structure of the satellite halo, which in turn depends on its baryonic content. The current state-of-the-art for modelling these intricate systems is through (magneto)hydrodynamical simulations, which must assume particular models for star formation, baryonic feedback, etc., and which require significant computational resources to resolve large numbers of dwarf galaxies. Moreover, such simulations are prone to numerical artefacts, such as artificial disruption~\citep{VanDenBosch18,VanDenBosch18a}, which affect characterizations of their satellite populations.

This paper proposes a new strategy for inferring the halo properties of satellite galaxies. The procedure relies on semi-analytic satellite generators to address several of the challenges listed above. Such generators combine empirical modelling of the galaxy--halo connection as parametrized from simulations, along with analytic modelling of tidal mass loss and dynamical friction, to find the orbital path and mass evolution of a satellite galaxy. The semi-analytic code can efficiently generate satellite populations for individual MW haloes and then incorporate halo-to-halo variance by generating many such iterations. There are three key advantages of this approach. First, the large statistical samples that can be generated allow one to find a population of satellites that closely resemble any satellite of interest based on observational data. Second, the comparatively low computational cost allows one to efficiently scan over known sources of uncertainty, especially with regards to parametrizations of the SMHM relation and feedback mechanisms, providing a means to effectively quantify systematic uncertainties on the predictions. Finally, the realizations of the semi-analytic model each carry detailed information about many internal and systemic properties of their satellites that can be leveraged in analyses. 

Semi-analytic models have long been employed for the study of dwarf galaxy formation, particularly in recovering the population of MW satellites as a whole. In practice, much of this literature centres around matching population statistics such as the luminosity function or mass function of the MW satellite system~\citep{Koposov09,Li10,Maccio10,Guo11,Font11,Brooks13,Starkenburg13,Barber14,Pullen14,Guo15,Lu16,Nadler19,Nadler23}. As an extension of this body of work, semi-analytic models can also be applied to the evolution of individual satellites as part of such a system~\citep{Taylor01,Penarrubia10,Hiroshima18,Ando20,Dekker22,Akita23}.

In this vein, the current study uses the semi-analytic model \SatGen{},\footnote{\href{https://github.com/JiangFangzhou/SatGen}{https://github.com/JiangFangzhou/SatGen}} which provides a statistical sample of MW-like satellite systems with adjustable parameters for the initialization and evolution of satellite galaxies, as discussed below. \SatGen{} is able to accurately reproduce distributions of satellite maximal circular velocity, $\vmax$, the radius at which this velocity occurs, $\rmax$, and the spatial distributions of observed satellite populations of the MW and M31~\citep{Jiang21}, as well as those produced by cosmological zoom-in simulations~\citep{Sawala16, Garrison-Kimmel17b}. These distributions are produced by integrating the orbits of satellite subhaloes, tracking the evolution of their density profiles and the galaxies they may host. As it does not require merger trees from an extant simulation, \SatGen{} can efficiently sample cosmologically-motivated assembly histories, which enables it to capture the dramatic halo-to-halo variance of satellite statistics. Further, \SatGen{} is calibrated to reproduce baryonic feedback seen in hydrodynamical simulations and self-consistently tracks tidal mass loss and the resultant density profile evolution during a satellite's orbit, as described in detail in \autoref{sec:2.1} and the references therein.

To model the halo of a particular satellite galaxy, we select satellite analogues from the overall \SatGen{} distribution in a principled way based on their similarity to the galaxy we wish to model. The result of this selection is impacted by both the choice of the galaxy--halo connection model and by the selection criteria, i.e. the incorporation of different sets of observables into the selection. The analysis finds good agreement with existing studies of the MW's classical satellites, recovering reasonable parameter values with physically-motivated uncertainties. 

The results of this work have broad applicability to the study of MW substructure. For any property of interest of a MW satellite, the method introduced here allows for inference of the allowed range of values consistent with arbitrary prior information. Throughout the text, we illustrate this advantage by giving examples of inference for both internal properties of the DM haloes of various MW satellite galaxies as well as distributions for orbital properties of these objects. The method presented in this paper can be used to constrain models of baryonic feedback, to probe the galaxy--halo connection, and to understand the spread of parameter values consistent with the observed properties of any individual satellite galaxy. In particular:
\begin{enumerate}
	\item Using our proposed technique on the Fornax dwarf galaxy provides interesting results. Specifically, we find that observations of the central density and stellar mass of Fornax are difficult to reproduce under assumptions of minimal baryonic feedback. A model including stronger DM core formation is required to produce both parameters simultaneously.
	\item We have compared our new method to similar inference methods based on the use of simple scaling relations, such as abundance matching. We find that while these techniques are able to recover reasonable estimates for the structural parameters of the MW's classical satellites, the inferred uncertainties do not properly reflect the complexities of non-linear satellite evolution and are typically underestimated.
	\item Using the techniques presented in this paper, we infer that it is unlikely that any of the classical satellites of the MW was part of a larger system at the time of accretion into the Galaxy, though the mode of accretion (i.e. whether directly accreted from the field or accreted as part of a group) significantly affects aspects of the orbit.
\end{enumerate}

The paper is organized as follows. \autoref{sec:2} overviews the methodology, reviewing \SatGen{} and discussing the statistical procedures used in the study. As a concrete example, \autoref{sec:3} applies this method to infer profile parameters for the MW's classical satellites -- specifically the radius $\rmax{}$ at which the maximum circular velocity, $\vmax{}$, is achieved. Also included is a discussion of the systematic uncertainties associated with this modelling, demonstrating that the resulting predictions in the $\rmax$ -- $\vmax$ plane correspond well to observational studies. \autoref{sec:4} provides other applications of the method by (i) correlating internal structure with orbital properties by considering the relation between pericentric distance and central density and (ii) discussing the likelihood that the classical satellites of the MW were contributed by the infall of a large system of satellites. \autoref{sec:5} summarizes the main findings of this study.

\section{Methodology}
\label{sec:2}
The method presented in this paper requires a large sample of realistic MW satellites. This statistical sample is generated using a semi-analytic model of halo formation, described in \autoref{sec:2.1}, that efficiently produces a population of satellites not subject to artificial disruption. Moreover, the model enables variation of the systematic uncertainties that contribute to structure formation, including baryonic feedback, the SMHM relation, and halo-to-halo variance. This diverse, physically-motivated sample of satellite systems provides the backbone for a weighting procedure that can be used to make predictions for particular dwarf galaxies of interest. The weighting procedure is described in \autoref{sec:2.2}.

\subsection{Semi-analytic satellite generation}
\label{sec:2.1} 

\subsubsection{SatGen overview}
In this work, we use the \SatGen{} semi-analytic halo model~\citep{Jiang21,Green22} and refer the reader to the original publications for more details on the implementation of the model and its calibration. The model generates satellite populations in two steps. First, it generates hierarchical merger trees according to the extended Press--Schechter theory~\citep{Parkinson07} and initializes the progenitor haloes for a target host of a particular mass (in this case, the target is a MW-like halo). The mergers prescribed by this theory source the growth of the MW host, and its mass evolves with time in accordance with these merger trees. Second, it evolves the orbit and internal structure of each progenitor halo according to physical prescriptions and empirical relations calibrated to high-resolution simulations. Ultimately, this produces a realistic population of surviving satellites around the target host.

Satellite progenitors at infall are assigned with cosmologically-motivated initial orbits~\citep{Li20}, density profiles~\citep{Zhao09,Freundlich20}, and stellar masses. The functional form for the halo profiles, introduced by~\citet{Dekel17}, belongs to the $\alpha\beta\gamma$ family of profiles~\citep{Zhao96}. \citet{Freundlich20} have shown that its four free parameters have the flexibility required to describe the DM halo's response to baryonic processes. A convenient parametrization of this profile uses the virial mass of the halo, a concentration parameter, the slope of the density profile at 1~per~cent of the virial radius, and the spherical virial overdensity.

The merger tree sets the virial mass and time of infall for each accreted object. The virial mass is defined in terms of a time-dependent virial overdensity that (for haloes in the field) follows the fit performed by~\citet{Bryan98}. \SatGen{} determines the two remaining parameters as follows: the concentration parameter is first calculated following the universal model of~\citet{Zhao09}, which is based on cosmological DM-only simulations with a wide range of cosmological parameters. Then, the stellar mass is determined using a SMHM relation, detailed below. The halo parameters are set such that they respond appropriately to baryonic effects: the inner slope of the density profile and the updated concentration parameter are calculated according to empirically-calibrated relations from hydrodynamical simulations~\citep[e.g.][]{Tollet16,Freundlich20}. This procedure fully determines the initial conditions of the galaxy and its DM halo. The merger tree algorithm then recurses, describing the assembly history of each satellite before falling into the MW.

\SatGen{} accounts for scatter in each of the aforementioned scaling relations. Notably, the SMHM relation is given a scatter of 0.2~dex in stellar mass, and the feedback prescriptions include additional scatter on the concentration parameter. An advantage of using a semi-analytic model is that one can sample effectively over this scatter, allowing for full exploration of reasonable satellite parameter space.

After initializing the DM and stellar properties of the satellites at infall, \SatGen{} evolves their orbit and structure within the dynamically evolving host potential, including a~\citet{Chandrasekhar43}-like treatment of dynamical friction. As the satellites orbit, their haloes lose mass in proportion to the mass outside of their tidal radius, on a time-scale set by the host's dynamical time. The satellites evolve along tidal tracks, which are empirical laws for the structural response of the satellites to tidal stripping and heating~\citep{Penarrubia10,Errani18,Green19}. Specifically, these tidal tracks determine baryonic mass loss, as well as the evolution of \vmax{} and \rmax{}, as a function of the halo mass loss and density profile shape.\footnote{With the mass parameter as a profile normalization, there are three parameters used for the shape. The tidal tracks determine two shape parameters (the concentration and virial overdensity), and the slope of the density profile at zero radius is taken to be fixed, which sets the third degree of freedom.} The resulting satellite mass functions are provided and validated in \citep{Jiang21}.

The combined result of these prescriptions is another important advantage of this semi-analytic orbit integration approach: the evolution of a satellite's internal structure is self-consistently accounted for with a formalism that is computationally cheap when compared to full numerical simulations. However, the formalism does lack some notable dynamical effects. Specifically, while \SatGen{} accounts for hierarchical structure formation, allowing for satellites to host (and potentially eject) their own satellites, it does not account for tidal effects beyond those of the immediate parent, nor does it account for gravitational interactions between satellites of the same order, or for the back-reaction of the satellites on the host potential. Further, apart from the disc included in the MW host, all potentials are spherically symmetric.

\subsubsection{Stellar mass--halo mass relation}
An important source of systematic uncertainty is the SMHM relation. \SatGen{} provides two possible calibrations for this relation, based on either the~\citet[][hereafter \RPSMHM{}]{Rodriguez-Puebla17} model or the~\citet[][hereafter \BSMHM{}]{Behroozi13} model. Both parametrize the stellar-to-halo mass ratio, $X = M_\star/M_\mathrm{vir}$, for haloes in the field in terms of $M_\mathrm{vir}$. On the $M_\star$ -- $M_\mathrm{vir}$ plane, these SMHM relations are effectively power laws (with some scatter) at the low-mass end, $M_\mathrm{vir} \lesssim 10^{10.5}~\mathrm{M}_\odot$ \citep{Behroozi10,Munshi21}. The index of this power law at the low-mass end (often denoted $\alpha$) contains information regarding the star-formation efficiency and stellar feedback in dwarf haloes. The default behaviour in \SatGen{} is to use the model presented in \RPSMHM{}, which has a fairly steep faint-end slope, $M_\star \propto M_\mathrm{vir}^{1.975}$ consistent with other recent studies -- see, e.g. \cite{Behroozi19} and references therein. However, this relation is very uncertain in the mass range of the classical satellites: \BSMHM{} fit a much shallower slope of $M_\star \propto M_\mathrm{vir}^{1.412}$ in this regime. With a shallower slope, galaxies of the same stellar mass can reside in much lighter haloes, which can significantly change the physics of galaxy evolution. In particular, the \BSMHM{} SMHM relation gives a factor of two enhancement to the satellite luminosity function relative to \RPSMHM{} in the regime of the classical satellites, $M_\star \sim 10^{5-7}~\mathrm{M}_\odot$.

It should be noted that the two calibrations for the SMHM relation provided in \SatGen{} do not probe the full range of theoretically-allowed SMHM relations. For example, the scatter about the mean is taken to be fixed in \SatGen{}. However, for low-mass galaxies, the stellar mass present in a particular halo depends strongly on the details of the halo mass at reionization and the growth experienced afterward; therefore, it is possible that the scatter in the SMHM relation grows with decreasing mass~\citep{Garrison-Kimmel17a,Nadler20,Munshi21,Danieli22,Kim24}. While we do not consider any models with growing scatter, in these cases it would be possible for smaller-mass haloes to be consistent with the stellar masses of the classical satellites considered in this work. Further, the shape of the SMHM relation may not be a power law. If the SMHM relation instead has an exponential suppression at low masses~\citep[e.g.][]{Fattahi18,Benitez-Llambay20}, then the brightest of the classical satellites are preferentially hosted by smaller-mass subhaloes compared to the \RPSMHM{} relation expectations, and, overall, the haloes which host dwarf galaxies are more clustered about $\vmax \sim 20-30$~km~$\mathrm{s}^{-1}$ due to the exponential suppression. Extending the SMHM relation beyond the mass scales of the classical satellites to ultrafaint dwarfs with $M_\star \lesssim 10^5~\mathrm{M}_\odot$, these theoretical uncertainties become even larger and the SMHM relation is even more poorly constrained.

\begin{figure*}
	\centering
	\includegraphics{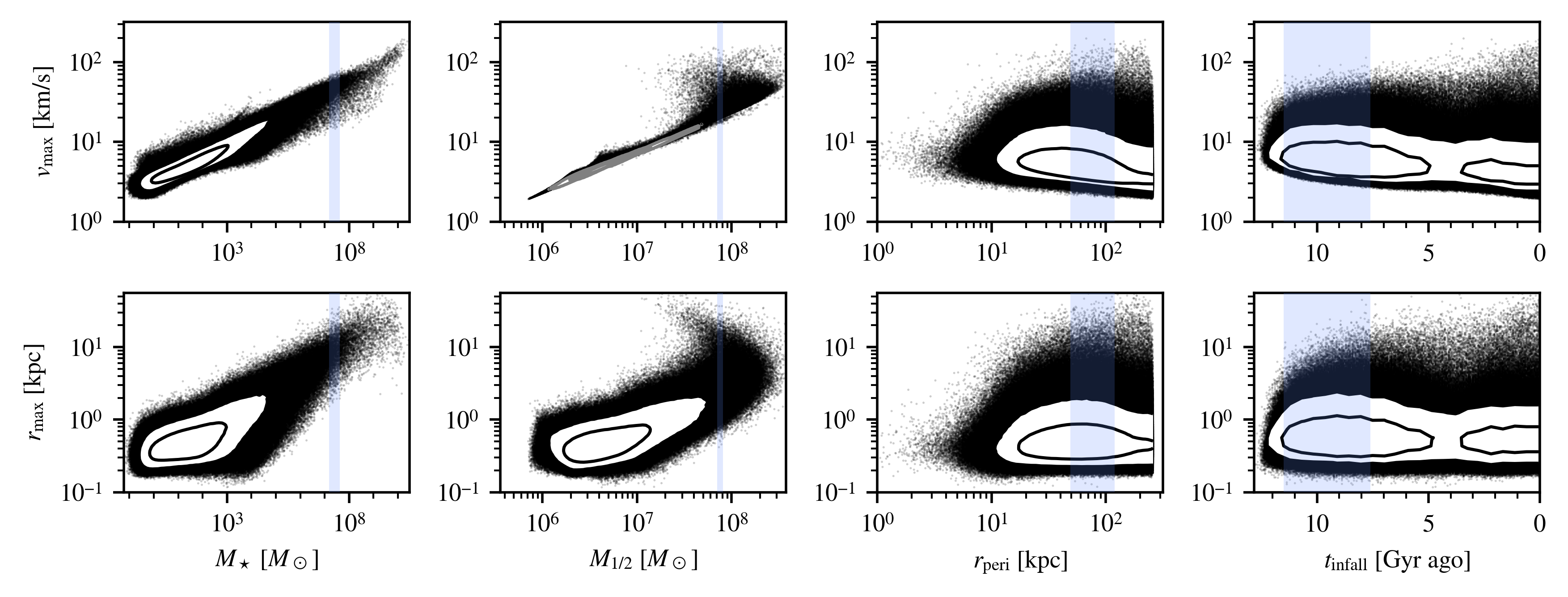}
	\caption{The satellites of all \SatGen{} realizations form a correlated probability distribution in a high-dimensional space, shown with contours enclosing 68 and 95~per~cent of satellites, while those outside the contours are shown as individual points. This figure shows the joint distribution of profile parameters $(\rmax,\,\vmax)$ against (i) the stellar mass $M_\star$, (ii) the mass $M_{1/2}$ contained within Fornax's half-light radius, (iii) the pericentre $r_\mathrm{peri}$, and (iv) the infall time $t_\mathrm{infall}$. Because many of these parameters correlate with $\rmax$ and $\vmax$, they can be used to infer these halo properties. As an example, the blue bands highlight the 68~per~cent confidence regions for the Fornax dwarf, collected in~\autoref{tab:1}. \SatGen{} satellites that lie in these bands contribute most strongly to the inference of Fornax's halo profile parameters, following the procedure detailed in \autoref{sec:2.2}. These satellites are generated with the SMHM relation of \RPSMHM{} and use the NIHAO-calibrated feedback emulator. The strong core formation of this model can be seen in the second column, where the large ratio of stellar mass to halo mass puffs up the satellites, leading to a population with large values for $\rmax$ and $\vmax$ with a relatively small $M_{1/2}$. The turnover in these panels is not present in the APOSTLE feedback emulator. Note that the sub-unity values in the $M_\star$ space reflect extrapolations of the semi-analytic tidal tracks of~\citet{Errani18}, which allow for tidal stripping to be tracked even below what is physically meaningful. The region of interest is well within the reasonable parts of this space, but the full distribution is shown for clarity.}
\label{fig:1}
\end{figure*}

\subsubsection{Baryonic feedback}
An additional source of systematic uncertainty is the baryonic feedback prescription, and \SatGen{} provides two possible calibrations for this relation as well. The strength of the baryonic feedback can have a significant effect on dwarf halo profiles, particularly in the inner regions of haloes that host large galaxies. In \SatGen{}, the feedback model is parametrized by $X = M_\star/M_\mathrm{vir}$ and affects (i) the logarithmic slope $s_{0.01}$ of the profile at 1~per~cent of the virial radius $r_\mathrm{vir}$, as well as (ii) the concentration, defined as $c_{-2} = r_\mathrm{vir}/r_{-2}$, where $r_{-2}$ is the radius at which the density profile is instantaneously a power law with index $-2$. Details regarding the functional form and asymptotic behaviour of these parametrizations are provided by~\citet{Jiang21}, but, in short, the two prescriptions for baryonic feedback provided in \SatGen{} reflect results from either the APOSTLE~\citep{Sawala16} or NIHAO~\citep{Wang15} simulations. The former exhibits much milder feedback, with the primary effect of the baryonic component on classical-mass satellites being the adiabatic contraction of the DM halo~\citep{Gnedin04}. NIHAO, on the other hand, has stronger baryonic feedback and allows for the formation of cores in the DM haloes. Both of these effects can be accounted for by modifying these slope and concentration parameters of the \SatGen{} haloes. Since these feedback relations depend on $X$, the strength of the feedback is sensitive to the choice of the SMHM relation. Further, as the tidal evolution depends on the shape of the density profile, the feedback prescription impacts both dark matter and baryonic mass loss as the satellite orbits.

In general, the baryonic feedback prescription will influence the resulting SMHM relation, and as such these choices are not independent of each other \citep{Santos-Santos22}. However, by varying the choice of SMHM relation and halo response to baryons independently, it is possible to isolate the effects of each. The \SatGen{} framework is flexible enough to account for correlations between the baryonic feedback model and SMHM relation through consistent choices for these two parameters, but the decision to use physically consistent choices is the prerogative of the user.

\subsubsection{Statistical sample}
The \SatGen{} sample used in this work consists of a primary data set of 4000 MW realizations produced with the NIHAO feedback emulator and \RPSMHM{} model, as well as 2000 MW realizations for the three other combinations of feedback prescriptions and the SMHM models to which it is compared. This gives a total of 10~000 satellite systems. The size of the data set ensures a thorough sampling of halo-to-halo variance in the assembly history. This is important for any study of classical satellites, as large satellites are somewhat rare; they sample the tail of the subhalo mass function. From the \SatGen{} output, we consider first-order satellites of the MW (i.e. ignoring satellites of satellites) that are within the MW's virial radius and that are gravitationally bound to the MW. We make two additional quality cuts. First, we remove satellites that have lost more than 99~per~cent of their virial mass to avoid satellites stripped beyond the range of calibration for the~\citet{Errani18} tidal tracks. This selection cut is physically motivated since it is unlikely that the satellites considered in this study have had such significant mass loss~\citep{Battaglia22}. Second, we remove satellites that have an ill-defined profile parameter $\alpha \geq 3$ (the $\alpha$ parameter is defined in terms of the concentration $c_{-2}$ and the slope $s_{0.01}$ described above, not to be confused with the notation used for the SMHM relation). If the profile is particularly cuspy or centrally concentrated,\footnote{The slope $s_{0.01}$ must be below 3 lest $\alpha$ be ill-defined. As the concentration $c_{-2}$ increases, this ceiling lowers. Further, if $c_{-2} > 18$, there is also a minimum value for $s_{0.01}$ that increases with $c_{-2}$. At $c_{-2} = 50$, the acceptable window is $1.4 < s_{0.01} < 2.5$. Typical values for $s_{0.01}$ are $1.5\pm 0.2$, but this can be pushed $\lesssim0.25$ when stellar feedback is effective.} the definition can require $\alpha \geq 3$, which leads to unphysical negative densities~\citep{Freundlich20}. The current implementation of \SatGen{} does not prevent this from happening, necessitating manual intervention. As a final cut, the mass resolution of the model is set to $10^{6.5}~\mathrm{M}_\odot$; as soon as a satellite is stripped enough to fall below this limit, its orbit is no longer tracked. Depending on the choice of feedback model, the SMHM relation, and dwarf of interest, there are typically hundreds to thousands of analogues for each classical satellite in the sample.

To aid in comparisons, we keep the host virial mass and disc mass consistent across all generated hosts. The virial mass is set to  $M_\mathrm{vir} = 10^{12}~\mathrm{M}_\odot$, which is roughly the inferred value for the MW across studies using different techniques~\citep{Wang20}; the spread across these studies is roughly a factor of two. Larger (smaller) MW host haloes would bias the accretion history to include larger (smaller) satellites. The disc mass is fixed at 
\vadjust{\penalty10000} $M_d = 0.05M_\text{vir} = 5\times 10^{10}~\mathrm{M}_\odot$, 
using a~\citet{Miyamoto75} potential with scale radius, $a$, set to reproduce the half-light radius of~\citet{Jiang19} and scale height, $b = a/12.5$, based on measurements of the MW's disc~\citep{Bland-Hawthorn16}. The disc mass is observationally uncertain to within $\osim 30$~per~cent, which can influence the survival of subhaloes, especially those with small pericentres; a more massive disc causes greater tidal mass loss. This effect is fairly insensitive to the shape of the disc potential~\citep{Green22}. Though \SatGen{} permits it, the generated host galaxies do not include a bulge component.

\begin{table*}
    \renewcommand{\arraystretch}{1.3}
	\caption{The observational values for the stellar mass $M_\star$, half-light mass $M_{1/2}$, 3D circularized half-light radius $r_{1/2}$, pericentric distance $r_\mathrm{peri}$, the infall time $t_\mathrm{infall}$, and the galactocentric distance $D_\mathrm{MW}$ used for the nine bright MW spheroidal dwarf satellites in this work.}
	\begin{tabularx}{\textwidth}{l@{\extracolsep{\fill}}c@{\extracolsep{\fill}}c@{\extracolsep{\fill}}c@{\extracolsep{\fill}}c@{\extracolsep{\fill}}c@{\extracolsep{\fill}}c}
	\toprule
	Name & $\log_{10}M_\star\,/\,\mathrm{M}_\odot$ &  $\log_{10}M_{1/2}\,/\,\mathrm{M}_\odot$ & $r_{1/2}$ [pc] & $r_\mathrm{peri}$ [kpc] & $t_\mathrm{infall}$ [Gyr] & $D_\mathrm{MW}$ [kpc] \\
		\midrule
	Canes Venatici I & $5.53_{-0.18}^{+0.18}$ & $7.26_{-0.05}^{+0.05}$ & $453.7_{-18.0}^{+18.0}$ & $84.5_{-37.2}^{+53.6}$ & $9.4_{-2.3}^{+0.9}$  & 217.8\\
	Carina & $5.78_{-0.18}^{+0.18}$ & $7.00_{-0.16}^{+0.16}$ & $331.8_{-18.8}^{+18.8}$ & $77.9_{-17.9}^{+24.1}$ & $9.9_{-2.7}^{+0.6}$ & 106.7\\
	Draco & $5.50_{-0.18}^{+0.18}$ & $7.15_{-0.03}^{+0.03}$ & $251.2_{-8.7}^{+8.7}$ & $58.0_{-9.5}^{+11.4}$ & $10.4_{-3.1}^{+2.4}$ & 76.0 \\
	Fornax & $7.40_{-0.22}^{+0.22}$ & $7.88_{-0.03}^{+0.03}$ & $836.3_{-17.1}^{+17.1}$ & $76.7_{-27.9}^{+43.1}$ & $10.7_{-3.1}^{+0.8}$ & 149.1 \\
	Leo I & $6.72_{-0.27}^{+0.27}$ & $7.26_{-0.04}^{+0.04}$ & $306.1_{-14.5}^{+14.5}$ & $47.5_{-24.0}^{+30.9}$ & $2.3_{-0.5}^{+0.6}$ & 256.0\\
	Leo II & $5.91_{-0.18}^{+0.18}$ & $6.88_{-0.05}^{+0.05}$ & $199.5_{-10.6}^{+10.6}$ & $61.4_{-34.7}^{+62.3}$ & $7.8_{-2.0}^{+3.3}$  & 235.7\\
	Sextans & $5.59_{-0.20}^{+0.20}$ & $7.47_{-0.04}^{+0.04}$ & $600.8_{-21.1}^{+23.4}$ & $82.2_{-4.3}^{+3.8}$ & $8.4_{-0.9}^{+2.7}$ & 89.2 \\
	Sculptor & $6.34_{-0.22}^{+0.22}$ & $7.36_{-0.03}^{+0.03}$ & $323.8_{-8.0}^{+8.0}$ & $44.9_{-3.9}^{+4.3}$ & $9.9_{-2.9}^{+1.7}$ & 86.1 \\
	Ursa Minor & $5.62_{-0.18}^{+0.18}$ & $7.21_{-0.04}^{+0.04}$ & $360.9_{-20.4}^{+20.4}$ & $55.7_{-7.0}^{+8.4}$ & $10.7_{-2.0}^{+1.7}$  & 78.0\\
	\bottomrule
	\end{tabularx}

	\begin{justify}	
	\emph{Note:} Stellar masses are derived from~\citet{Munoz18}, using a mass-to-light ratio of 1.2~$\mathrm{M}_\odot/\mathrm{L}_\odot$ with an added uncertainty of 0.16~dex~\citep{Woo08}. Half-light masses are derived from the~\citet{Munoz18} data using the~\citet{Wolf10} estimator, accounting for the circularization of $r_{1/2}$ following~\citet{Sanders16}. Pericentres are taken from the LMC-perturbed potential of~\citet{Pace22}, infall times from~\citet{Fillingham19}, and distances from~\citet{Munoz18}.
	\end{justify}
\label{tab:1}
\end{table*}
\subsection{Statistical framework}
\label{sec:2.2}
This work applies a generic statistical framework to infer the values of a galaxy parameter $\xb$ given a set of observed parameters, $\thetab$. A summary of this framework follows. For concreteness, the summary is grounded in the context of inferring the halo parameters \vmax{} and \rmax{} of the Fornax dwarf galaxy from various observational data, although the approach applies more generally to any other satellite galaxy and to additional halo or orbital parameters.

To illustrate this example, \autoref{fig:1} shows the distribution of the satellite halo parameters $\vmax$ and $\rmax$ for a sample of 4000 MW realizations in \SatGen{}, plotted as a function of the stellar mass $M_\star$, the mass $M_{1/2}$ enclosed within Fornax's observed half-light radius $r_{1/2} \approx 836$~pc, the pericentric distance $r_\mathrm{peri}$, and the infall time $t_\mathrm{infall}$. Correlations between these parameters are evident, and the parameters that correlate most strongly with $\vmax$ and $\rmax$ are the masses $M_\star$ and $M_{1/2}$, as expected from empirical and physical arguments~\citep{Wechsler18}. In general, one hopes to leverage these correlations to robustly infer difficult-to-measure properties of a satellite galaxy, such as those related to its DM halo or its merger history. The correlations are themselves subject to systematic uncertainties that should also be taken into account.

One simple way to perform this inference is to take a parameter with a known observational value, e.g. the choice $\thetab = M_\star$, and find the satellites within the \SatGen{} sample that match this value. The selected satellites then provide a distribution of values for parameters that are more poorly constrained observationally, e.g. the choice $\xb = \vmax$. In terms of \autoref{fig:1}, this is equivalent to selecting a preferred region of the abscissa (associated with the observed parameter range of a given dwarf) and projecting the resulting probability on to the ordinate axis. Ideally, the procedure should be flexible enough to work with any number of observed and inferred parameters, properly accounting for correlations between them. 

Quantitatively, let the distribution $\fobs(\xb,\,\thetab)$ describe the probability that satellite galaxy $i$ is described by parameters $\xb$ and $\thetab$. Marginalizing over the observed parameters $\thetab$ yields
\begin{equation}
	\fobs(\xb) = \int\! \fobs(\xb,\thetab)\,\mathrm{d}\thetab = \int\! \fobs(\xb | \thetab) \fobs(\thetab)\,\mathrm{d}\thetab \,,
\label{eq:1}
\end{equation}
where $\fobs(\thetab)$ is a marginal distribution of the observed parameter $\thetab$, chosen to be a two-sided Gaussian distribution centred at the observed mean, with widths set by estimated observational uncertainties on $\thetab$.\footnote{Systematic uncertainties on the observed $\thetab$, though not considered in this study, can be accounted for by choosing another distribution for $\fobs(\thetab)$.} The conditional distribution $\fobs(\xb | \thetab)$ is unknown; fortunately, the statistical sample of \SatGen{} realizations provides a means of estimating it. 

From \SatGen{}, one can obtain the theoretical probability distribution $\fpred(\xb,\,\thetab)$ that describes the properties of \emph{all} satellites in MW-like hosts. Note that this covers a broader range of parameter space than is reasonable for satellite $i$. For example, the CDM theory of structure formation predicts an increasingly larger number of satellites at lower masses, so the \SatGen{} procedure will naturally produce an abundance of satellites that sit below current observational thresholds, where the parameters $(\xb,\,\thetab)$ take on values inconsistent with the properties of the observed satellite. This does not pose a challenge so long as the \SatGen{} satellites with $\thetab$ within the observational range have a distribution of $\xb$ values that provides a realistic description of the MW satellites; it is only this conditional distribution that matters. Importantly, this means that the shape of the full $\fpred(\xb,\thetab)$ distribution need not be physically consistent nor accurate, so long as the \SatGen{} haloes which match the observed $\thetab{}$ for a dwarf of interest have physically consistent and accurate $\xb$.

Therefore, under the assumption that $\fpred(\xb|\thetab)$ approximates $\fobs(\xb|\thetab)$ in the region of parameter space where $\fobs(\thetab)$ is maximized, \autoref{eq:1} becomes
\begin{align}
	\fobs(\xb) &\approx \int\!\fpred(\xb | \thetab) \fobs(\thetab)\,\mathrm{d}\thetab = \int\! \fpred(\xb,\thetab)\frac{\fobs(\thetab)}{\fpred(\thetab)}\,\mathrm{d}\thetab \,,
\label{eq:2}
\end{align}
where $\fpred(\xb,\,\thetab)$ and $\fpred(\thetab)$ are the corresponding joint and marginal distributions obtained from the \SatGen{} realizations. The applicability of this approximation depends on the choice of $\xb$ and $\thetab$: one must choose parameters such that a \SatGen{} satellite with the desired $\thetab$ will have a value of $\xb$ appropriately consistent with observations. While the integral is performed over the entire population of \SatGen{} satellites, the factor of $\fobs(\thetab)/\fpred(\thetab)$ weighs satellites by how closely they match $\fobs(\thetab)$, regardless of the number density of \SatGen{} satellites at that $\thetab$. This ensures that the final inference of $\fobs(\xb)$ is set by the satellites that match $\fobs(\thetab)$, without contamination from, e.g. a large number of low-mass satellites. It is important to note that this approach relies on a thorough sampling of $\fpred$ near the $\thetab$ region of interest. In the event that the distribution is poorly sampled, there may be only a few \SatGen{} satellites with $\thetab$ near the maximum of $\fobs(\thetab)$, and therefore only these few contribute strongly to the inference of $\fobs(\xb)$.

\begin{table*}
	\caption{This table contains the number of \SatGen{} satellites comprising 68~per~cent (95~per~cent) of the weight in the $M_\star - M_{1/2}$ combined inference for the models considered in this work.}
	\begin{tabularx}{\textwidth}{l@{\extracolsep{\fill}}*{4}{c@{\extracolsep{\fill}}}}
	\toprule
	Name & NIHAO, \RPSMHM & APOSTLE, \RPSMHM & NIHAO, \BSMHM & APOSTLE, \BSMHM\\
	\midrule
	Canes Venatici I & 1341 (4083) & 183 (706) & 4988 (13571) & 4546 (12922) \\
	Carina & 931 (7069) & {\textbf{19}} (786) & 9464 (22914) & 3160 (15444) \\
	Draco & 4160 (11420) & 1784 (5209) & 1567 (4161) & 3473 (9234) \\
	Fornax & 269 (832) & {\textbf{2}} ({\textbf{4}}) & 382 (1079) & {\textbf{1}} ({\textbf{2}}) \\
	Leo I & 1351 (3766) & {\textbf{14}} (216) & 726 (2199) & 660 (2506) \\
	Leo II & 2884 (7944) & 393 (1341) & 1880 (5203) & 3144 (8699) \\
	Sextans & 1376 (4449) & 217 (940) & 4952 (13468) & 4760 (13312) \\
	Sculptor & 1397 (3863) & 244 (867) & 672 (1846) & 1502 (4148) \\
	Ursa Minor & 1715 (5235) & 276 (970) & 3045 (8163) & 3618 (9866) \\
	\bottomrule
	\end{tabularx}

	\begin{justify}	
	\emph{Note:} The models vary over all combinations of the feedback emulator (NIHAO versus APOSTLE) and the SMHM relation (\RPSMHM{} versus \BSMHM{}). A lower number indicates that there are few satellites in the \SatGen{} sample that appropriately reproduce the $M_\star$ and $M_{1/2}$ measurements for the dwarf, leading to a small number of high-weighted satellites driving the 68 or 95~per~cent containment regions of the inference. For example, in the APOSTLE emulator with \RPSMHM{}~SMHM relation, there are two satellites that most closely resemble Fornax that furnish $\geq 68$~per~cent of the weight used to perform inferences, since there are few \SatGen{} satellites able to match both the $M_\star$ and the $M_{1/2}$ of Fornax in this model. We highlight the dwarfs for which fewer than 25~\SatGen{} satellites contribute to the inference. This threshold is an arbitrary choice, but it indicates those dwarfs for which the inference is subject to low statistics, despite the large sample of MWs (4000 for the NIHAO \RPSMHM{} data set and 2000 for the other three). In all figures, the inferences corresponding to bolded numbers in this table are distinguished (via dashed-line contours) from the inferences driven by more than 25~\SatGen{} satellites.
	\end{justify}
\label{tab:2}
\end{table*}

Returning to the concrete example, consider the Fornax dwarf galaxy, whose mass and orbital parameters are provided in \autoref{tab:1}. The 68~per~cent confidence regions for these parameters are also indicated by the blue bands in \autoref{fig:1}. In all the cases shown, the observed marginal distribution $\fobs(\thetab)$, which roughly corresponds to the width of the blue bands, is significantly smaller than the distribution $\fpred(\thetab)$ obtained from \SatGen{}, which corresponds to the spread of black points and contours. Only those satellites that fall in or near the blue band will contribute significantly to the weight $\fobs(\thetab)/\fpred(\thetab)$ in \autoref{eq:2}, shaping the resulting prediction for $\fobs(\xb)$ with $\xb = (\rmax, \vmax)$. While \autoref{fig:1} uses the NIHAO emulator and the \RPSMHM{} SMHM relation, the \SatGen{} sample can be easily run with different models to quantify the systematic effects on the inferred parameters. The next section works through this procedure for inferring \rmax{} and \vmax{} for Fornax and the other classical satellites.

\section{Proof-of-Concept: Density Profile Inference}
\label{sec:3}
The technique described in \autoref{sec:2} is very general and allows for the inference of arbitrary satellite parameters using any set of observables. This section applies the procedure to the specific case of modelling the profile parameters $(\rmax,\, \vmax)$, starting with the Fornax dwarf galaxy and later generalizing to other dwarfs. We examine a number of potential parameters that may constrain the halo profile and ultimately derive an inferred distribution for the profile parameters using the satellite's stellar mass and its total mass within its half-light radius, discussing the systematic uncertainties on these results. These results agree well with analyses of stellar kinematics of MW dwarfs, as well as with a simple inference based on scaling relations such as abundance matching, though the uncertainties in these alternate methods are typically underestimated. \autoref{sec:4} explores other galaxy properties beyond $(\rmax,\, \vmax)$ that can be inferred using this procedure.

\subsection{Profile inference for Fornax-like galaxies}
\label{sec:3.1}

\autoref{fig:1} shows four observables that can potentially correlate with the halo profile of a Fornax-like galaxy: the stellar mass $M_\star$, the mass $M_{1/2}$ within the observed half-light radius $r_{1/2}$, the pericentric distance $r_\mathrm{peri}$, and the infall time $t_\mathrm{infall}$. The figure panels illustrate the strength of these correlations across the sample of \SatGen{} realizations of MW-like galaxies. In general, the orbital parameters $r_\mathrm{peri}$ and $t_\mathrm{infall}$ do not correlate very strongly with either $\rmax$ or $\vmax$ and are thus not expected to have much constraining power. In contrast, the parameters $M_{1/2}$ and $M_{\star}$ exhibit strong correlations with the halo parameters $\rmax$ and $\vmax$, and therefore they are expected to provide substantial constraining power when inferring, e.g. the joint distribution function of $\rmax$ and $\vmax$.\footnote{Note that other parameters may be better suited for a different choice of \xb{}.} In particular, the correlation between $\vmax$ and $M_\star$ is monotonic with little scatter, reflecting the fact that larger galaxies reside in larger haloes. $M_{1/2}$ is also tightly correlated with $\vmax$, although the relation exhibits a turnover at the higher-mass end. This turnover reflects the feedback prescription: the NIHAO emulator efficiently cores satellites with a large stellar-to-halo mass ratio, which increases for larger subhaloes according to the SMHM~relation. As such, at the $M_{1/2}$ of Fornax (indicated by the blue shaded region), there is a population of both low-\vmax{} haloes and high-\vmax{} haloes, where the former population is comprised of smaller galaxies with low $M_\star$ and $M_{1/2}$ and the latter population is comprised of galaxies with larger $M_\star$ that correspondingly are cored to lower $M_{1/2}$. In the APOSTLE emulator, where feedback is less efficient, this effect is not present, and the $M_{1/2}$ -- $\vmax$ relation remains monotonic across the entire parameter space. 

\begin{figure*}
	\centering
	\includegraphics{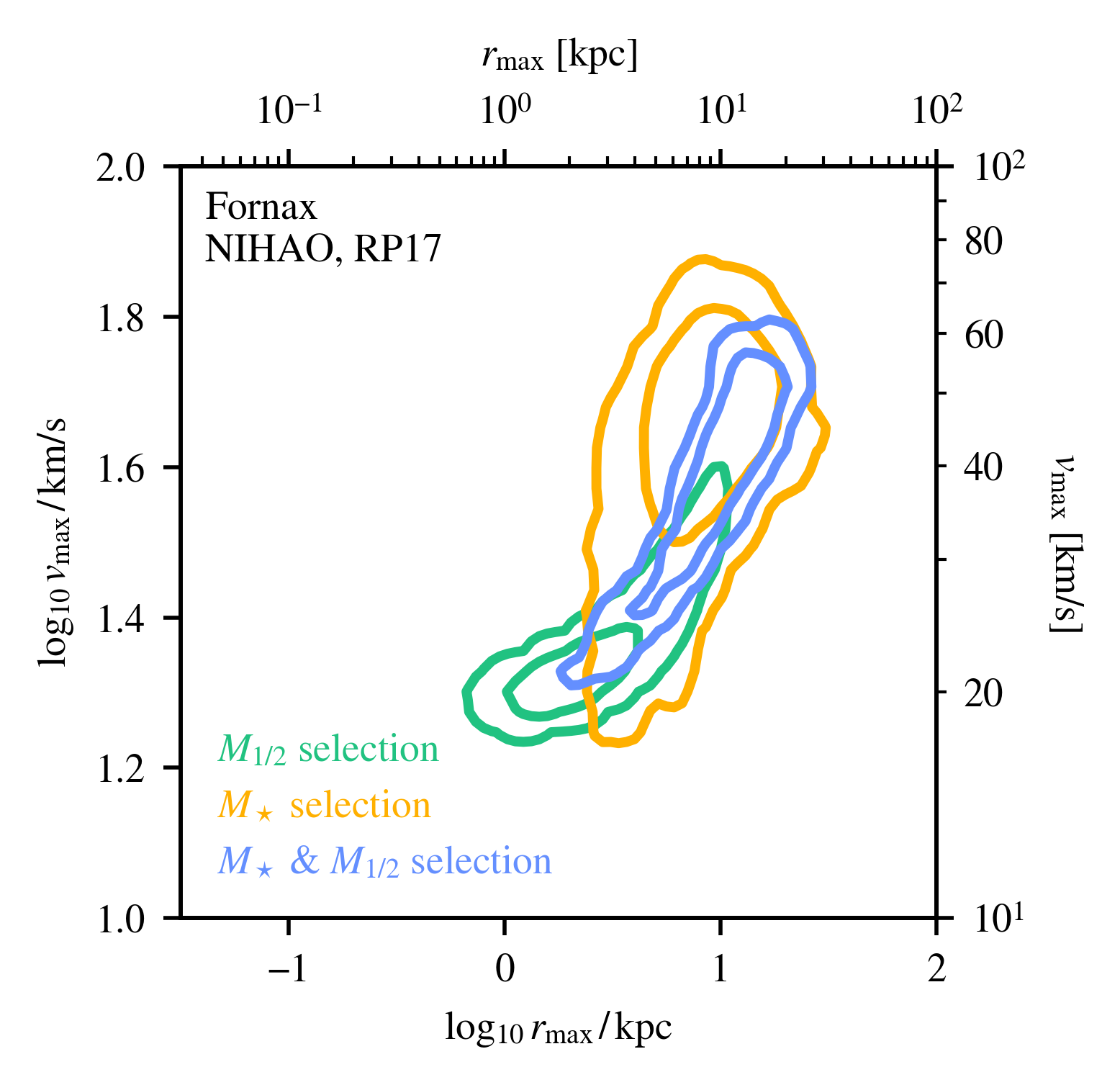}
 	\includegraphics{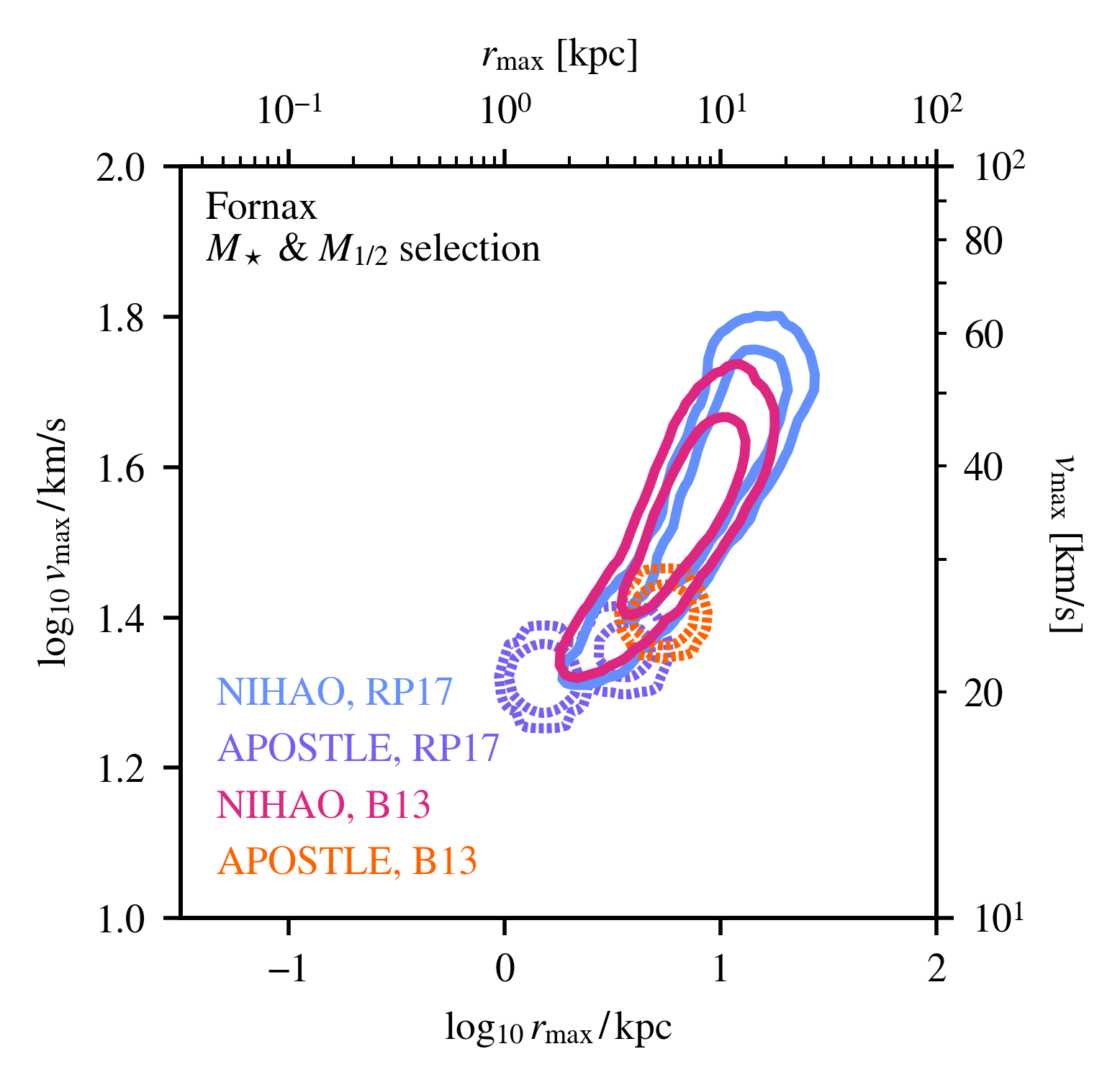}
	\caption{(Left) The inferred profile parameters $\rmax$ and $\vmax$ for the Fornax dwarf according to the procedure described in~\autoref{sec:2.2}. The contours enclose 68 and 95~per~cent confidence regions based on the mass within the observed half-light radius ($M_{1/2}$, in green), the stellar mass ($M_\star$, in gold), and the combination of both parameters simultaneously (in blue). Observationally, Fornax has a fairly low $M_{1/2}$, leading this inference to be consistent both with the massive cored satellites and with smaller uncored satellites. Fornax's stellar mass, on the other hand, is quite large, and selects only this former population. This selection is further refined by requiring that satellites simultaneously satisfy the observed $M_{1/2}$ and $M_{\star}$ values. The pericentric distance $r_\mathrm{peri}$ and infall time $t_\mathrm{infall}$ do not correlate strongly with $\vmax$ or $\rmax$ and are thus unable to constrain the space from the overall distribution of satellites; as such, they are not shown here. (Right)~The combined inference from the left plot is shown under various models for baryonic feedback and the SMHM relation. The blue and magenta `NIHAO' contours assume strong feedback, while the purple and orange `APOSTLE' contours use a weaker model of baryonic feedback. The blue and purple contours assume the \RPSMHM{} SMHM relation, which has a steeper faint-end slope than \BSMHM{}, taken for the magenta and orange contours. Notably, without the strong core formation of the NIHAO feedback emulator, the satellites consistent with both $M_\star$ and $M_{1/2}$ are much smaller, with lower $\vmax$ and $\rmax$.
	In both panels, the probability density is defined as a function of the base-ten logarithm of $\rmax$ and $\vmax$, not the quantities themselves, though $\rmax$ and $\vmax$ are provided on the top and right axes, respectively, for convenience. This convention is followed throughout this work. Furthermore, in this plot and throughout this work, we plot contours that are statistics-limited as dashed lines rather than complete contours. See \autoref{tab:2} for more information.}
\label{fig:2}
\end{figure*}

The technique of \autoref{sec:2.2} leverages these correlations to infer the distribution of $\xb = (\rmax,\,\vmax)$. The left panel of \autoref{fig:2} shows 68 and 95 per~cent containment regions for $\fobs(\xb{})$ in the 2D $\rmax$ -- $\vmax$ plane for the Fornax dwarf galaxy, highlighting the differences in choice of the known parameter $\thetab$. Gold and green contours correspond to $\thetab=M_\star$ and $\thetab=M_{1/2}$, respectively, while blue contours correspond to the joint $\thetab = (M_\star,\,M_{1/2})$.\footnote{The kernel density estimation used to generate contours for weighted distributions is accelerated by placing the lowest-weighted points in a histogram and using kernels only for the highest-weighted points: typically, the histogram comprises a sub--per cent component of the total weight, though it contains a majority of the points.} Using $\thetab{} = M_\star$, the inferred $\xb$ values are $\rmax = {8.21}_{-3.11}^{+4.54}$~kpc and $\vmax = {45.36}_{-11.96}^{+9.30}$~km/s, occupying a large region of the parameter space shown in \autoref{fig:2}. In contrast, the tight scatter in the individual relation between $M_{1/2}$ and $\vmax$ leads to a smaller preferred region of parameter space when $\thetab = M_{1/2}$. However, as mentioned above, the fixed $M_{1/2}$ window corresponding to the Fornax measurement contains both smaller satellites and larger satellites that have been cored to reach that $M_{1/2}$ value. This results in generally smaller inferred values of $\vmax = {21.09}_{-1.54}^{+2.70}$~km/s with a tail that extends to the larger cored systems. Said differently, the low $M_{1/2}$ value of Fornax points to the likely existence of a relatively small DM halo. However, another possibility allowed by strong feedback (such as that of the NIHAO emulator) is a much larger DM halo that has been efficiently cored to explain the low $M_{1/2}$ measurement. The inference derived using the joint $\thetab = (M_\star,\,M_{1/2})$ corresponds to $\rmax = {10.36}_{-4.56}^{+4.25}$~kpc and $\vmax = {40.53}_{-11.36}^{+8.80}$~km/s; selecting on $M_\star$ and $M_{1/2}$ simultaneously removes the smaller region of parameter space preferred by $M_{1/2}$ alone while improving the uncertainty of using $M_\star$ alone.

The analysis performed in this section is contingent on the assumption that the population of \SatGen{} satellites accurately models the population of MW satellites, but there are a number of systematic uncertainties that influence the underlying distribution. Two important sources of modelling uncertainty are the feedback model and the SMHM~relation, including their combined effects on each other. As described previously, \SatGen{} can be tuned to emulate baryonic feedback from either the NIHAO~\citep{Wang15} or APOSTLE~\citep{Sawala16} simulations, and it can incorporate an SMHM relation based on either \RPSMHM{} or \BSMHM{}. Modifying either of these aspects of the model leads to different overall populations of satellites and thus potentially different inferences for $\rmax$ and $\vmax$. The right-hand panel of \autoref{fig:2} shows how the combined $M_\star$ -- $M_{1/2}$ inference varies for all four combinations of the two SMHM and two feedback prescriptions. Note that dashed contours in this figure correspond to inferences that are set by a small number of highly weighted \SatGen{} satellites, and thus should be interpreted with caution. Specifically, if fewer than 25 \SatGen{} satellites comprise $\geq 68$~per~cent of the total weight, then the 68~per~cent containment contour is dashed, and similarly for the 95~per~cent contour. 

A main distinction between the four inferred distributions is the difference between the two corresponding to the NIHAO feedback emulator and the two corresponding to the APOSTLE feedback emulator. Qualitatively, the NIHAO emulator (which provides stronger feedback than APOSTLE) pushes the contours to larger values of both $\vmax$ and $\rmax$. This occurs for reasons that depend on both the $M_\star$ and the $M_{1/2}$ requirements. First, Fornax has a sizable $M_\star$, which can produce strong feedback in general, and even more so with the NIHAO emulator. Therefore, simultaneously matching Fornax's large $M_\star$ and relatively low $M_{1/2}$ corresponds to either smaller and cuspier haloes that occur naturally with APOSTLE, or larger and more cored haloes, which occur naturally with NIHAO. However, this difference between NIHAO and APOSTLE is also partially driven by low statistics. Due to the lack of cored satellites in the APOSTLE emulator, it is unlikely for a satellite to satisfy both the observed $M_\star$ and $M_{1/2}$ simultaneously. Finally, for a given stellar mass, the \BSMHM{} SMHM relation prefers satellites with less massive DM haloes than does the \RPSMHM{} SMHM relation. However, varying the SMHM relation betwen these two choices has only a mild effect on the shapes and positions of the contours for both feedback prescriptions.

More extreme SMHM relations can have significant effects, however; for example, if the scatter in the \RPSMHM{}~SMHM relation is taken to be 2~dex rather than the fiducial 0.2~dex, more \SatGen{} haloes are populated with enough stellar mass for baryonic feedback to become relevant. In the NIHAO emulator, this yields two new populations: (i) cored haloes well into the ultrafaint luminosity regime, with $M_\star\gtrsim 10^{4}~\mathrm{M}_\odot$, and (ii) adiabatically-contracted haloes at the higher-luminosity end of the classical satellite regime, with $M_\star \gtrsim 10^{6}~\mathrm{M}_\odot$. Both of these populations are relevant at the stellar mass of Fornax. With this high scatter, the inferred $\rmax-\vmax$ curve takes on a U-shape, with the minimum $\vmax$ set by the cored halo population at $\rmax\sim$~0.8~kpc, $\vmax\sim$~20~km/s, and the minimum $\rmax$ set by the contracted population consistent with Fornax, with $\rmax\sim$~80~pc, $\vmax\sim$~35~km/s.

\begin{table*}
    \renewcommand{\arraystretch}{1.3}
	\caption{The \SatGen{} $\rmax$ -- $\vmax$ inference for the nine bright MW spheroidal dwarfs.}
	\begin{tabularx}{\textwidth}{l@{\extracolsep{\fill}}*{8}{c@{\extracolsep{\fill}}}}
        \toprule
	& \multicolumn{2}{c}{NIHAO, \RPSMHM} &\multicolumn{2}{c}{APOSTLE, \RPSMHM}&\multicolumn{2}{c}{NIHAO, \BSMHM}&\multicolumn{2}{c}{APOSTLE, \BSMHM}\vspace{-\baselineskip}\\
	Name & \rmax{} [kpc] & \vmax{} [km/s] & \rmax{} [kpc] & \vmax{} [km/s] & \rmax{} [kpc] & \vmax{} [km/s] & \rmax{} [kpc] & \vmax{} [km/s] \\
	\midrule
	Canes Venatici I & ${2.68}_{-1.74}^{+3.09}$ & ${16.97}_{-3.50}^{+5.39}$ & ${1.02}_{-0.46}^{+5.35}$ & ${13.94}_{-1.29}^{+7.18}$ & ${2.01}_{-0.63}^{+0.94}$ & ${16.55}_{-1.81}^{+1.95}$ & ${1.84}_{-0.64}^{+0.91}$ & ${15.00}_{-1.33}^{+1.59}$ \\
	Carina & ${4.48}_{-3.26}^{+4.36}$ & ${21.60}_{-9.23}^{+4.06}$ & ${6.67}_{-5.68}^{+1.83}$ & ${22.74}_{-9.01}^{+2.86}$ & ${2.49}_{-0.84}^{+1.16}$ & ${17.74}_{-3.03}^{+3.42}$ & ${2.59}_{-1.06}^{+1.41}$ & ${15.38}_{-2.74}^{+2.89}$ \\
	Draco & ${2.72}_{-0.97}^{+1.22}$ & ${23.49}_{-3.36}^{+3.47}$ & ${2.77}_{-1.30}^{+1.44}$ & ${22.30}_{-3.85}^{+3.87}$ & ${1.99}_{-0.69}^{+1.05}$ & ${21.63}_{-2.46}^{+2.82}$ & ${1.74}_{-0.61}^{+0.96}$ & ${19.60}_{-1.88}^{+2.31}$ \\
	Fornax & ${10.36}_{-4.56}^{+4.25}$ & ${40.53}_{-11.36}^{+8.80}$ & ${1.74}_{-0.39}^{+2.01}$ & ${21.41}_{-3.03}^{+3.53}$ & ${7.03}_{-2.64}^{+3.29}$ & ${34.65}_{-7.50}^{+8.48}$ & ${5.40}_{-0.84}^{+0.99}$ & ${25.21}_{-3.03}^{+3.44}$ \\
	Leo I & ${5.39}_{-1.86}^{+2.39}$ & ${35.98}_{-7.97}^{+7.76}$ & ${7.40}_{-3.19}^{+2.67}$ & ${35.80}_{-9.19}^{+5.83}$ & ${3.79}_{-1.33}^{+1.98}$ & ${31.24}_{-5.78}^{+6.56}$ & ${4.13}_{-1.51}^{+2.26}$ & ${26.00}_{-4.46}^{+5.58}$ \\
	Leo II & ${3.62}_{-1.47}^{+1.69}$ & ${25.97}_{-5.63}^{+4.99}$ & ${4.38}_{-2.22}^{+1.71}$ & ${26.56}_{-7.70}^{+4.79}$ & ${2.40}_{-0.87}^{+1.26}$ & ${22.67}_{-3.60}^{+3.86}$ & ${2.19}_{-0.69}^{+1.12}$ & ${19.21}_{-2.71}^{+3.04}$ \\
	Sextans & ${2.49}_{-1.49}^{+3.42}$ & ${17.25}_{-2.71}^{+5.03}$ & ${1.05}_{-0.44}^{+4.16}$ & ${15.04}_{-1.13}^{+4.55}$ & ${2.10}_{-0.68}^{+0.99}$ & ${17.07}_{-1.58}^{+1.87}$ & ${1.88}_{-0.66}^{+0.96}$ & ${15.87}_{-1.17}^{+1.44}$ \\
	Sculptor & ${4.35}_{-1.53}^{+1.80}$ & ${31.74}_{-6.02}^{+5.84}$ & ${5.02}_{-2.93}^{+2.30}$ & ${29.81}_{-8.34}^{+5.99}$ & ${3.07}_{-1.06}^{+1.74}$ & ${28.62}_{-4.32}^{+5.02}$ & ${2.83}_{-0.95}^{+1.50}$ & ${24.02}_{-3.08}^{+3.66}$ \\
	Ursa Minor & ${3.58}_{-2.20}^{+1.83}$ & ${21.10}_{-5.35}^{+4.08}$ & ${2.63}_{-1.95}^{+3.17}$ & ${17.78}_{-3.77}^{+6.58}$ & ${2.02}_{-0.64}^{+1.02}$ & ${18.60}_{-2.10}^{+2.45}$ & ${1.88}_{-0.62}^{+0.91}$ & ${16.63}_{-1.55}^{+1.82}$ \\
	\bottomrule
	\end{tabularx}
	\begin{justify}	
	\emph{Note:} From left to right, this table contains the name of the satellite, the inferred \rmax{}, and the inferred \vmax{}, with 16th, 50th, and 84th percentiles reported, for each of the four models considered. The four models vary over all combinations of the feedback emulator (NIHAO versus APOSTLE) and the SMHM relation (\RPSMHM{} versus \BSMHM{}). For a comparison to observations, see Figs.~\hyperref[fig:3]{3} and \hyperref[fig:A1]{A1}. These inferences are performed independently as one-dimensional inferences, where the quantiles are well-defined and are not subject to the choice of logarithmic versus non-logarithmic coordinates described in the caption of \autoref{fig:2}.
	\end{justify}
\label{tab:3}
\end{table*}
\begin{figure*}
	\centering
	\includegraphics{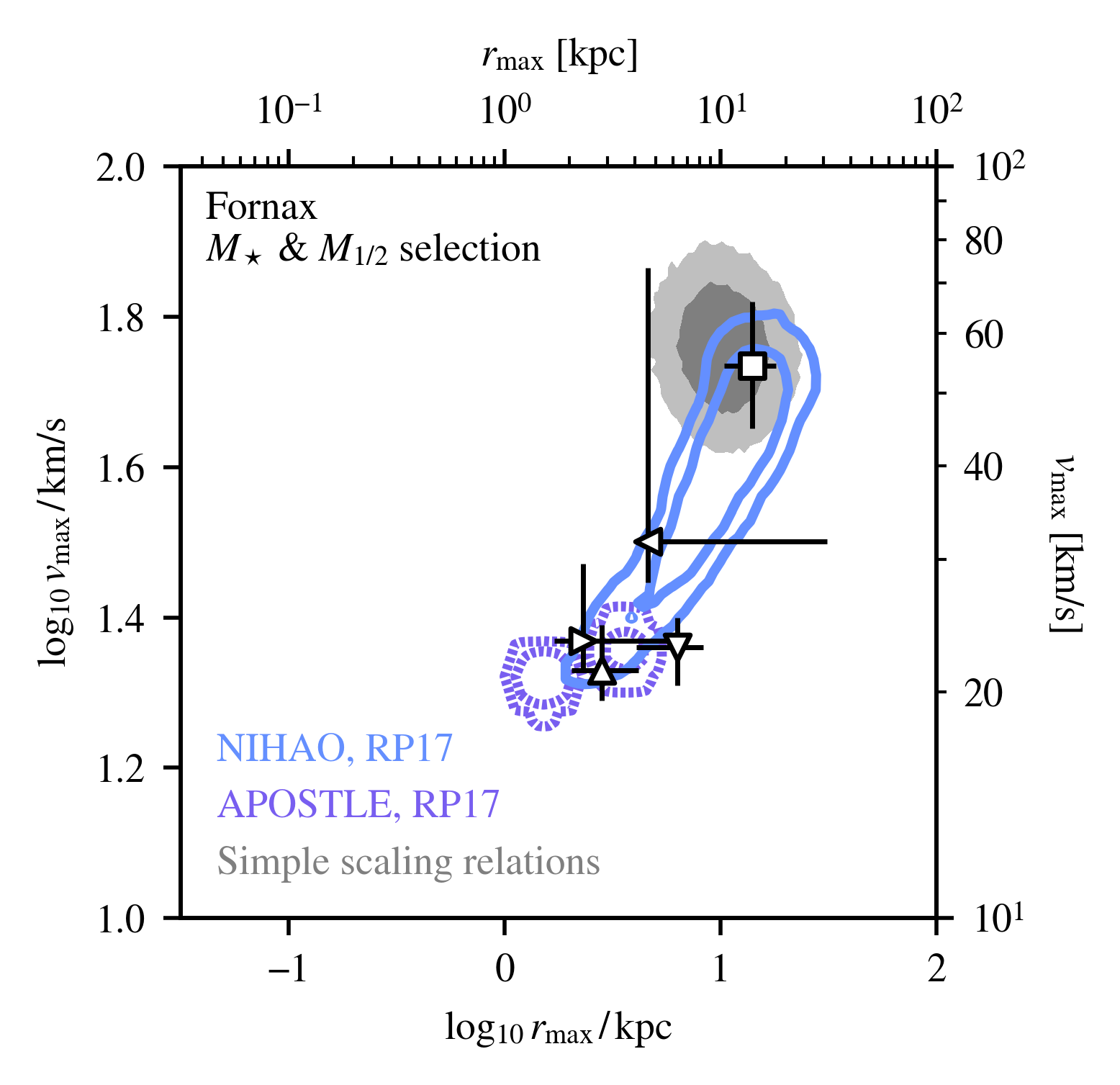}
	\includegraphics{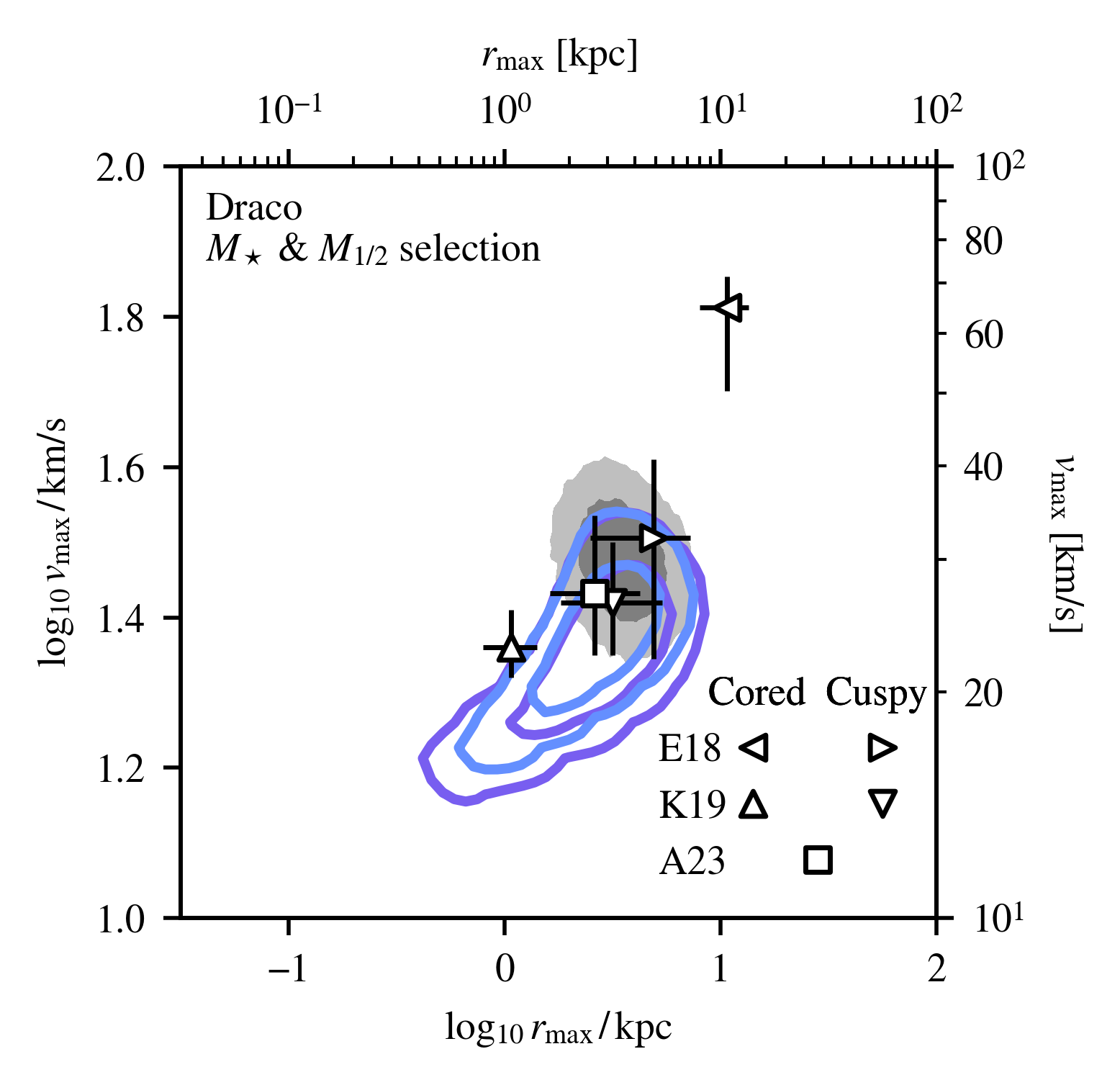}
	\caption{This figure shows the results of the \SatGen{} analysis for the Fornax (left) and Draco (right) dwarfs, assuming the \RPSMHM{}~SMHM relation and varying over feedback emulation, as compared with other results from the literature. Data points are kinematic measurements, including \ERV{}, shown in leftward- and rightward-pointing triangles; \KRV{}, shown in upward- and downward-pointing triangles; and \ARV{}, shown as a square. For the first two models, the direction of the triangle indicates the assumption of a cored or cuspy DM profile, respectively, while \ARV{} models the DM halo with a profile where the inner slope is allowed to vary; the other two analyses model the DM halo with both a cuspy profile and a cored profile. Additionally, we compare these results to the prediction obtained by the simple scaling relations of the \RPSMHM{}~SMHM relation and the concentration--mass relation from~\citet{Moline23}. The \SatGen{} inference is generally in agreement with these analyses, particularly due to the highly correlated and non-Gaussian errors in the observational analyses. For results on the other dwarfs, see \autoref{fig:A1}.}
\label{fig:3}
\end{figure*}

\subsection{Comparisons to observations}
\label{sec:3.2}
In the previous section, the analysis and systematics were described for the case of Fornax. Below, the scope is broadened to the rest of the classical satellites. \autoref{tab:3} provides the median inferred $\rmax$ and $\vmax$ values (together with 65~per~cent containment intervals) for the nine bright MW spheroidal dwarfs considered in this work and for all four combinations of feedback prescriptions and SMHM relations. The feedback models and choice of SMHM relation affect the broader population in much the same way as described for the Fornax example above, although there is less extreme tension between $M_{1/2}$ and $M_{\star}$ for these satellites and thus higher statistics for all four combinations. The 2D $\rmax - \vmax$ plots for these nine systems are shown in \autoref{fig:A1} for $\thetab = (M_{\star},\, M_{1/2})$, assuming \RPSMHM{} for both the NIHAO and APOSTLE emulators. Additionally, the specific examples of the Fornax and Draco dwarfs are shown in \autoref{fig:3}.\footnote{Since the $\vmax{}$ of a satellite correlates strongly with its mass at infall, it may be of interest to the reader to view the distribution of masses instead. To this end, the inferred infall masses are shown in \autoref{fig:A3} for the \RPSMHM{} SMHM~relation, plotted against the $z = 0$ stellar mass.} It is important to note that the inferences for each MW dwarf are derived independently from one another, with each inference derived from the total population of \SatGen{} satellites from all realizations. These inferences do not enforce consistency with, e.g. the subhalo mass function of individual satellite systems, which would restrict the allowed haloes for a particular dwarf based on the halo inferred for other dwarfs.

As a point of comparison to the contours inferred by using \SatGen{}, one can rely on a combination of empirical scaling relations as a simple way to directly infer a similar distribution in the \rmax{} -- \vmax{} plane. In particular, the total stellar mass $M_\star$ has long been used in abundance matching studies to infer a DM halo mass from stellar observations~\citep{Kravtsov04}. Along with a concentration--mass relation, this is enough to specify a standard two-parameter halo such as the Navarro--Frenk--White~(NFW) profile~\citep{Navarro97}. The grey shaded regions in \autoref{fig:3} correspond to our inference of the distribution using such a technique. This simple inference agrees fairly well with the colored contours (those that are not statistics-limited), as is also illustrated for the full sample of classical satellites in \autoref{fig:A1}. 

These simple grey distributions assume the \RPSMHM{} model as well as the~\citet{Moline23} position-dependent concentration--mass relation for subhaloes, which provides a simple fit to account for the orbital evolution of a satellite's density profile. The SMHM relation is assumed not to change after infall. However, this is inaccurate for a number of reasons. Most importantly, tidal stripping removes DM halo mass from the outside inwards, while stellar mass is generally maintained (except in the most disrupted haloes). Consequently, the simple scaling relations (grey colored region) miss allowed regions of lower-mass haloes, which are properly included when using the \SatGen{} formalism (colored contours). Since the \SatGen{} inference also includes information on the $M_{1/2}$ of the satellites, the most probable regions are shifted from the simple scaling relation prediction, particularly for the case of Fornax, which has an $M_{1/2}$ that tends to prefer lower-mass haloes, as mentioned above. On the other hand, for other satellites (e.g. Draco) the $M_{1/2}$ and $M_\star$ \SatGen{} inferences agree both with each other and  with the simple scaling relation-based inference.

However, the spread in the distributions in all cases are significantly different: \SatGen{} provides realistic, correlated errors, while the simple scaling relation inferences are somewhat more na\"ive in their shape and extent. The uncertainties in the simple scaling relation inference technique come from the intrinsic scatter in the SMHM relation and in the mass-concentration relation. In determining the grey regions in the figure, the~\citet{Moline23}\footnote{In terms of \rmax{}, \vmax{}, and the Hubble parameter $H_0$, the concentration used is $c_V = 2\vmax^2(\rmax H_0)^{-2}$ \citep{Diemand07}.} mass-concentration relation has been given a scatter of 0.33~dex, based on conclusions drawn in the earlier~\citet{Moline17} model. This simple, uncorrelated treatment of uncertainties tends to underestimate the extent and shape of the uncertainties found through the \SatGen{} technique -- such simple estimations ignore the complex non-linear evolution satellites undergo.

The two types of $\rmax{}$ -- $\vmax{}$ inferences described above (using \SatGen{} and simple scaling relations) can also be compared to previous studies in the literature that use internal kinematic data to constrain $\rmax{}$ and $\vmax{}$. The points with error-bars in \autoref{fig:3} show the results of analyses performed by~\citet[][hereafter \ERV{}]{Errani18}, \citet[][\KRV{}]{Kaplinghat19}, and~\citet[][\ARV{}]{Andrade23}. The data points generally align with the \SatGen{} inference, and their spread tends to correlate between \rmax{} and \vmax{} in roughly the same way. The scatter between these observational measurements is often large, suggesting sensitivity to untreated systematic uncertainties in the observational models. This is especially true for Leo~II and Ursa~Minor (see \autoref{fig:A1}), where these observational fits have large scatter and poor agreement with each other, particularly in $\rmax{}$.

The large scatter is likely due, at least in part, to the different approaches used to model the DM halo profile and the stellar velocity anisotropy and density. The profile shapes assumed in each analysis vary somewhat: \ERV{} and \KRV{} both perform Jeans analyses under the assumption of two different DM density profiles, one of which is cored with a flat inner slope and one of which is cuspy with an inner profile of $\rho\propto r^{-1}$, though the exact functional forms used in each study differ. \ARV{} uses a distribution function modelling approach that assumes a profile shape with a flexible inner slope, allowing for the fitting procedure to choose an optimal value. Evidently, the different approaches and assumptions of each study greatly affect the results, which can lead to disagreement with the flexible \SatGen{}-based inference.

By sampling over feedback models, the \SatGen{} analysis allows for variation in the inner slope. In general, the slopes of the inferred profiles from \SatGen{} lie between $\rho\propto r^{-0.5}$ and $r^{-2}$, with the notable exception of Fornax, which is consistent with even shallower slopes in the NIHAO feedback emulator. For each dwarf, the NIHAO feedback emulator prefers a more cored slope than the APOSTLE emulator, which is to be expected. \autoref{fig:A2} shows the distribution of inferred inner slopes for each of the classical dwarfs considered in this study.

\begin{figure*}
	\centering
	\includegraphics{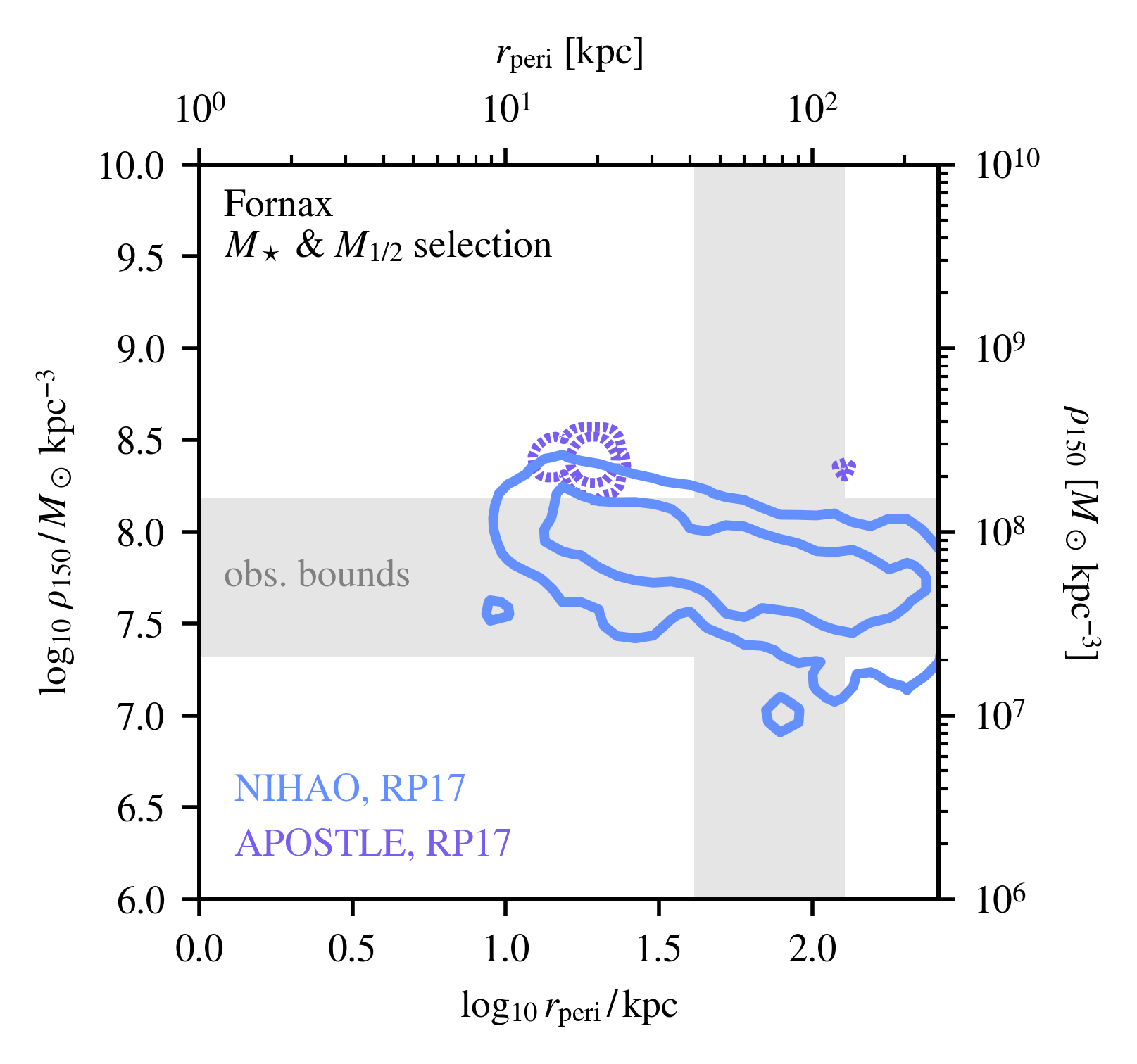}
	\includegraphics{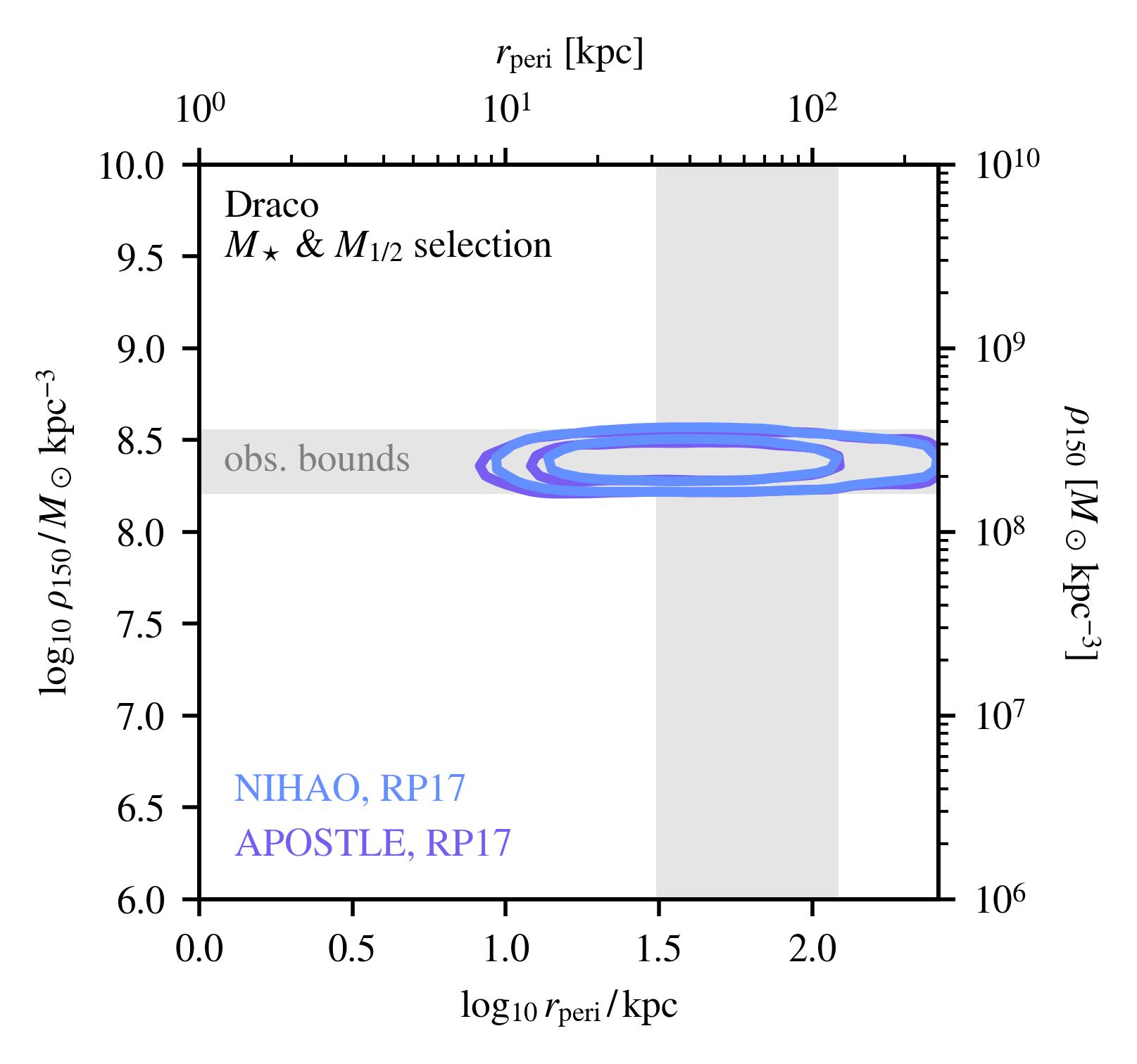}
	\caption{68 and 95~per~cent confidence intervals for the pericentre $r_\mathrm{peri}$ and central density $\rho_\mathrm{150}$ of Fornax (left) and Draco (right), based on the combined $M_\star$ and $M_{1/2}$-based inference. The bands represent the maximum and minimum one-sigma observational results for~\citet{Fritz18}, \citet{Battaglia22}, and~\citet{Pace22} for $r_\mathrm{peri}$; and~\citet{Read19}, \KRV{}, and~\citet{Hayashi20} for $\rho_{150}$, marginalizing over all MW potentials in the pericentre modelling and all assumed profiles (including both cusps and cores) in the density modelling. The \SatGen{} results assume the \RPSMHM{} model and are shown for both the NIHAO and APOSTLE feedback emulators (in blue and purple, respectively). In the case of Fornax, the satellites able to match both masses in the NIHAO feedback emulator also match the observed low central density and fairly large pericentre -- Fornax-like satellites do not remain Fornax-like with low pericentres. Conversely, the APOSTLE feedback emulator requires the tidal stripping enhancement of a low pericentre to match both mass constraints, which brings it in slight tension with observations. Unlike Fornax, the high central density of Draco can be achieved in either feedback emulator, and satellites that match the observed masses also satisfy observational limits on the central density and pericentre.}
\label{fig:4}
\end{figure*}

\section{Inference of Additional Properties}
\label{sec:4}
As shown in \autoref{sec:3}, the method presented in this work allows for the inference of the structural properties of DM haloes based on the stellar mass of the satellite galaxies they host, and this inference can be performed efficiently over many models of baryonic physics and galaxy formation. The inference also agrees with other observational models of the structural properties, even though it does not utilize this information to perform the inference. Below, we extend the applications of the method beyond internal properties of the classical satellites, leveraging the assembly history data provided by the \SatGen{} model. In particular, we infer the relation between the central densities and pericentres of the classical satellites, as well as their association with a larger group of subhaloes at their accretion time.

It should be noted that \SatGen{} does not account for all the physics that a full simulation does. For example, though Press--Schechter theory reproduces assembly histories observed in cosmological simulations, it is not conditioned on the environment (e.g. Local Group-like structures), which may bias the assembly history of the MW. Additionally, the MWs considered here all have $M_\mathrm{vir} = 10^{12}~\mathrm{M}_\odot$ at the present day, which limits the range of possible accretion histories. Gravitational potentials are taken to be spherically symmetric, except for the MW's axisymmetric disc potential. Further, satellites orbit under the influence of their immediate parent object alone, where satellites of satellites do not experience the MW host's gravitational potential. Further, there is no prescription for satellite--satellite gravitational interaction, nor for the reflex motion of the satellite system. Both of these behaviours are known to exist in simulations of MW analogues with large infalling systems like the LMC and can have detectable effects on the satellite population -- see, e.g. \citet{Vasiliev23} and references therein. Recognizing these caveats, one may still consider satellite orbital effects and their connection to internal dynamics.

\subsection{Central density--pericentre relations}
\label{sec:4.1}
An advantage of the semi-analytic orbit integration approach is that it not only provides access to the internal properties of the \SatGen{} satellites, but also to their orbital properties. This allows for the inference of reasonable orbital distributions of these satellites. This inference is grounded in a formalism that does not suffer from the numerical artefacts impacting studies of numerical simulations: the semi-analytic approach allows one to track the satellites in arbitrary environments, without issues such as artificial disruption or low resolution. Additionally, simulations struggle to resolve satellites in high-density environments, e.g. at their orbital pericentres, but this is also where dynamical effects are most important. \SatGen{} self-consistently accounts for these effects, allowing for investigation of the interplay between orbital trajectory and resultant subhalo profile.

\SatGen{} allows for the presence of a disc potential in the orbit integration, which causes more tidal mass loss near pericentre, an effect that is enhanced for smaller pericentres and lower concentrations~\citep{Green22}. The halo profile reacts to the DM mass loss as the subhalo evolves on the~\citet{Errani18} tidal tracks, causing the structural parameters to evolve throughout the orbit. We choose to examine this evolution in terms of the central density at 150~pc, denoted $\rho_{150}$, which may posses some dependence on the orbital trajectory, particularly in the lowest-concentration haloes that are most affected by tides. This is particularly interesting for the case of Fornax, which many lines of argument suggest to have a low-density dark matter core (e.g. \citet{Cole12,Jardel12,Kowalczyk19}, though see~\citet{Boldrini19,Meadows20,Genina22} for some caveats to these arguments). As is demonstrated below, the interplay of central density and orbital pericentre, $r_\mathrm{peri}$, can provide critical insight into the formation of Fornax-like haloes in a manner that is sensitive to the choice of feedback model.

\autoref{fig:4} shows 68 and 95~per~cent containment regions for $\fobs(\xb{})$ in the 2D $r_\mathrm{peri}$ -- $\rho_{150}$ plane for Fornax and Draco (in the language used in \autoref{sec:2.2}, $\xb = (r_\mathrm{peri},\,\rho_{150}$)). This figure shows results for both feedback models and for the \RPSMHM{} SMHM relation. In principle, one could use $\thetab = (M_\star,\,M_{1/2},\,t_\mathrm{infall})$, however we find that $t_\mathrm{infall}$ is too poorly observationally constrained to provide meaningful information and thus choose $\thetab = (M_\star,\,M_{1/2})$. The contours are compared to measurements from~\citet{Fritz18}, \citet{Battaglia22}, and~\citet{Pace22} for $r_\mathrm{peri}$; and~\citet{Read19}, \KRV{}, and~\citet{Hayashi20} for $\rho_{150}$, shown as grey bands in the figure. The width of each of the grey bands corresponds to the maximum and minimum values consistent with observations to within $1\sigma$, marginalizing over all measurements including the choice of cusp or core in $\rho_{150}$ modelling or the choice of potential in modelling the pericentre (note that two of the $r_\mathrm{peri}$ models include an LMC-like partner with the MW potential, and this has a strong influence on the satellites' orbits). Results for all classical satellites are presented in \autoref{fig:A4} and equivalent plots using the \BSMHM{} SMHM relation are presented in \autoref{fig:A5}.

\autoref{fig:4} reveals that the NIHAO feedback emulator cores Fornax-like dwarfs enough to shift the central density by nearly an order of magnitude relative to the APOSTLE feedback emulator. Correspondingly, the pericentre distribution of the Fornax-like satellites in the NIHAO emulator is shifted toward larger values than APOSTLE, suggesting that only extremely significant DM mass loss (with large initial masses) can lead to APOSTLE satellites simultaneously matching the observed $M_\star$ and $M_{1/2}$ values. This in particular provides a sharp insight into the interplay of feedback and orbital evolution: while the NIHAO feedback emulator forms Fornax-like satellites through coring, the APOSTLE model forms them through tidally stripping large-$M_\star$ galaxies until their $M_{1/2}$ is sufficiently low, which is a leading factor in Fornax analogues being so rare in the APOSTLE emulator (the APOSTLE contours for Fornax are dashed because of low statistics, cf.~\autoref{tab:2}). While both formation channels are possible, Fornax-like haloes produced via tidal stripping, as in the APOSTLE emulator, are rarer than those produced by coring in the NIHAO emulator. Due to the unique status of Fornax's mass parameters (i.e. the large $M_\star$ but relatively low $M_{1/2}$), feedback differences are most pronounced in this system. Due to these pronounced differences, it is clear that the NIHAO and APOSTLE emulators form Fornax-like dwarfs in differing ways; however, the distributions for analogues of the smaller dwarfs (\autoref{fig:A4} and \autoref{fig:A5}) generally do not exhibit as strong a difference between the feedback emulators.

While the inferences made on the classical dwarfs' density profiles (\autoref{sec:3}) are broadly consistent with observational models, the inference of Fornax's central density and pericentre in particular provide constraining information on the semi-analytic models examined here. Due to the discrepancy shown in \autoref{fig:4}, it seems that the APOSTLE feedback emulator requires a particular orbital history to simultaneously explain the combined observations of Fornax's $\rho_{150}$, $r_\mathrm{peri}$, $M_\star$, and $M_{1/2}$, with the caveat that the \SatGen{} model makes many simplifying assumptions regarding the orbital properties, enumerated at the start of the section. Even given these assumptions, such a multidimensional combined analysis is difficult to perform without semi-analytic techniques that allow for efficient scanning over the many sources of uncertainty present in galaxy formation. The semi-analytic technique reveals how rare it is to produce a dwarf such as Fornax under these modelling assumptions.

\begin{figure}
	\centering
	\includegraphics{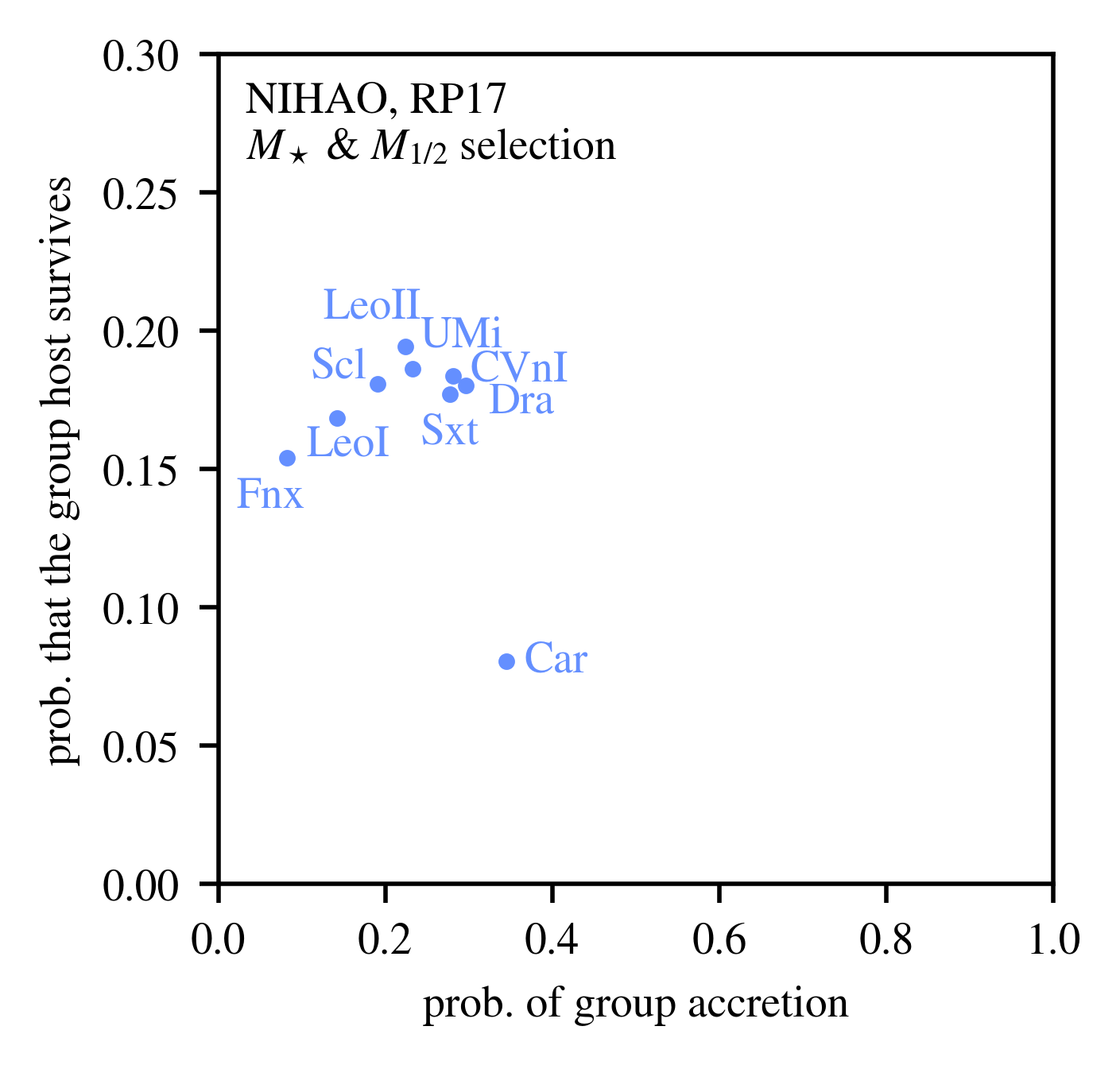}
	\caption{The probability that any given classical satellite is accreted as part of a group rather than directly from the field (as inferred though the $\thetab = (M_\star,\,M_{1/2})$ selection). The vertical axis shows the probability that, given a group-accretion scenario, the original group host persists as a satellite of the MW. While the dwarfs presented here are unlikely to have been substructure at infall, in the case that they \emph{were}, it is unlikely that the group host survived to the present day.}
\label{fig:5}
\end{figure}

\subsection{Comparison of accretion modes}
\label{sec:4.2}
Lastly, we use \SatGen{} to examine the likelihood that each of the classical satellites were directly accreted on to the MW, as opposed to entering as part of a larger group. \SatGen{} resolves substructure due to hierarchical formation of the infalling satellite groups, and during orbital evolution, a satellite group can release some of its substructure into the MW. This release has a chance to occur whenever substructure passes outside the virial radius of its parent, with probability inversely proportional to the local dynamical time of the parent. Intuitively, the dynamical time at the location of the parent is the time-scale on which the parent is tidally stripped; release of higher-order substructure is akin to tidal stripping.

A comparison of the chances of various accretion modes is shown in \autoref{fig:5} -- the abscissa shows the probability that a satellite is accreted with a group (as opposed to being accreted directly from the field), and the ordinate axis shows the probability that the group host survives to the present day in the case that the satellite has a parent. The satellites are selected to match the properties of the classical MW dwarfs using the $\thetab = (M_\star,\,M_{1/2})$ inference.\footnote{While $M_{1/2}$ does not contribute much information to this inference, it is used in the selection for consistency.} The \SatGen{} analysis suggests that the classical satellites overwhelmingly form in the field, with a 23~per~cent chance, on average, of entering the MW within a larger substructure. In general, \SatGen{} predicts that satellites that match the classical dwarfs in $M_\star$ and $M_{1/2}$ and come in with a group only have a 16~per~cent chance of the group surviving, corresponding to an overall 4~per~cent absolute chance of having a surviving group host.

Group-accretion events are known to have an impact on the resulting satellite distribution, both in terms of the count of satellites and in their spatial distribution~\citep{Smercina21,DSouza21}. \autoref{fig:6} illustrates how the properties of the \SatGen{} satellites vary depending on their origin. For example, the top right-hand panel shows the inferred distance of closest approach of each satellite to the MW's centre (denoted $r_\mathrm{peri}$, but see caveat below), both for satellites accreted directly~(blue) and for those accreted as part of a larger group~(red). The bottom right-hand panel shows the inferred mass loss in terms of the satellite's present-day virial mass, $M_\mathrm{vir}$, relative to its peak mass, $M_\mathrm{peak}$. The latter is defined when the satellite exits the field and is accreted on to the host -- either the MW, in the case of direct accretion, or the group host, in the case of group accretion.

As shown in the top right-hand panel of \autoref{fig:6}, the pericentre distribution peaks at roughly the same value for both the direct and group-accretion cases; however the former has an extended tail towards larger $r_\mathrm{peri}$. This is also observed as an extended tail at large $M_\mathrm{vir}/M_\mathrm{peak}$ in the bottom right-hand panel of the figure. The extended tails in the direct-accretion satellites arise from the fact that these satellites are considered first-order as soon as they cross within the MW's virial radius and are thus always included in our sample. In contrast, for the group-accretion scenario, the group host is the first-order satellite at infall, while objects bound to it are considered higher-order satellites of the MW. This higher-order structure must be released from the group to be included in the satellite sample considered here. As a result of this distinction, there is a population of recently accreted satellites that is present in the direct-infall population but not in the group-infall population. These satellites have little mass loss and have not completed a pericentric passage by $z=0$. Their location at $z=0$ is then their closest approach to the MW, but it is not truly their orbital pericentre. On the other hand, since groups tend to shed their satellites most efficiently near their own pericentres, the satellites contributed by such groups typically have experienced much closer approaches. In the lower panel, the mass loss of group-accreted satellites is enhanced, since they will undergo tidal stripping even before reaching the MW. Even when restricting the direct-infall population to those that have completed at least one pericentric passage, the group-accretion case prefers more recent infall times, smaller pericentres, and greater mass loss.

As a further exploration of the tidal mass loss, the left-hand panel of \autoref{fig:6} shows the  infall time into the MW, $t_\mathrm{infall}$, and the inferred mass at this time, $M_\mathrm{infall}$, for the two satellite populations. Due to the time spent evolving in the potential of their host, group-accretion satellites have a lower $M_\mathrm{infall}$ than those that are directly accreted, an effect exacerbated for satellites with more recent $t_\mathrm{infall}$ that evolved for longer within the group. The time of infall itself is also affected by the orbital dynamics of the accreting group. At very recent infall times, accretion is dominated by directly-accreting first-order satellites, since any infalling groups have not yet been stripped enough to release substructure into the MW. At slightly larger infall times there is a lull in the direct accretion, which corresponds to satellites that are at or near the apocentre of their orbit~\citep[e.g.][]{Ludlow09,Yun19,Bakels21,Engler21}. While dynamical friction efficiently pulls the groups toward low pericentres, directly-accreted satellites can maintain higher orbital energies and even exit the virial radius, becoming `splashback' satellites. Because we select only satellites with present-day distance $D_\mathrm{MW}$ within the MW's virial radius $R_\mathrm{vir}$, splashback satellites are removed from our sample. \autoref{fig:7} demonstrates this explicitly, by plotting the distribution of infall times for both direct and group-accretion satellites, as well as the splashback satellites with $D_\mathrm{MW} > R_\mathrm{vir}$. The bimodality of $t_\mathrm{infall}$ is clear in the blue direct-accretion distribution, but adding in the population of splashback galaxies (black) fills in the gap between the two infall times. For completeness, the group-accreted satellites are also included in the probability distribution~(red), since these three populations together comprise the entire set of surviving satellites.

\begingroup
\setlength{\textfloatsep}{0pt}
\setlength{\intextsep}{0pt}
\begin{figure*}
	\centering
	\includegraphics{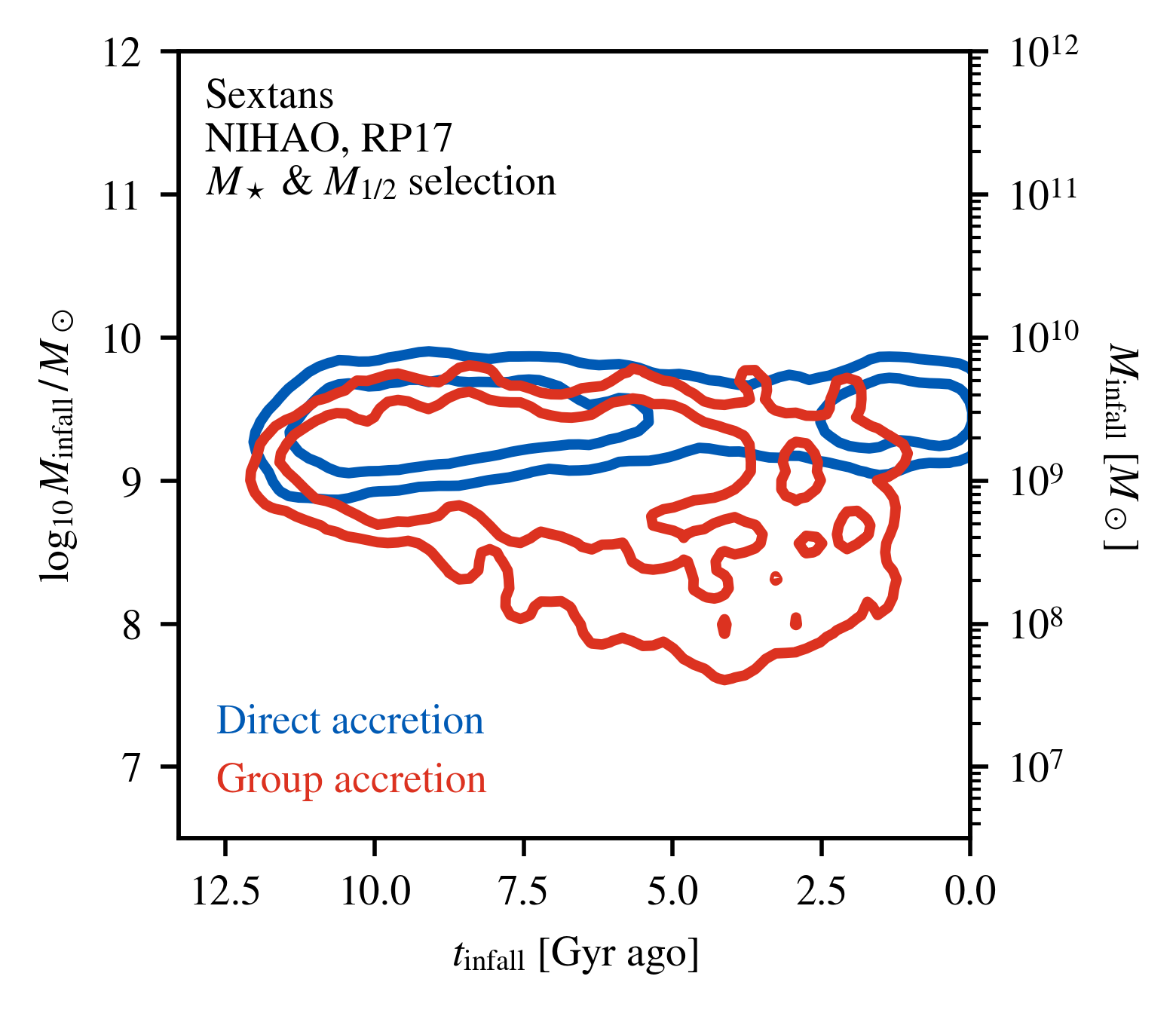}
	\includegraphics{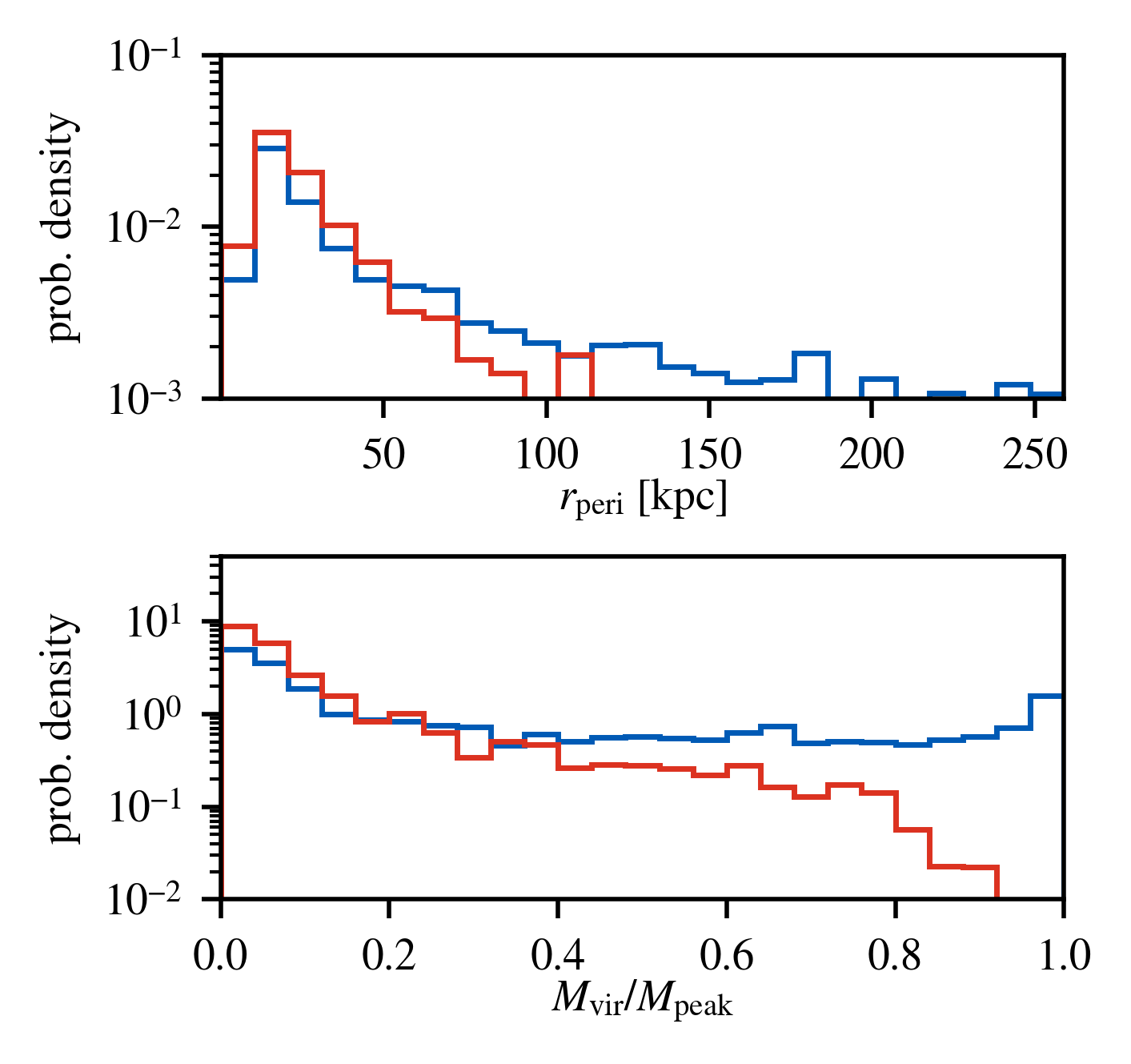}
	\caption{A demonstration of how a satellite galaxy's properties depend on  whether it is accreted directly onto the MW, shown in blue, or as part of a larger group, shown in red. (Left:) The inferred infall time and virial mass at infall as derived from directly- and group-accreted satellites based on their $M_\star$ and $M_{1/2}$, where the infall time is calculated as the time since the satellite (or the satellite group) crossed the virial radius of the MW. The directly-accreted satellites show a bimodality in this parameter space, which can be understood due to the population of splashback galaxies (see \autoref{fig:7} and discussion in the text). (Right:) The one-dimensional inferred probability distributions for $r_\mathrm{peri}$ (in the upper panel) and $M_\mathrm{vir}/M_\mathrm{peak}$ (in the lower panel). Satellites accreted as part of a group generally have lower pericentres and greater mass loss than directly-accreted satellites, a trend enhanced by the fact that there are no recently-accreted haloes in this population; for directly-accreted satellites that have yet to complete a full orbit, their closest approach to the MW is recorded as their pericentre, and these do not lose much mass to tidal stripping. The mass loss that takes place in the group, prior to MW infall, is evident in the left-hand panel; the group-accretion contour extends to lower masses than the direct-accretion contour. These trends are clear in Sextans, so it is used as an example, though \autoref{fig:A6} contains corresponding plots for each of the other classical satellites.}
\label{fig:6}
\vspace{-\baselineskip}
\end{figure*}
\begin{figure}
	\centering
	\includegraphics{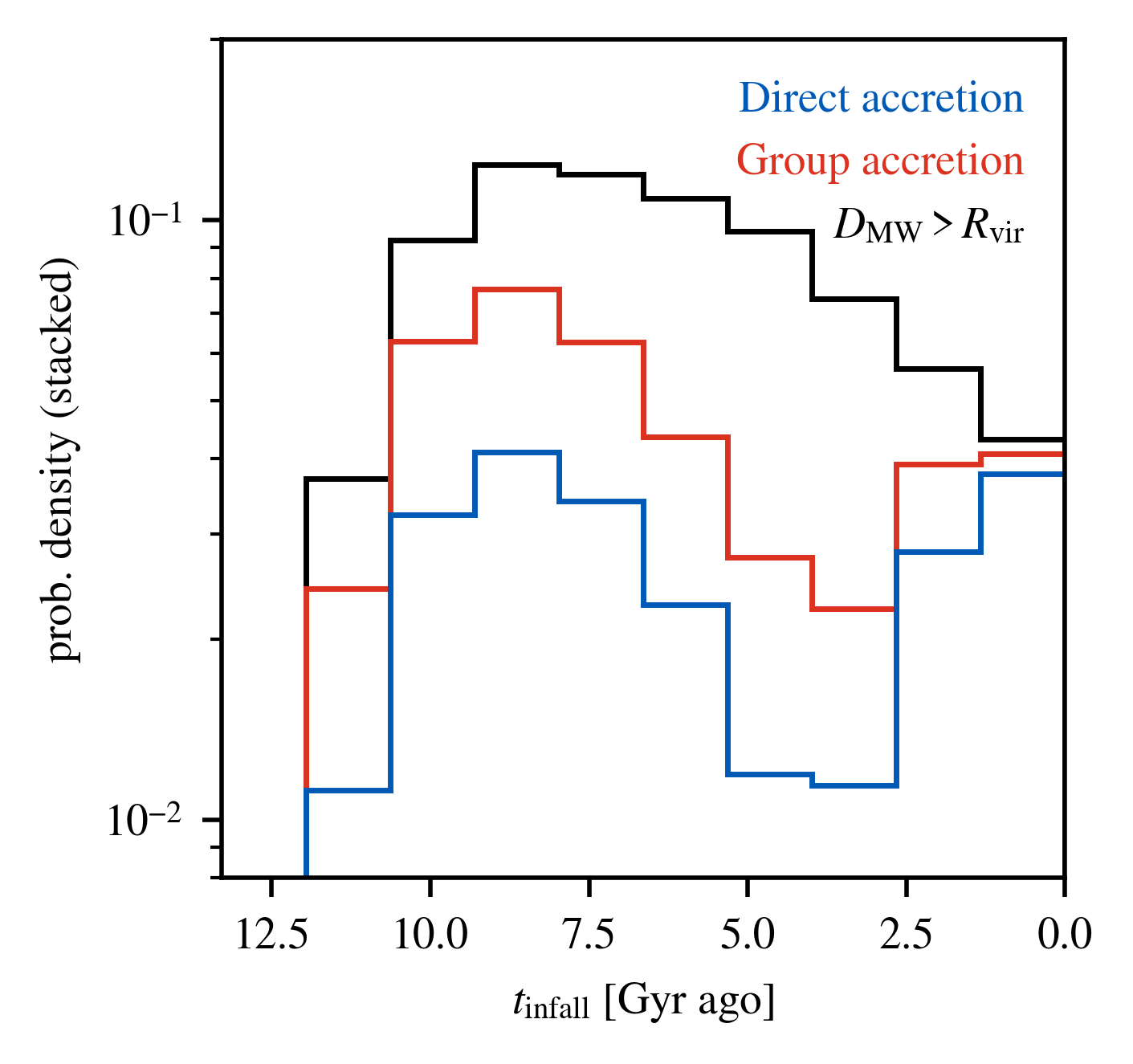}
	\caption{The distribution of infall times of all surviving satellites, split into three components, which are shown stacked atop each other. Of the satellites that pass the cuts enumerated in \autoref{sec:2.1}, those that are directly accreted are shown in blue; including those accreted as part of a group yields the red curve. Atop these is the set of `splashback galaxies,' i.e. satellites that are located at a distance from the MW, $D_\mathrm{MW}$, that is beyond its virial radius, $R_\mathrm{vir}$. Many satellites that fell in $\gtrsim 3$~Gyr ago are currently approaching their apocentres and may be instantaneously outside the virial radius of the MW at the present day. The requirement that $D_\mathrm{MW}\leq R_\mathrm{vir}$ therefore removes these splashback galaxies from our sample, leading to an apparent bimodality in the infall times of the surviving satellites in \autoref{fig:6}.}
	\label{fig:7}
\end{figure}
\endgroup 

While this study considers group accretion in general, these results may also be applied to the specific case of LMC-like groups. The LMC has a mass $\gtrsim 10^{11}~\mathrm{M}_\odot$ \citep{Vasiliev23}, so an interesting comparison is to restrict attention to satellites in MW realizations that have an LMC-like satellite, i.e. one which survives to $z = 0$ with a mass $M_\mathrm{vir} \geq 10^{11}~\mathrm{M}_\odot$.\footnote{Note that this is not conditioned on orbital properties of the LMC-like satellite itself; these LMCs may not share similar orbits to the LMC orbiting our Galaxy. In fact, only 21 per cent of the LMCs in this sample have completed a pericentric passage.} 
LMC-like satellites are found in $\osim 10$~per~cent of the MW systems, although an additional $\osim1.6$~per~cent have an LMC that has exited the MW's virial radius. We find that the classical dwarfs in \SatGen{} have a $\osim 0.5$ per cent probability of originating from a surviving LMC, consistent with orbital studies of these dwarfs~\citep{Sales11,Erkal20,Patel20,CorreaMagnus22,Battaglia22,Pace22}. While generally the classical satellites are inferred to have infall times $t_\mathrm{infall}\sim 7$~Gyr ago,\footnote{With the exception of Fornax, for which $t_\mathrm{infall} \sim 5$~Gyr ago.}
this shifts to more recent values of $t_\mathrm{infall} \lesssim 4$~Gyr ago when considering only haloes that originated in LMC-like group accretion events, since LMC-mass host haloes take time to assemble and therefore would not be present in the early Universe. In accordance with the lower inferred $t_\mathrm{infall}$, the inferred pericentres for the classical satellites are shifted to larger distances, although the inferred $r_\mathrm{peri}$ in the LMC-contributed cases are often higher than other group-accretion scenarios with the same $t_\mathrm{infall}$. Satellites from an LMC-like group are also inferred to be closer to their $M_\mathrm{peak}$ than in the general group-accretion case. However, these trends should be understood with the warnings that (i) LMC infall events are rare and thus statistics-limited, and (ii) the \SatGen{} model assumptions enumerated at the start of the section will also generically impact these results.

\section{Conclusions}
\label{sec:5}
This work presents a novel method for inferring properties of MW dwarf galaxies using semi-analytic models of satellite systems. The foundation of the approach is a statistical sample of MW satellites generated using the semi-analytic model \SatGen{}. This large sample, which almost completely captures cosmological halo-to-halo variance, can be produced quickly with \SatGen{}, in contrast to \emph{N}-body or hydrodynamical codes. One can then select in a principled way the satellites that have properties matching well-constrained observational parameters of MW dwarfs, such as their stellar mass $M_{\star}$ or total mass $M_{1/2}$ within the half-light radius. The distribution of selected \SatGen{} satellites constitutes an inference for unknown properties of the MW dwarf itself. 

We applied this method to three sets of inferences for the nine bright `classical dwarfs' of the MW. In particular, the procedure was used to:
\begin{enumerate}
	\item infer the structural parameters $(\rmax,\,\vmax)$ describing the DM haloes of the dwarfs, accounting for systematic uncertainties present in this inference due to the SMHM relation and feedback modelling. The results of this study, which are compiled in \autoref{tab:3}, generally agree with Jeans analyses and predictions from empirical scaling relations, but include a more comprehensive modelling of systematic uncertainties and a realistic, correlated uncertainty band on the inferred halo parameters.
	\item infer the correlations between orbital parameters of a dwarf galaxy and its internal properties. Importantly, we showed that the correlation between a galaxy's central density and pericentre provides a means to constrain uncertainties in the feedback modelling. For example, the only satellites that simultaneously matched the observed $M_\star$ and $M_{1/2}$ of Fornax in the APOSTLE feedback emulator (which prefers cuspier haloes) had lower pericentre and higher central density than expected from observations. On the other hand, the NIHAO feedback emulator, which cores satellite galaxies more efficiently, yielded a population of high-$M_\star$, low-$M_{1/2}$ galaxies in much better agreement with the observed parameters for Fornax. This preference for the stronger feedback model is evinced not only by the agreement of the central densities but also in the statistical prevalence: there are very few satellites in the APOSTLE emulator sample that accurately reproduce both the $M_\star$ and $M_{1/2}$ of Fornax, but the NIHAO emulator produces many more, by two orders of magnitude.
	\item infer the likelihood that a given dwarf is associated with the LMC or other larger structure at the time of its accretion, and to infer the differences between this case and the case of direct accretion from the field. Based solely on the present-day $M_\star$ and $M_{1/2}$ of the classical satellites, we found no convincing evidence that any were associated with larger systems, in agreement with detailed orbital modelling studies. In the event that they were, the inferred properties are more consistent with smaller pericentres and greater mass loss.
\end{enumerate}

In each of the above scenarios, the space of MW satellites was constrained using only observations of $M_\star$ and $M_{1/2}$ and found to be in good agreement with analyses that leveraged more difficult observations of internal kinematics or systemic motions. Despite the successes of the method, it is important to recognize assumptions that have been made to enable efficient computation. While \SatGen{} is well-calibrated to reproduce results from idealized and cosmological simulations, it still relies on a number of simplifying assumptions and empirical fits. For example, satellites orbit only under the influence of their immediate parent; satellites of the same host have no impact on each other; and there is no prescription for the reflex motion of the parent due to LMC-like satellites. Further, DM potentials are modelled as perfectly spherical and follow a particular functional form, with the only non-spherical component being the MW disc.

As our understanding of galaxy formation grows, semi-analytic models will continue to improve in accuracy and complexity, allowing for sharper analyses across a broader domain of parameters. One particularly promising direction is the use of semi-analytic satellite generators to constrain feedback prescriptions. As our results demonstrate, correlations between orbital parameters and internal properties of a dwarf galaxy can be used to distinguish between different feedback emulators (specifically, NIHAO and APOSTLE). An even more powerful and generic approach would be to parametrize the feedback model within \SatGen{}, and to then directly constrain the feedback parameters on data using a full likelihood analysis. With analytic models for gas ejection~\citep[e.g.][]{Li23}, \SatGen{} could have a more physically-motivated parametrization for the response of the DM halo to gas ejection, rather than simply changing the inner slope and concentration. Such developments are on the horizon for semi-analytic models like \SatGen{}, and will allow for even more robust study of baryonic feedback and the galaxy--halo connection.
 
\vspace{-\baselineskip}
\section*{Acknowledgements}
The authors gratefully acknowledge Kassidy Kollmann, Sandip Roy, and Adriana Dropulic for helpful conversations. ML and DF are supported by the Department of Energy~(DOE) under Award Number DE-SC0007968. ML and OS are supported by the Binational Science Foundation (grant No. 2018140). ML also acknowledges support from the Simons Investigator in Physics Award. OS is also supported by the NSF (grant~No.~PHY-2210498) and acknowledges support from the Yang Institute for Theoretical Physics. ML and OS acknowledge the Simons Foundation for support. MK is supported by the NSF (grant~No.~PHY-2210283). This work was performed in part at Aspen Center for Physics, which is supported by National Science Foundation grant PHY-2210452.

The work presented in this paper was also performed on computational resources managed and supported by Princeton Research Computing. This research made extensive use of the publicly available codes
\texttt{IPython}~\citep{Perez07}, 
\texttt{matplotlib}~\citep{Hunter07}, 
\texttt{scikit-learn}~\citep{Pedregosa11},
\texttt{Jupyter}~\citep{Kluyver16}, 
\texttt{NumPy}~\citep{Harris20}, 
\texttt{SciPy}~\citep{Virtanen20}, 
and
\texttt{astropy}~\citep{AstropyCollaboration22}.
\vspace{-\baselineskip}

\section*{Data Availability}
The data underlying this article are available via Zenodo at \href{https://zenodo.org/doi/10.5281/zenodo.10068111}{\texttt{https://zenodo.org/doi/10.5281/zenodo.10068111}}. The code used to perform the inferences in this work is publicly available in the GitHub repository \href{https://github.com/folsomde/Semianalytic_Inference}{\texttt{https://github.com/folsomde/Semianalytic\char`_Inference}}. 
\vspace{-\baselineskip}

\bibliographystyle{mnras}
\DeclareRobustCommand{\VAN}[2]{#2}
\bibliography{main}

\begin{thebibliography}{}
\makeatletter
\relax
\def\mn@urlcharsother{\let\do\@makeother \do\$\do\&\do\#\do\^\do\_\do\%\do\~}
\def\mn@doi{\begingroup\mn@urlcharsother \@ifnextchar [ {\mn@doi@}
  {\mn@doi@[]}}
\def\mn@doi@[#1]#2{\def\@tempa{#1}\ifx\@tempa\@empty \href
  {http://dx.doi.org/#2} {doi:#2}\else \href {http://dx.doi.org/#2} {#1}\fi
  \endgroup}
\def\mn@eprint#1#2{\mn@eprint@#1:#2::\@nil}
\def\mn@eprint@arXiv#1{\href {http://arxiv.org/abs/#1} {{\tt arXiv:#1}}}
\def\mn@eprint@dblp#1{\href {http://dblp.uni-trier.de/rec/bibtex/#1.xml}
  {dblp:#1}}
\def\mn@eprint@#1:#2:#3:#4\@nil{\def\@tempa {#1}\def\@tempb {#2}\def\@tempc
  {#3}\ifx \@tempc \@empty \let \@tempc \@tempb \let \@tempb \@tempa \fi \ifx
  \@tempb \@empty \def\@tempb {arXiv}\fi \@ifundefined
  {mn@eprint@\@tempb}{\@tempb:\@tempc}{\expandafter \expandafter \csname
  mn@eprint@\@tempb\endcsname \expandafter{\@tempc}}}

\bibitem[\protect\citeauthoryear{{Akita} \& {Ando}}{{Akita} \&
  {Ando}}{2023}]{Akita23}
{Akita} K.,  {Ando} S.,  2023, \mn@doi [JCAP] {10.1088/1475-7516/2023/11/037},
  2023, 037

\bibitem[\protect\citeauthoryear{Ando, {Geringer-Sameth}, Hiroshima, Hoof,
  Trotta  \& Walker}{Ando et~al.}{2020}]{Ando20}
Ando S.,  {Geringer-Sameth} A.,  Hiroshima N.,  Hoof S.,  Trotta R.,   Walker
  M.~G.,  2020, \mn@doi [Phys. Rev. D] {10.1103/PhysRevD.102.061302}, 102,
  061302

\bibitem[\protect\citeauthoryear{Andrade, Kaplinghat  \& Valli}{Andrade
  et~al.}{2024}]{Andrade23}
Andrade K.~E.,  Kaplinghat M.,   Valli M.,  2024, \mn@doi [MNRAS]
  {10.1093/mnras/stae1716}, 532, 4157

\bibitem[\protect\citeauthoryear{{Astropy Collaboration} et~al.,}{{Astropy
  Collaboration} et~al.}{2022}]{AstropyCollaboration22}
{Astropy Collaboration} et~al., 2022, \mn@doi [ApJ] {10.3847/1538-4357/ac7c74},
  \href {https://ui.adsabs.harvard.edu/abs/2022ApJ...935..167A} {935, 167}

\bibitem[\protect\citeauthoryear{Bakels, Ludlow  \& Power}{Bakels
  et~al.}{2021}]{Bakels21}
Bakels L.,  Ludlow A.~D.,   Power C.,  2021, \mn@doi [MNRAS]
  {10.1093/mnras/staa3979}, 501, 5948

\bibitem[\protect\citeauthoryear{Barber, Starkenburg, Navarro, McConnachie  \&
  Fattahi}{Barber et~al.}{2014}]{Barber14}
Barber C.,  Starkenburg E.,  Navarro J.~F.,  McConnachie A.~W.,   Fattahi A.,
  2014, \mn@doi [MNRAS] {10.1093/mnras/stt1959}, 437, 959

\bibitem[\protect\citeauthoryear{Battaglia, Taibi, Thomas  \& Fritz}{Battaglia
  et~al.}{2022}]{Battaglia22}
Battaglia G.,  Taibi S.,  Thomas G.~F.,   Fritz T.~K.,  2022, \mn@doi [A\&A]
  {10.1051/0004-6361/202141528}, 657, A54

\bibitem[\protect\citeauthoryear{Behroozi, Conroy  \& Wechsler}{Behroozi
  et~al.}{2010}]{Behroozi10}
Behroozi P.~S.,  Conroy C.,   Wechsler R.~H.,  2010, \mn@doi [ApJ]
  {10.1088/0004-637X/717/1/379}, 717, 379

\bibitem[\protect\citeauthoryear{Behroozi, Wechsler  \& Conroy}{Behroozi
  et~al.}{2013}]{Behroozi13}
Behroozi P.~S.,  Wechsler R.~H.,   Conroy C.,  2013, \mn@doi [ApJ]
  {10.1088/0004-637X/770/1/57}, 770, 57

\bibitem[\protect\citeauthoryear{Behroozi, Wechsler, Hearin  \&
  Conroy}{Behroozi et~al.}{2019}]{Behroozi19}
Behroozi P.,  Wechsler R.~H.,  Hearin A.~P.,   Conroy C.,  2019, \mn@doi
  [MNRAS] {10.1093/mnras/stz1182}, 488, 3143

\bibitem[\protect\citeauthoryear{{Benitez-Llambay} \& Frenk}{{Benitez-Llambay}
  \& Frenk}{2020}]{Benitez-Llambay20}
{Benitez-Llambay} A.,  Frenk C.,  2020, \mn@doi [MNRAS]
  {10.1093/mnras/staa2698}, 498, 4887

\bibitem[\protect\citeauthoryear{{Bland-Hawthorn} \& Gerhard}{{Bland-Hawthorn}
  \& Gerhard}{2016}]{Bland-Hawthorn16}
{Bland-Hawthorn} J.,  Gerhard O.,  2016, \mn@doi [ARA\&A]
  {10.1146/annurev-astro-081915-023441}, 54, 529

\bibitem[\protect\citeauthoryear{Boldrini, Mohayaee  \& Silk}{Boldrini
  et~al.}{2019}]{Boldrini19}
Boldrini P.,  Mohayaee R.,   Silk J.,  2019, \mn@doi [MNRAS]
  {10.1093/mnras/stz573}, 485, 2546

\bibitem[\protect\citeauthoryear{Brooks, Kuhlen, Zolotov  \& Hooper}{Brooks
  et~al.}{2013}]{Brooks13}
Brooks A.~M.,  Kuhlen M.,  Zolotov A.,   Hooper D.,  2013, \mn@doi [ApJ]
  {10.1088/0004-637X/765/1/22}, 765, 22

\bibitem[\protect\citeauthoryear{Bryan \& Norman}{Bryan \&
  Norman}{1998}]{Bryan98}
Bryan G.~L.,  Norman M.~L.,  1998, \mn@doi [ApJ] {10.1086/305262}, 495, 80

\bibitem[\protect\citeauthoryear{Chandrasekhar}{Chandrasekhar}{1943}]{Chandrasekhar43}
Chandrasekhar S.,  1943, \mn@doi [ApJ] {10.1086/144517}, 97, 255

\bibitem[\protect\citeauthoryear{Cole, Dehnen, Read  \& Wilkinson}{Cole
  et~al.}{2012}]{Cole12}
Cole D.~R.,  Dehnen W.,  Read J.~I.,   Wilkinson M.~I.,  2012, \mn@doi [MNRAS]
  {10.1111/j.1365-2966.2012.21885.x}, 426, 601

\bibitem[\protect\citeauthoryear{Conroy, Wechsler  \& Kravtsov}{Conroy
  et~al.}{2006}]{Conroy06}
Conroy C.,  Wechsler R.~H.,   Kravtsov A.~V.,  2006, \mn@doi [ApJ]
  {10.1086/503602}, 647, 201

\bibitem[\protect\citeauthoryear{Correa~Magnus \& Vasiliev}{Correa~Magnus \&
  Vasiliev}{2022}]{CorreaMagnus22}
Correa~Magnus L.,  Vasiliev E.,  2022, \mn@doi [MNRAS]
  {10.1093/mnras/stab3726}, 511, 2610

\bibitem[\protect\citeauthoryear{D'Souza \& Bell}{D'Souza \&
  Bell}{2021}]{DSouza21}
D'Souza R.,  Bell E.~F.,  2021, \mn@doi [MNRAS] {10/gph5x4}, 504, 5270

\bibitem[\protect\citeauthoryear{{Danieli}, {Greene}, {Carlsten}, {Jiang},
  {Beaton}  \& {Goulding}}{{Danieli} et~al.}{2023}]{Danieli22}
{Danieli} S.,  {Greene} J.~E.,  {Carlsten} S.,  {Jiang} F.,  {Beaton} R.,
  {Goulding} A.~D.,  2023, \mn@doi [ApJ] {10.3847/1538-4357/acefbd}, 956, 6

\bibitem[\protect\citeauthoryear{Dekel, Ishai, Dutton  \& Maccio}{Dekel
  et~al.}{2017}]{Dekel17}
Dekel A.,  Ishai G.,  Dutton A.~A.,   Maccio A.~V.,  2017, \mn@doi [MNRAS]
  {10.1093/mnras/stx486}, 468, 1005

\bibitem[\protect\citeauthoryear{Dekker, Ando, Correa  \& Ng}{Dekker
  et~al.}{2022}]{Dekker22}
Dekker A.,  Ando S.,  Correa C.~A.,   Ng K. C.~Y.,  2022, \mn@doi [Phys. Rev.
  D] {10.1103/PhysRevD.106.123026}, 106, 123026

\bibitem[\protect\citeauthoryear{Diemand, Kuhlen  \& Madau}{Diemand
  et~al.}{2007}]{Diemand07}
Diemand J.,  Kuhlen M.,   Madau P.,  2007, \mn@doi [ApJ] {10.1086/520573}, 667,
  859

\bibitem[\protect\citeauthoryear{Engler et~al.,}{Engler
  et~al.}{2021}]{Engler21}
Engler C.,  et~al., 2021, \mn@doi [MNRAS] {10.1093/mnras/staa3505}, 500, 3957

\bibitem[\protect\citeauthoryear{Erkal \& Belokurov}{Erkal \&
  Belokurov}{2020}]{Erkal20}
Erkal D.,  Belokurov V.~A.,  2020, \mn@doi [MNRAS] {10.1093/mnras/staa1238},
  495, 2554

\bibitem[\protect\citeauthoryear{Errani, Pe{\~n}arrubia  \& Walker}{Errani
  et~al.}{2018}]{Errani18}
Errani R.,  Pe{\~n}arrubia J.,   Walker M.~G.,  2018, \mn@doi [MNRAS]
  {10.1093/mnras/sty2505}, 481, 5073

\bibitem[\protect\citeauthoryear{Fattahi, Navarro, Frenk, Oman, Sawala  \&
  Schaller}{Fattahi et~al.}{2018}]{Fattahi18}
Fattahi A.,  Navarro J.~F.,  Frenk C.~S.,  Oman K.~A.,  Sawala T.,   Schaller
  M.,  2018, \mn@doi [MNRAS] {10.1093/mnras/sty408}, 476, 3816

\bibitem[\protect\citeauthoryear{Fillingham et~al.,}{Fillingham
  et~al.}{2019}]{Fillingham19}
Fillingham S.~P.,  et~al., 2019, preprint, \mn@doi{10.48550/arXiv.1906.04180}

\bibitem[\protect\citeauthoryear{Font et~al.,}{Font et~al.}{2011}]{Font11}
Font A.~S.,  et~al., 2011, \mn@doi [MNRAS] {10.1111/j.1365-2966.2011.19339.x},
  417, 1260

\bibitem[\protect\citeauthoryear{Freundlich et~al.,}{Freundlich
  et~al.}{2020}]{Freundlich20}
Freundlich J.,  et~al., 2020, \mn@doi [MNRAS] {10.1093/mnras/staa2790}, 499,
  2912

\bibitem[\protect\citeauthoryear{Fritz, Battaglia, Pawlowski, Kallivayalil,
  {van der Marel}, Sohn, Brook  \& Besla}{Fritz et~al.}{2018}]{Fritz18}
Fritz T.~K.,  Battaglia G.,  Pawlowski M.~S.,  Kallivayalil N.,  {van der
  Marel} R.,  Sohn S.~T.,  Brook C.,   Besla G.,  2018, \mn@doi [A\&A]
  {10.1051/0004-6361/201833343}, 619, A103

\bibitem[\protect\citeauthoryear{{Garrison-Kimmel}, Bullock, {Boylan-Kolchin}
  \& Bardwell}{{Garrison-Kimmel} et~al.}{2017a}]{Garrison-Kimmel17a}
{Garrison-Kimmel} S.,  Bullock J.~S.,  {Boylan-Kolchin} M.,   Bardwell E.,
  2017a, \mn@doi [MNRAS] {10.1093/mnras/stw2564}, 464, 3108

\bibitem[\protect\citeauthoryear{{Garrison-Kimmel} et~al.,}{{Garrison-Kimmel}
  et~al.}{2017b}]{Garrison-Kimmel17b}
{Garrison-Kimmel} S.,  et~al., 2017b, \mn@doi [MNRAS] {10.1093/mnras/stx1710},
  471, 1709

\bibitem[\protect\citeauthoryear{Genina, Read, Fattahi  \& Frenk}{Genina
  et~al.}{2022}]{Genina22}
Genina A.,  Read J.~I.,  Fattahi A.,   Frenk C.~S.,  2022, \mn@doi [MNRAS]
  {10.1093/mnras/stab3526}, 510, 2186

\bibitem[\protect\citeauthoryear{Gnedin, Kravtsov, Klypin  \& Nagai}{Gnedin
  et~al.}{2004}]{Gnedin04}
Gnedin O.~Y.,  Kravtsov A.~V.,  Klypin A.~A.,   Nagai D.,  2004, \mn@doi [ApJ]
  {10.1086/424914}, 616, 16

\bibitem[\protect\citeauthoryear{Green \& {\VAN{Van den}{van den}}~Bosch}{Green
  \& {\VAN{Van den}{van den}}~Bosch}{2019}]{Green19}
Green S.~B.,  {\VAN{Van den}{van den}}~Bosch F.~C.,  2019, \mn@doi [MNRAS]
  {10.1093/mnras/stz2767}, 490, 2091

\bibitem[\protect\citeauthoryear{Green, {\VAN{Van den}{van den}}~Bosch  \&
  Jiang}{Green et~al.}{2022}]{Green22}
Green S.~B.,  {\VAN{Van den}{van den}}~Bosch F.~C.,   Jiang F.,  2022, \mn@doi
  [MNRAS] {10.1093/mnras/stab3130}, 509, 2624

\bibitem[\protect\citeauthoryear{Guo et~al.,}{Guo et~al.}{2011}]{Guo11}
Guo Q.,  et~al., 2011, \mn@doi [MNRAS] {10.1111/j.1365-2966.2010.18114.x}, 413,
  101

\bibitem[\protect\citeauthoryear{Guo, Cooper, Frenk, Helly  \& Hellwing}{Guo
  et~al.}{2015}]{Guo15}
Guo Q.,  Cooper A.~P.,  Frenk C.,  Helly J.,   Hellwing W.~A.,  2015, \mn@doi
  [MNRAS] {10.1093/mnras/stv1938}, 454, 550

\bibitem[\protect\citeauthoryear{Harris et~al.,}{Harris
  et~al.}{2020}]{Harris20}
Harris C.~R.,  et~al., 2020, \mn@doi [Nature] {10.1038/s41586-020-2649-2}, 585,
  357

\bibitem[\protect\citeauthoryear{Hayashi, Chiba  \& Ishiyama}{Hayashi
  et~al.}{2020}]{Hayashi20}
Hayashi K.,  Chiba M.,   Ishiyama T.,  2020, \mn@doi [ApJ]
  {10.3847/1538-4357/abbe0a}, 904, 45

\bibitem[\protect\citeauthoryear{Hiroshima, Ando  \& Ishiyama}{Hiroshima
  et~al.}{2018}]{Hiroshima18}
Hiroshima N.,  Ando S.,   Ishiyama T.,  2018, \mn@doi [Phys. Rev. D]
  {10.1103/PhysRevD.97.123002}, 97, 123002

\bibitem[\protect\citeauthoryear{Hunter}{Hunter}{2007}]{Hunter07}
Hunter J.~D.,  2007, \mn@doi [Comput. Sci. Eng.] {10.1109/MCSE.2007.55}, 9, 90

\bibitem[\protect\citeauthoryear{Jardel \& Gebhardt}{Jardel \&
  Gebhardt}{2012}]{Jardel12}
Jardel J.~R.,  Gebhardt K.,  2012, \mn@doi [ApJ] {10.1088/0004-637X/746/1/89},
  746, 89

\bibitem[\protect\citeauthoryear{Jiang et~al.,}{Jiang et~al.}{2019}]{Jiang19}
Jiang F.,  et~al., 2019, \mn@doi [MNRAS] {10.1093/mnras/stz1952}, 488, 4801

\bibitem[\protect\citeauthoryear{Jiang, Dekel, Freundlich, {\VAN{Van den}{van
  den}}~Bosch, Green, Hopkins, Benson  \& Du}{Jiang et~al.}{2021}]{Jiang21}
Jiang F.,  Dekel A.,  Freundlich J.,  {\VAN{Van den}{van den}}~Bosch F.~C.,
  Green S.~B.,  Hopkins P.~F.,  Benson A.,   Du X.,  2021, \mn@doi [MNRAS]
  {10.1093/mnras/staa4034}, 502, 621

\bibitem[\protect\citeauthoryear{Kaplinghat, Valli  \& Yu}{Kaplinghat
  et~al.}{2019}]{Kaplinghat19}
Kaplinghat M.,  Valli M.,   Yu H.-B.,  2019, \mn@doi [MNRAS]
  {10.1093/mnras/stz2511}, 490, 231

\bibitem[\protect\citeauthoryear{Kim et~al.,}{Kim et~al.}{2024}]{Kim24}
Kim S.~Y.,  et~al., 2024, preprint, \mn@doi{10.48550/arXiv.2408.15214}

\bibitem[\protect\citeauthoryear{Kluyver et~al.,}{Kluyver
  et~al.}{2016}]{Kluyver16}
Kluyver T.,  et~al., 2016, in Loizides F.,  Scmidt B.,  eds, Positioning and
  Power in Academic Publishing: {{Players}}, Agents and Agendas. {IOS Press},
  pp 87--90, \url {https://eprints.soton.ac.uk/403913/}

\bibitem[\protect\citeauthoryear{Koposov, Yoo, Rix, Weinberg, Macci{\`o}  \&
  Escud{\'e}}{Koposov et~al.}{2009}]{Koposov09}
Koposov S.~E.,  Yoo J.,  Rix H.-W.,  Weinberg D.~H.,  Macci{\`o} A.~V.,
  Escud{\'e} J.~M.,  2009, \mn@doi [ApJ] {10.1088/0004-637X/696/2/2179}, 696,
  2179

\bibitem[\protect\citeauthoryear{Kowalczyk, {del Pino}, {\L}okas  \&
  Valluri}{Kowalczyk et~al.}{2019}]{Kowalczyk19}
Kowalczyk K.,  {del Pino} A.,  {\L}okas E.~L.,   Valluri M.,  2019, \mn@doi
  [MNRAS] {10.1093/mnras/sty3100}, 482, 5241

\bibitem[\protect\citeauthoryear{Kravtsov, Berlind, Wechsler, Klypin,
  Gottl{\"o}ber, Allgood  \& Primack}{Kravtsov et~al.}{2004}]{Kravtsov04}
Kravtsov A.~V.,  Berlind A.~A.,  Wechsler R.~H.,  Klypin A.~A.,  Gottl{\"o}ber
  S.,  Allgood B.,   Primack J.~R.,  2004, \mn@doi [ApJ] {10.1086/420959}, 609,
  35

\bibitem[\protect\citeauthoryear{Li, De~Lucia  \& Helmi}{Li
  et~al.}{2010}]{Li10}
Li Y.-S.,  De~Lucia G.,   Helmi A.,  2010, \mn@doi [MNRAS]
  {10.1111/j.1365-2966.2009.15803.x}, 401, 2036

\bibitem[\protect\citeauthoryear{Li, Zhao, Jing, Han  \& Dong}{Li
  et~al.}{2020}]{Li20}
Li Z.-Z.,  Zhao D.-H.,  Jing Y.~P.,  Han J.,   Dong F.-Y.,  2020, \mn@doi [ApJ]
  {10.3847/1538-4357/abc481}, 905, 177

\bibitem[\protect\citeauthoryear{Li, Dekel, Mandelker, Freundlich  \& Fran{\c
  c}ois}{Li et~al.}{2023}]{Li23}
Li Z.~Z.,  Dekel A.,  Mandelker N.,  Freundlich J.,   Fran{\c c}ois T.~L.,
  2023, \mn@doi [MNRAS] {10.1093/mnras/stac3233}, 518, 5356

\bibitem[\protect\citeauthoryear{Lu, Benson, Mao, Tonnesen, Peter, Wetzel,
  {Boylan-Kolchin}  \& Wechsler}{Lu et~al.}{2016}]{Lu16}
Lu Y.,  Benson A.,  Mao Y.-Y.,  Tonnesen S.,  Peter A. H.~G.,  Wetzel A.~R.,
  {Boylan-Kolchin} M.,   Wechsler R.~H.,  2016, \mn@doi [ApJ]
  {10.3847/0004-637X/830/2/59}, 830, 59

\bibitem[\protect\citeauthoryear{Ludlow, Navarro, Springel, Jenkins, Frenk  \&
  Helmi}{Ludlow et~al.}{2009}]{Ludlow09}
Ludlow A.~D.,  Navarro J.~F.,  Springel V.,  Jenkins A.,  Frenk C.~S.,   Helmi
  A.,  2009, \mn@doi [ApJ] {10.1088/0004-637X/692/1/931}, 692, 931

\bibitem[\protect\citeauthoryear{Macci{\`o}, Kang, Fontanot, Somerville,
  Koposov  \& Monaco}{Macci{\`o} et~al.}{2010}]{Maccio10}
Macci{\`o} A.~V.,  Kang X.,  Fontanot F.,  Somerville R.~S.,  Koposov S.,
  Monaco P.,  2010, \mn@doi [MNRAS] {10.1111/j.1365-2966.2009.16031.x}, 402,
  1995

\bibitem[\protect\citeauthoryear{Meadows, Navarro, {Santos-Santos},
  {Ben{\'i}tez-Llambay}  \& Frenk}{Meadows et~al.}{2020}]{Meadows20}
Meadows N.,  Navarro J.~F.,  {Santos-Santos} I.,  {Ben{\'i}tez-Llambay} A.,
  Frenk C.,  2020, \mn@doi [MNRAS] {10.1093/mnras/stz3280}, 491, 3336

\bibitem[\protect\citeauthoryear{Miyamoto \& Nagai}{Miyamoto \&
  Nagai}{1975}]{Miyamoto75}
Miyamoto M.,  Nagai R.,  1975, PASJ, 27, 533

\bibitem[\protect\citeauthoryear{Molin{\'e}, {S{\'a}nchez-Conde},
  {Palomares-Ruiz}  \& Prada}{Molin{\'e} et~al.}{2017}]{Moline17}
Molin{\'e} {\'A}.,  {S{\'a}nchez-Conde} M.~A.,  {Palomares-Ruiz} S.,   Prada
  F.,  2017, \mn@doi [MNRAS] {10.1093/mnras/stx026}, 466, 4974

\bibitem[\protect\citeauthoryear{Molin{\'e} et~al.,}{Molin{\'e}
  et~al.}{2023}]{Moline23}
Molin{\'e} {\'A}.,  et~al., 2023, \mn@doi [MNRAS] {10.1093/mnras/stac2930},
  518, 157

\bibitem[\protect\citeauthoryear{Moster, Naab  \& White}{Moster
  et~al.}{2013}]{Moster13}
Moster B.~P.,  Naab T.,   White S. D.~M.,  2013, \mn@doi [MNRAS]
  {10.1093/mnras/sts261}, 428, 3121

\bibitem[\protect\citeauthoryear{Mu{\~n}oz, C{\^o}t{\'e}, Santana, Geha, Simon,
  Oyarz{\'u}n, Stetson  \& Djorgovski}{Mu{\~n}oz et~al.}{2018}]{Munoz18}
Mu{\~n}oz R.~R.,  C{\^o}t{\'e} P.,  Santana F.~A.,  Geha M.,  Simon J.~D.,
  Oyarz{\'u}n G.~A.,  Stetson P.~B.,   Djorgovski S.~G.,  2018, \mn@doi [ApJ]
  {10.3847/1538-4357/aac16b}, 860, 66

\bibitem[\protect\citeauthoryear{Munshi, Brooks, Applebaum, Christensen, Quinn
  \& Sligh}{Munshi et~al.}{2021}]{Munshi21}
Munshi F.,  Brooks A.~M.,  Applebaum E.,  Christensen C.~R.,  Quinn T.,   Sligh
  S.,  2021, \mn@doi [ApJ] {10.3847/1538-4357/ac0db6}, 923, 35

\bibitem[\protect\citeauthoryear{Nadler, Mao, Green  \& Wechsler}{Nadler
  et~al.}{2019}]{Nadler19}
Nadler E.~O.,  Mao Y.-Y.,  Green G.~M.,   Wechsler R.~H.,  2019, \mn@doi [ApJ]
  {10.3847/1538-4357/ab040e}, 873, 34

\bibitem[\protect\citeauthoryear{Nadler et~al.,}{Nadler
  et~al.}{2020}]{Nadler20}
Nadler E.~O.,  et~al., 2020, \mn@doi [ApJ] {10.3847/1538-4357/ab846a}, 893, 48

\bibitem[\protect\citeauthoryear{Nadler et~al.,}{Nadler
  et~al.}{2023}]{Nadler23}
Nadler E.~O.,  et~al., 2023, \mn@doi [ApJ] {10.3847/1538-4357/acb68c}, 945, 159

\bibitem[\protect\citeauthoryear{Navarro, Frenk  \& White}{Navarro
  et~al.}{1997}]{Navarro97}
Navarro J.~F.,  Frenk C.~S.,   White S. D.~M.,  1997, \mn@doi [ApJ]
  {10.1086/304888}, 490, 493

\bibitem[\protect\citeauthoryear{Neto et~al.,}{Neto et~al.}{2007}]{Neto07}
Neto A.~F.,  et~al., 2007, \mn@doi [MNRAS] {10.1111/j.1365-2966.2007.12381.x},
  381, 1450

\bibitem[\protect\citeauthoryear{Pace, Erkal  \& Li}{Pace
  et~al.}{2022}]{Pace22}
Pace A.~B.,  Erkal D.,   Li T.~S.,  2022, \mn@doi [ApJ]
  {10.3847/1538-4357/ac997b}, 940, 136

\bibitem[\protect\citeauthoryear{Parkinson, Cole  \& Helly}{Parkinson
  et~al.}{2007}]{Parkinson07}
Parkinson H.,  Cole S.,   Helly J.,  2007, \mn@doi [MNRAS]
  {10.1111/j.1365-2966.2007.12517.x}, 383, 557

\bibitem[\protect\citeauthoryear{Patel et~al.,}{Patel et~al.}{2020}]{Patel20}
Patel E.,  et~al., 2020, \mn@doi [ApJ] {10.3847/1538-4357/ab7b75}, 893, 121

\bibitem[\protect\citeauthoryear{Pedregosa et~al.,}{Pedregosa
  et~al.}{2011}]{Pedregosa11}
Pedregosa F.,  et~al., 2011, JMLR, 12, 2825

\bibitem[\protect\citeauthoryear{Pe{\~n}arrubia, Navarro, McConnachie  \&
  Martin}{Pe{\~n}arrubia et~al.}{2009}]{Penarrubia09}
Pe{\~n}arrubia J.,  Navarro J.~F.,  McConnachie A.~W.,   Martin N.~F.,  2009,
  \mn@doi [ApJ] {10.1088/0004-637X/698/1/222}, 698, 222

\bibitem[\protect\citeauthoryear{Pe{\~n}arrubia, Benson, Walker, Gilmore,
  McConnachie  \& Mayer}{Pe{\~n}arrubia et~al.}{2010}]{Penarrubia10}
Pe{\~n}arrubia J.,  Benson A.~J.,  Walker M.~G.,  Gilmore G.,  McConnachie
  A.~W.,   Mayer L.,  2010, \mn@doi [MNRAS] {10.1111/j.1365-2966.2010.16762.x},
  406, 1290

\bibitem[\protect\citeauthoryear{P{\'e}rez \& Granger}{P{\'e}rez \&
  Granger}{2007}]{Perez07}
P{\'e}rez F.,  Granger B.~E.,  2007, \mn@doi [Comput. Sci. Eng.]
  {10.1109/MCSE.2007.53}, 9, 21

\bibitem[\protect\citeauthoryear{Pullen, Benson  \& Moustakas}{Pullen
  et~al.}{2014}]{Pullen14}
Pullen A.~R.,  Benson A.~J.,   Moustakas L.~A.,  2014, \mn@doi [ApJ]
  {10.1088/0004-637X/792/1/24}, 792, 24

\bibitem[\protect\citeauthoryear{Read, Walker  \& Steger}{Read
  et~al.}{2019}]{Read19}
Read J.~I.,  Walker M.~G.,   Steger P.,  2019, \mn@doi [MNRAS]
  {10.1093/mnras/sty3404}, 484, 1401

\bibitem[\protect\citeauthoryear{{Rodr{\'i}guez-Puebla}, Primack, {Avila-Reese}
   \& Faber}{{Rodr{\'i}guez-Puebla} et~al.}{2017}]{Rodriguez-Puebla17}
{Rodr{\'i}guez-Puebla} A.,  Primack J.~R.,  {Avila-Reese} V.,   Faber S.~M.,
  2017, \mn@doi [MNRAS] {10.1093/mnras/stx1172}, 470, 651

\bibitem[\protect\citeauthoryear{Sales, Navarro, Cooper, White, Frenk  \&
  Helmi}{Sales et~al.}{2011}]{Sales11}
Sales L.~V.,  Navarro J.~F.,  Cooper A.~P.,  White S. D.~M.,  Frenk C.~S.,
  Helmi A.,  2011, \mn@doi [MNRAS] {10.1111/j.1365-2966.2011.19514.x}, 418, 648

\bibitem[\protect\citeauthoryear{Sanders \& Evans}{Sanders \&
  Evans}{2016}]{Sanders16}
Sanders J.~L.,  Evans N.~W.,  2016, \mn@doi [ApJ]
  {10.3847/2041-8205/830/2/L26}, 830, L26

\bibitem[\protect\citeauthoryear{{Santos-Santos}, Sales, Fattahi  \&
  Navarro}{{Santos-Santos} et~al.}{2022}]{Santos-Santos22}
{Santos-Santos} I. M.~E.,  Sales L.~V.,  Fattahi A.,   Navarro J.~F.,  2022,
  \mn@doi [MNRAS] {10.1093/mnras/stac2057}, 515, 3685

\bibitem[\protect\citeauthoryear{Sawala et~al.,}{Sawala
  et~al.}{2016}]{Sawala16}
Sawala T.,  et~al., 2016, \mn@doi [MNRAS] {10.1093/mnras/stw145}, 457, 1931

\bibitem[\protect\citeauthoryear{{Smercina}, {Bell}, {Samuel}  \&
  {D'Souza}}{{Smercina} et~al.}{2022}]{Smercina21}
{Smercina} A.,  {Bell} E.~F.,  {Samuel} J.,   {D'Souza} R.,  2022, \mn@doi
  [ApJ] {10.3847/1538-4357/ac5d56}, 930, 69

\bibitem[\protect\citeauthoryear{Starkenburg et~al.,}{Starkenburg
  et~al.}{2013}]{Starkenburg13}
Starkenburg E.,  et~al., 2013, \mn@doi [MNRAS] {10.1093/mnras/sts367}, 429, 725

\bibitem[\protect\citeauthoryear{Taylor \& Babul}{Taylor \&
  Babul}{2001}]{Taylor01}
Taylor J.~E.,  Babul A.,  2001, \mn@doi [ApJ] {10.1086/322276}, 559, 716

\bibitem[\protect\citeauthoryear{Tollet et~al.,}{Tollet
  et~al.}{2016}]{Tollet16}
Tollet E.,  et~al., 2016, \mn@doi [MNRAS] {10.1093/mnras/stv2856}, 456, 3542

\bibitem[\protect\citeauthoryear{{\VAN{Van den}{van den}}~Bosch \&
  Ogiya}{{\VAN{Van den}{van den}}~Bosch \& Ogiya}{2018}]{VanDenBosch18a}
{\VAN{Van den}{van den}}~Bosch F.~C.,  Ogiya G.,  2018, \mn@doi [MNRAS]
  {10.1093/mnras/sty084}, 475, 4066

\bibitem[\protect\citeauthoryear{{\VAN{Van den}{van den}}~Bosch, Ogiya, Hahn
  \& Burkert}{{\VAN{Van den}{van den}}~Bosch et~al.}{2018}]{VanDenBosch18}
{\VAN{Van den}{van den}}~Bosch F.~C.,  Ogiya G.,  Hahn O.,   Burkert A.,  2018,
  \mn@doi [MNRAS] {10.1093/mnras/stx2956}, 474, 3043

\bibitem[\protect\citeauthoryear{{Vasiliev}}{{Vasiliev}}{2023}]{Vasiliev23}
{Vasiliev} E.,  2023, \mn@doi [Galaxies] {10.3390/galaxies11020059}, 11, 59

\bibitem[\protect\citeauthoryear{Virtanen et~al.,}{Virtanen
  et~al.}{2020}]{Virtanen20}
Virtanen P.,  et~al., 2020, \mn@doi [Nat. Methods] {10.1038/s41592-019-0686-2},
  \href {https://rdcu.be/b08Wh} {17, 261}

\bibitem[\protect\citeauthoryear{Vogelsberger, Marinacci, Torrey  \&
  Puchwein}{Vogelsberger et~al.}{2020}]{Vogelsberger20}
Vogelsberger M.,  Marinacci F.,  Torrey P.,   Puchwein E.,  2020, \mn@doi [Nat.
  Rev. Phys.] {10.1038/s42254-019-0127-2}, 2, 42

\bibitem[\protect\citeauthoryear{Wang, Dutton, Stinson, Macci{\`o}, Penzo,
  Kang, Keller  \& Wadsley}{Wang et~al.}{2015}]{Wang15}
Wang L.,  Dutton A.~A.,  Stinson G.~S.,  Macci{\`o} A.~V.,  Penzo C.,  Kang X.,
   Keller B.~W.,   Wadsley J.,  2015, \mn@doi [MNRAS] {10.1093/mnras/stv1937},
  454, 83

\bibitem[\protect\citeauthoryear{Wang, Han, Cautun, Li  \& Ishigaki}{Wang
  et~al.}{2020}]{Wang20}
Wang W.,  Han J.,  Cautun M.,  Li Z.,   Ishigaki M.~N.,  2020, \mn@doi [Sci.
  Chin. Phys. Mech. Astron.] {10.1007/s11433-019-1541-6}, 63, 109801

\bibitem[\protect\citeauthoryear{Wechsler \& Tinker}{Wechsler \&
  Tinker}{2018}]{Wechsler18}
Wechsler R.~H.,  Tinker J.~L.,  2018, \mn@doi [Annu. Rev. Astron. Astrophys.]
  {10.1146/annurev-astro-081817-051756}, 56, 435

\bibitem[\protect\citeauthoryear{Wolf, Martinez, Bullock, Kaplinghat, Geha,
  Mu{\~n}oz, Simon  \& Avedo}{Wolf et~al.}{2010}]{Wolf10}
Wolf J.,  Martinez G.~D.,  Bullock J.~S.,  Kaplinghat M.,  Geha M.,  Mu{\~n}oz
  R.~R.,  Simon J.~D.,   Avedo F.~F.,  2010, \mn@doi [MNRAS]
  {10.1111/j.1365-2966.2010.16753.x}, 406, 1220

\bibitem[\protect\citeauthoryear{Woo, Courteau  \& Dekel}{Woo
  et~al.}{2008}]{Woo08}
Woo J.,  Courteau S.,   Dekel A.,  2008, \mn@doi [MNRAS]
  {10.1111/j.1365-2966.2008.13770.x}, 390, 1453

\bibitem[\protect\citeauthoryear{Yun et~al.,}{Yun et~al.}{2019}]{Yun19}
Yun K.,  et~al., 2019, \mn@doi [MNRAS] {10.1093/mnras/sty3156}, 483, 1042

\bibitem[\protect\citeauthoryear{Zhao}{Zhao}{1996}]{Zhao96}
Zhao H.,  1996, \mn@doi [MNRAS] {10.1093/mnras/278.2.488}, 278, 488

\bibitem[\protect\citeauthoryear{Zhao, Jing, Mo  \& B{\"o}rner}{Zhao
  et~al.}{2009}]{Zhao09}
Zhao D.~H.,  Jing Y.~P.,  Mo H.~J.,   B{\"o}rner G.,  2009, \mn@doi [ApJ]
  {10.1088/0004-637X/707/1/354}, 707, 354

\makeatother
\end{thebibliography}
\vspace{-\baselineskip}

\appendix
\section{Supplementary figures}
\label{appendix}
\begin{figure}
	\centering
    \setlength\fboxrule{0.8pt}
  	\hspace{3.3em}\fbox{\parbox{119.8pt}{{\color{NIHAO}NIHAO, RP17}\qquad{\color{APOSTLE}APOSTLE, RP17}}}

	\includegraphics{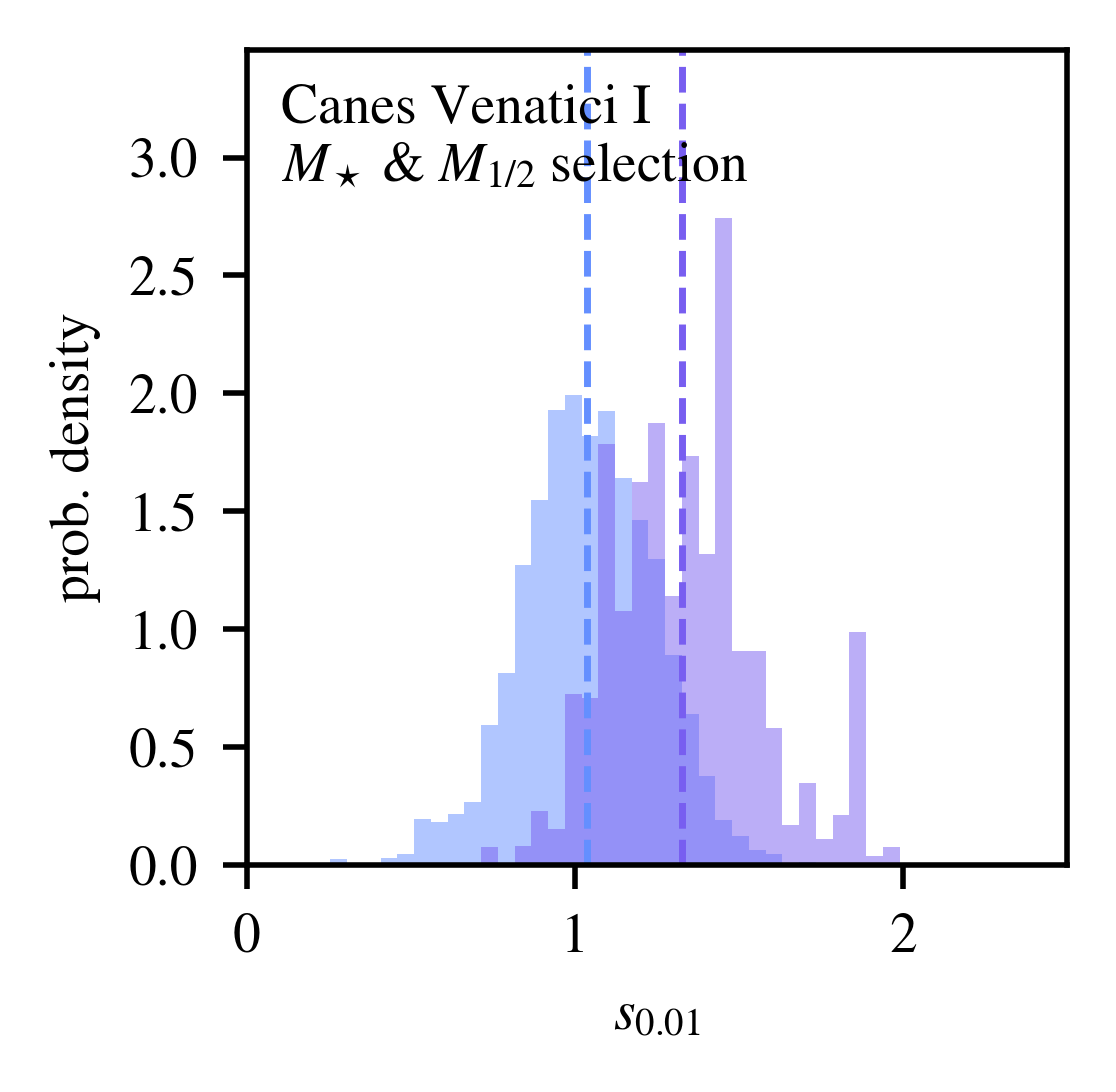}
	\includegraphics{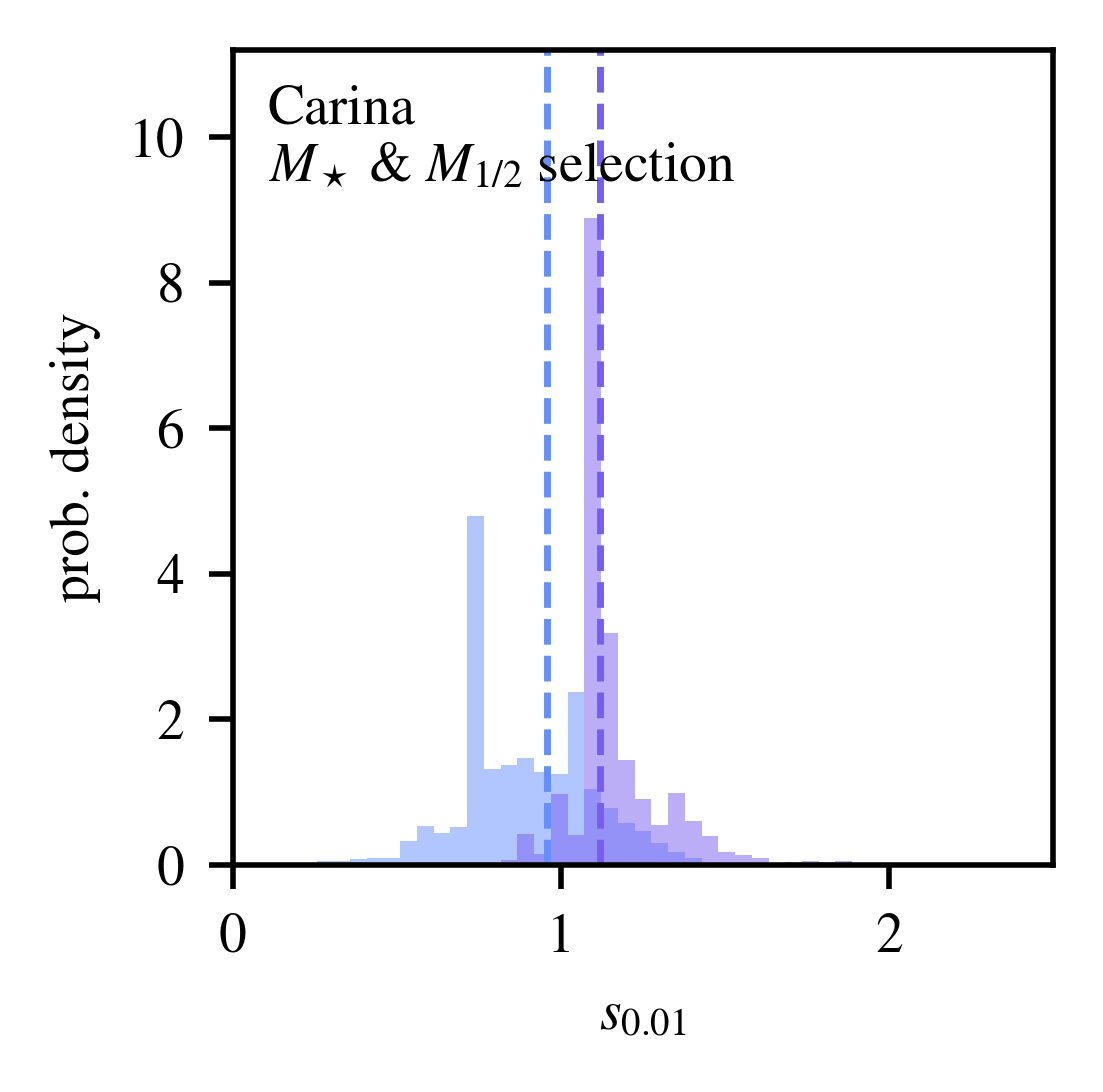}
	\includegraphics{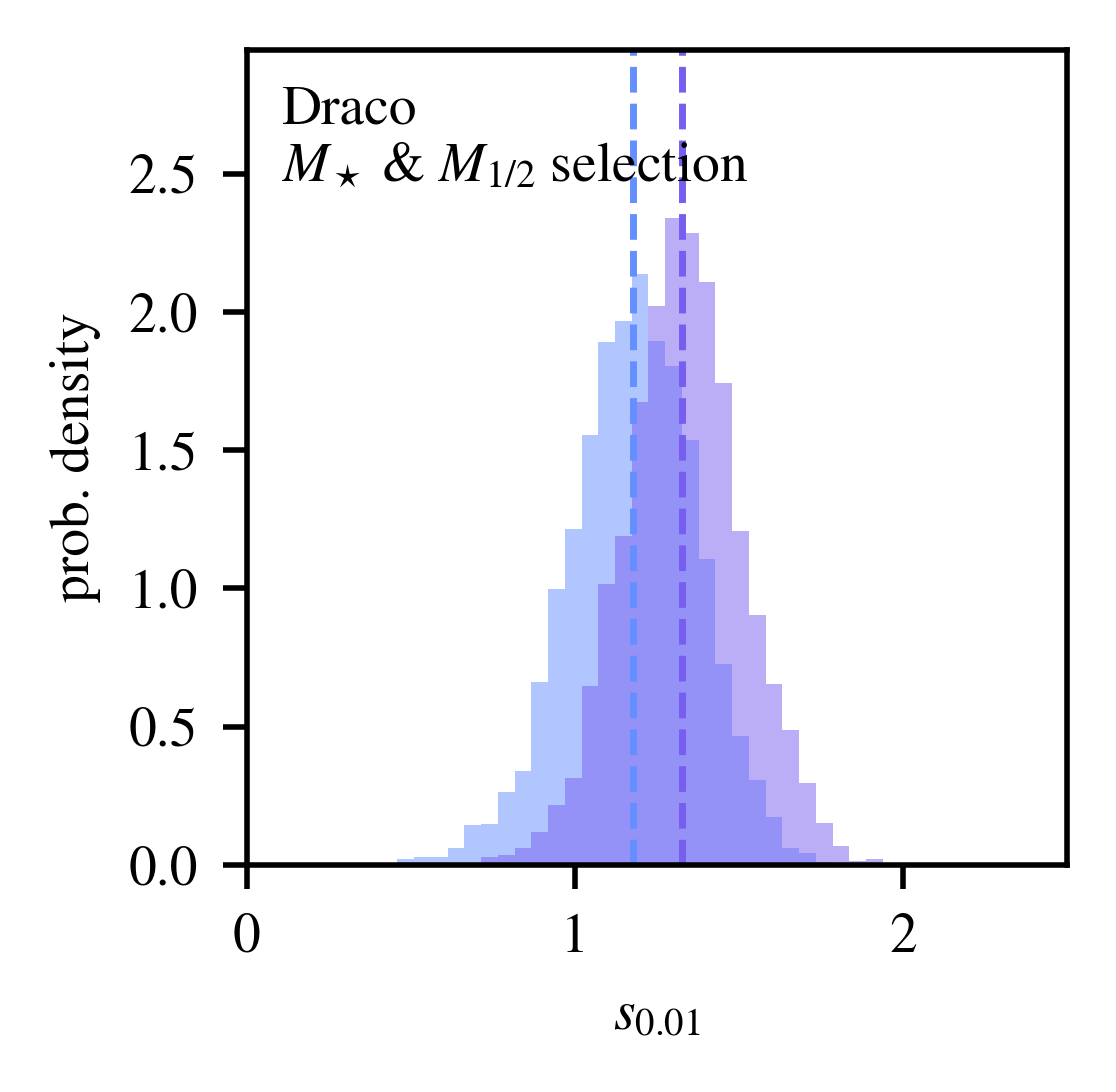}
	\includegraphics{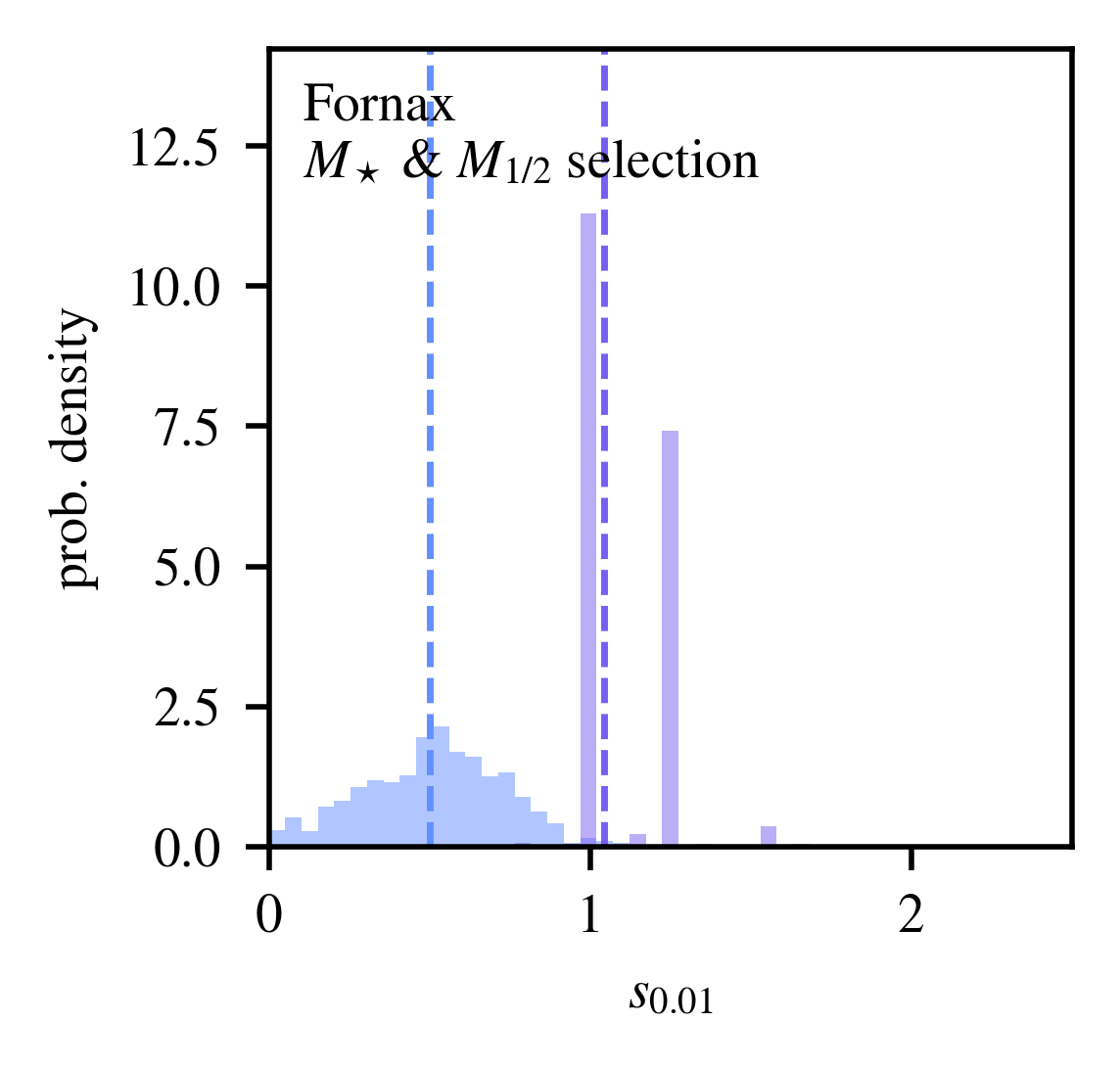}
	\includegraphics{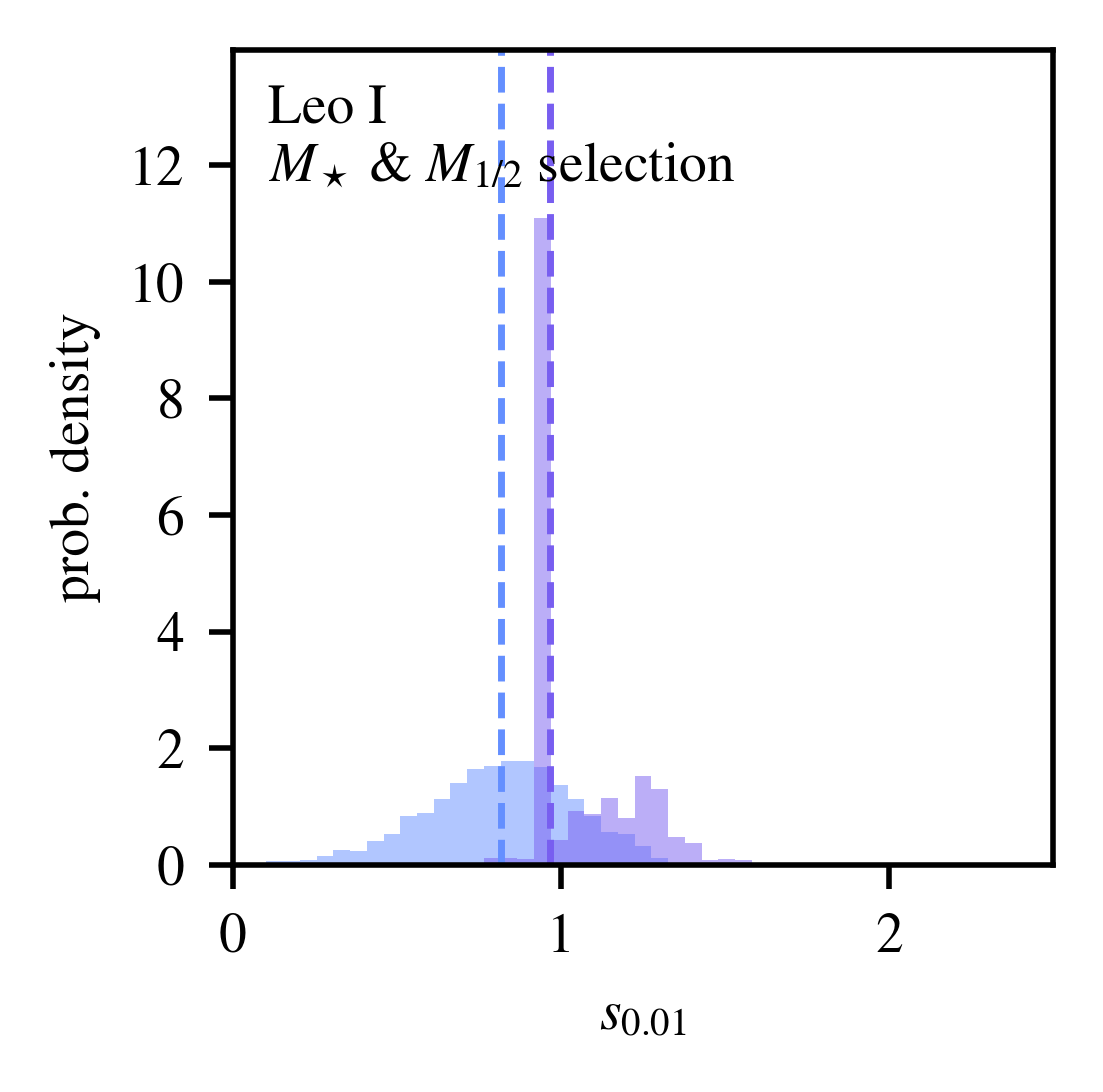}
	\includegraphics{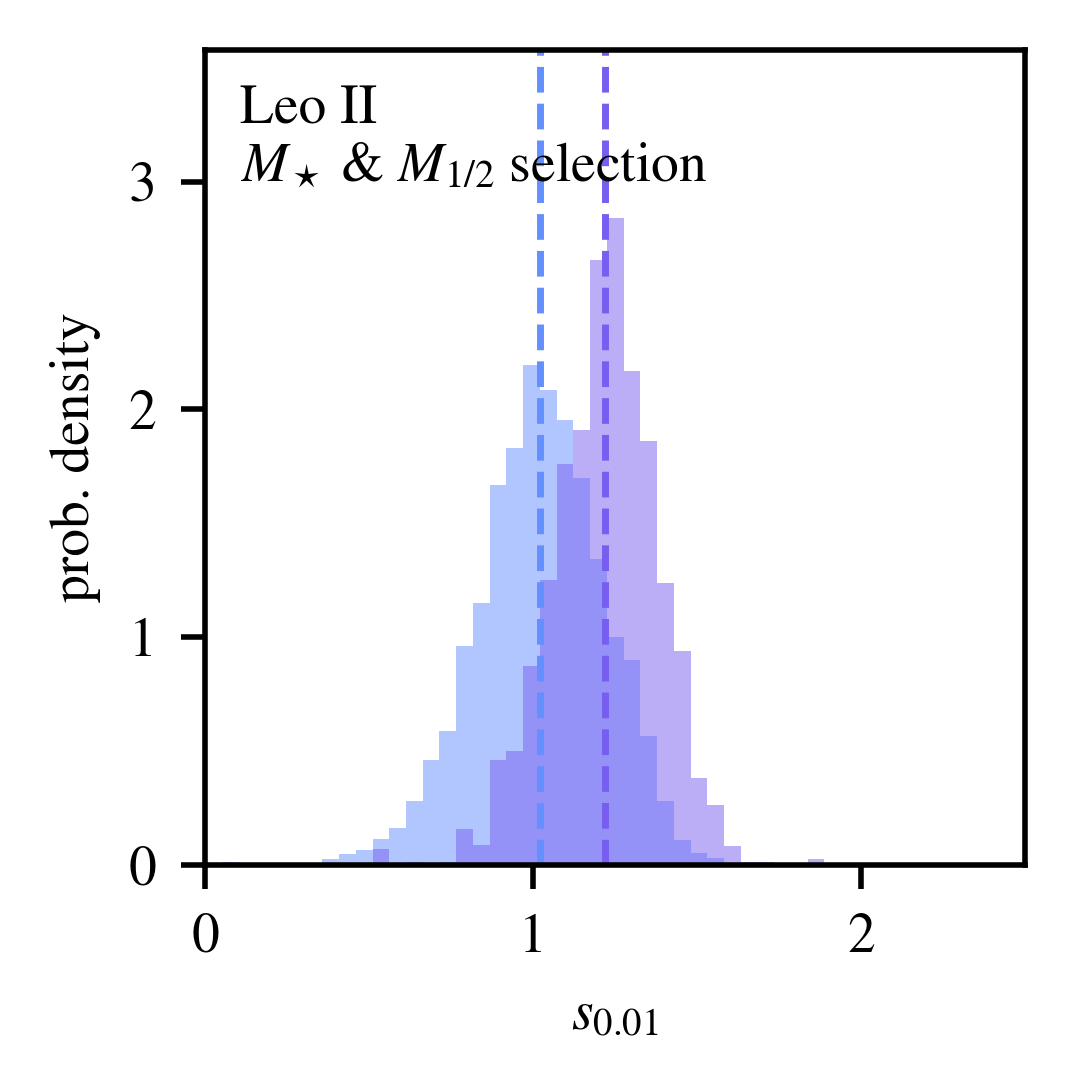}
	\includegraphics{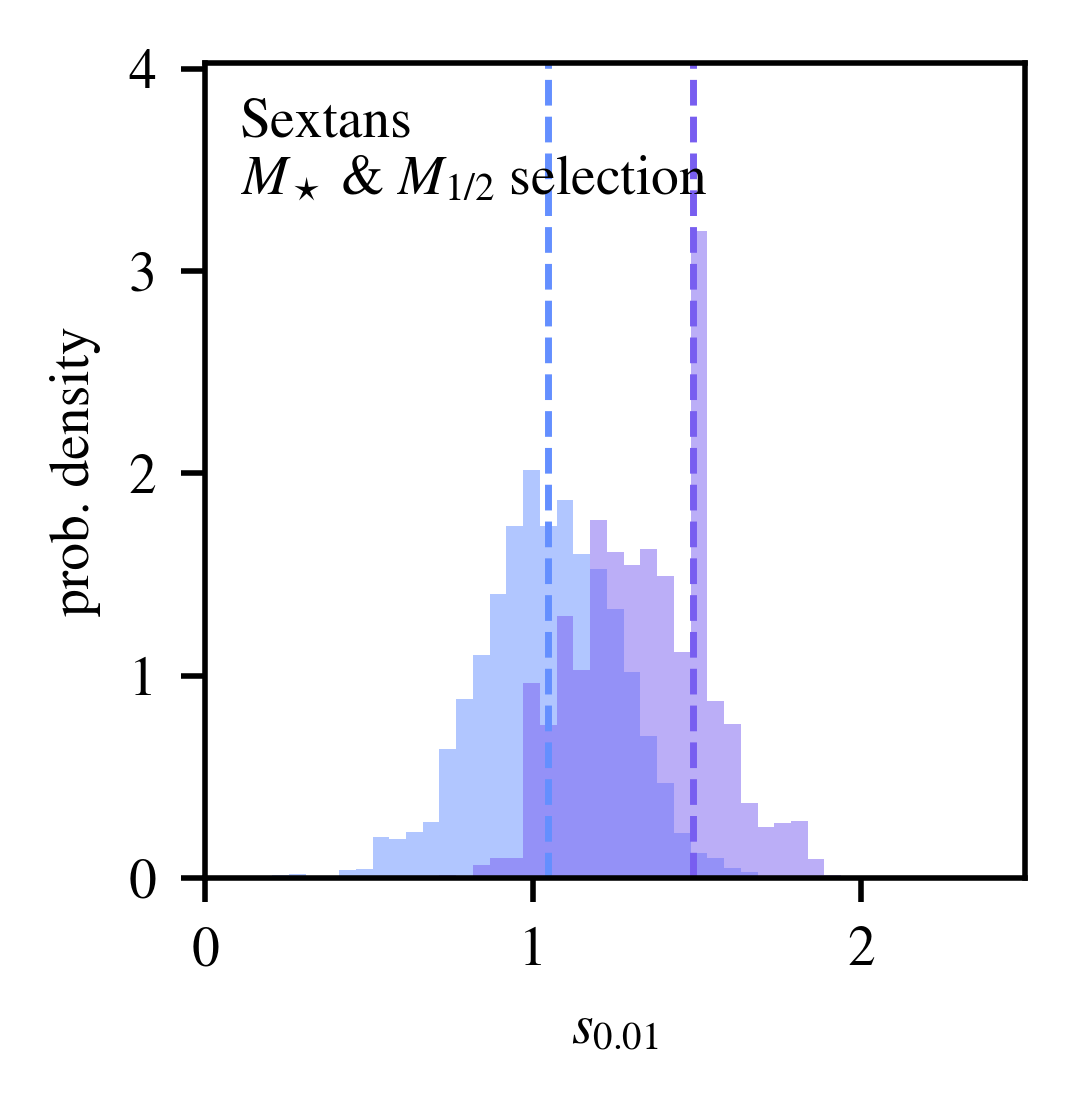}
	\includegraphics{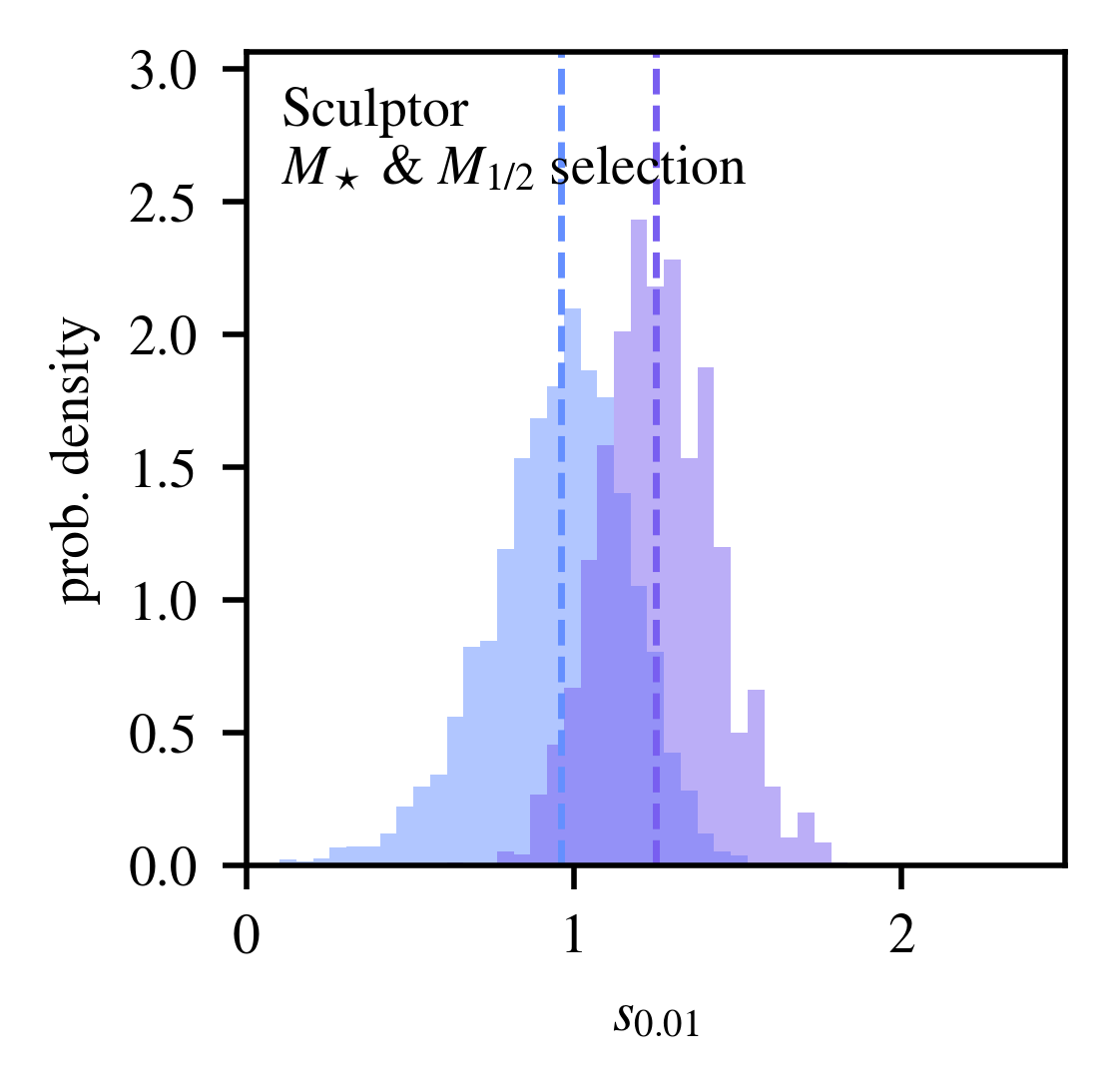}
	\includegraphics{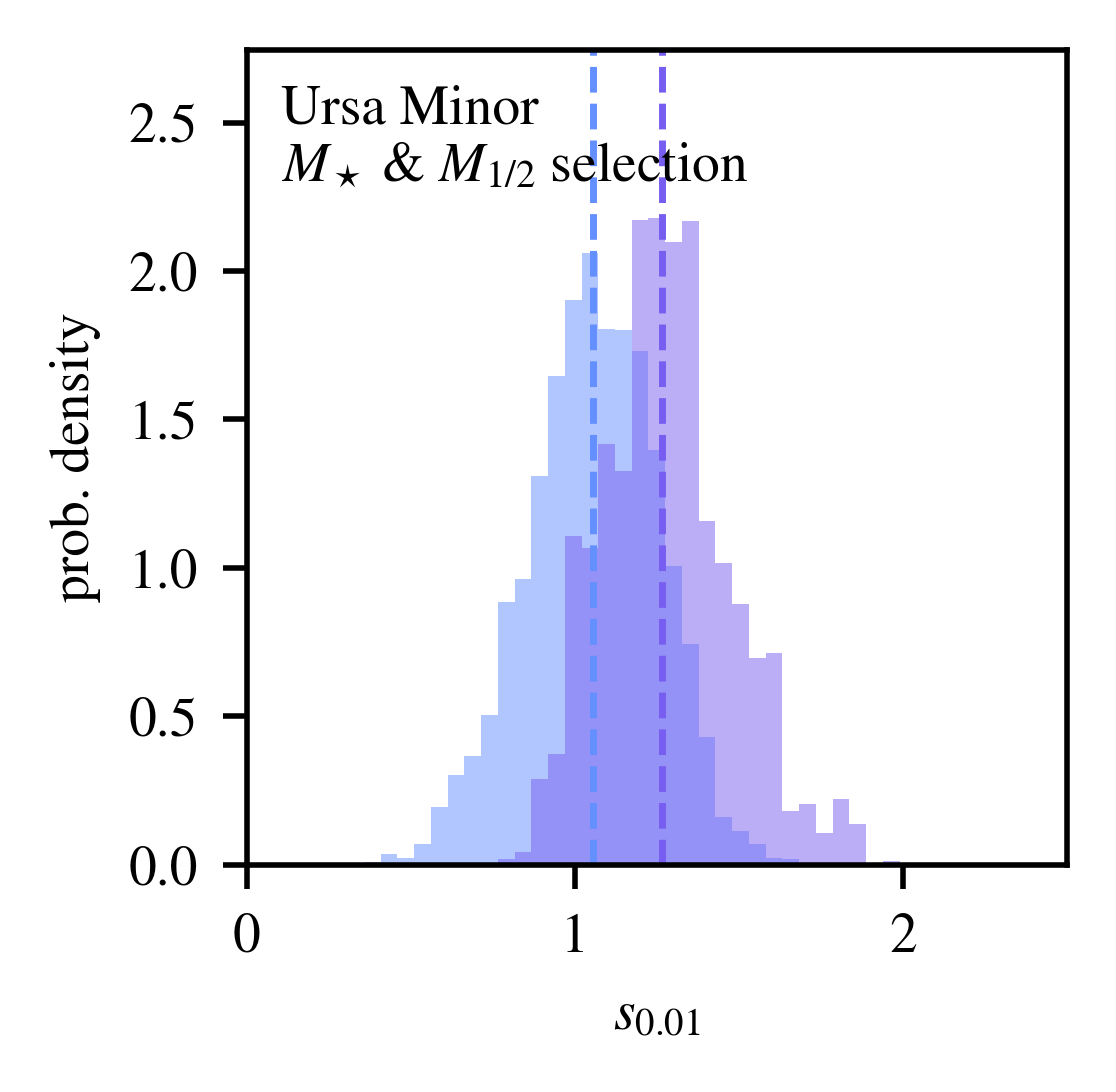}
	\caption{Inferred inner slopes for all classical satellites, assuming the \RPSMHM{} SMHM relation and varying over feedback emulation. The slope is parametrized by $s_{0.01}$, which is the slope at 1 per cent of the virial radius. For an NFW profile, this will range from 1 to about 1.5 depending on the concentration, while particularly cored profiles will approach zero. The dashed lines show the median inferred slope, though the full distributions are shown for clarity. Fornax and Sextans are particularly interesting for the large difference between feedback models.}
\label{fig:A2}
\end{figure}

\begin{figure}
	\centering
	\includegraphics{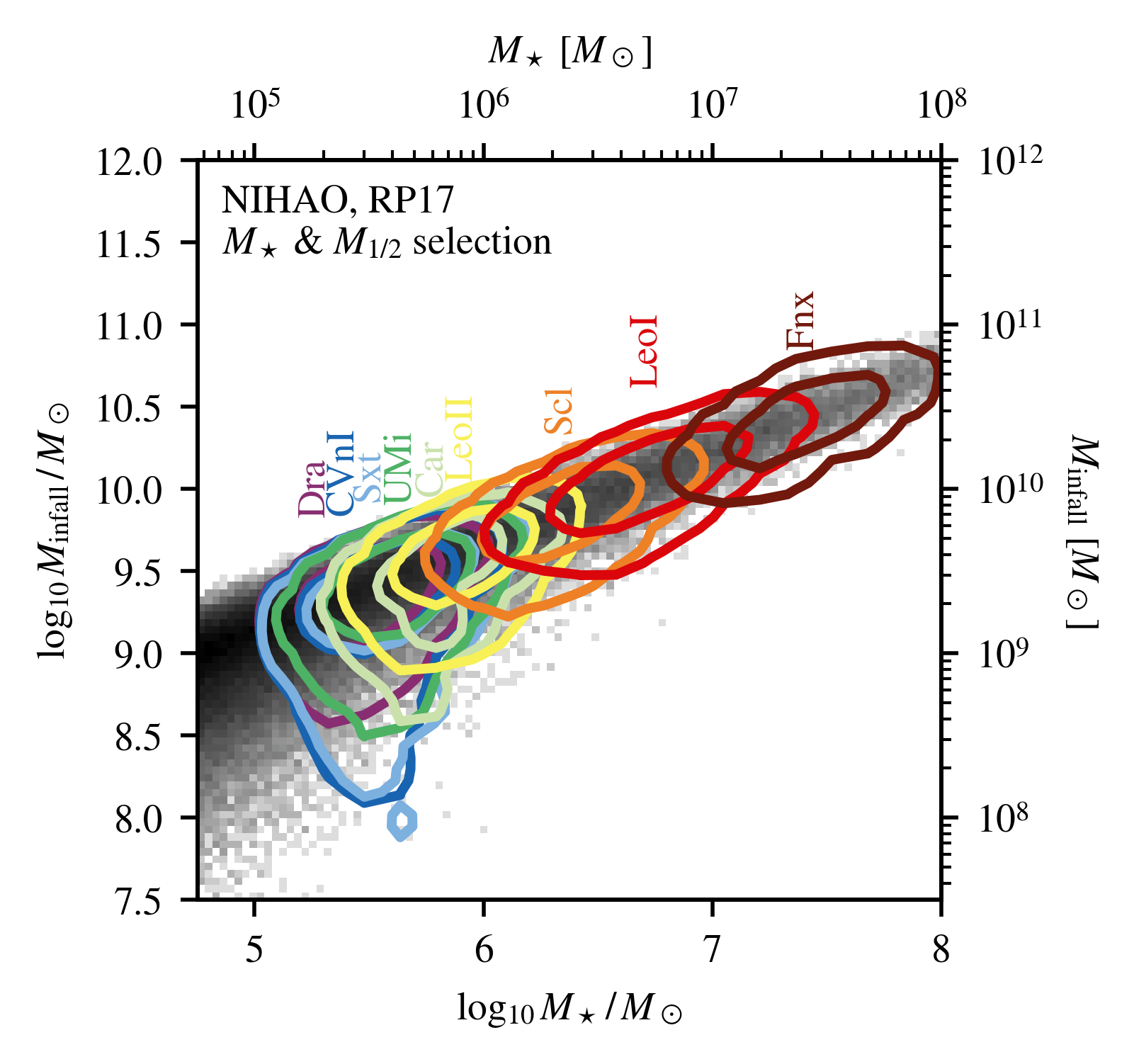}
	\includegraphics{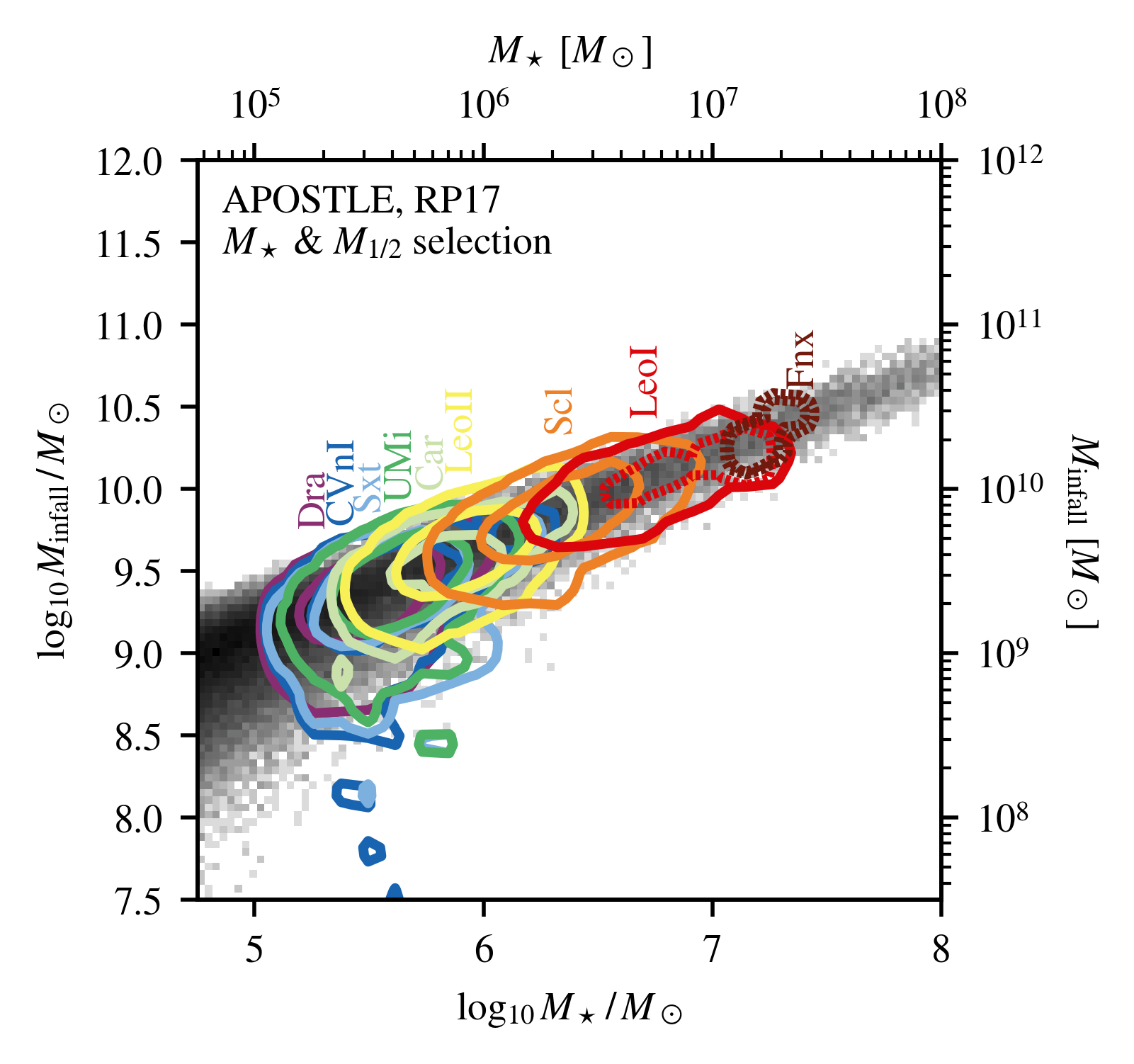}
	\caption{68 and 95 per cent confidence intervals for the $z = 0$ stellar mass, $M_\star$, and the virial mass at the time of accretion into the MW, $M_\mathrm{infall}$, for each of the nine bright spheroidal dwarfs using the combined $M_\star$ and $M_{1/2}$-based inference. The grey histogram shows the overall distribution of \SatGen{} satellites in the NIHAO feedback emulator (left) and in the APOSTLE feedback emulator (right). This is in large part set by the SMHM relation, though (i) $M_\mathrm{infall}$ is modified from the SMHM relation in satellites accreted as part of a group system, as they undergo tidal mass loss prior to infall, and (ii) $M_\star$ evolves along the~\citet{Errani18} tidal tracks, which prescribes the stellar mass lost in haloes with significant DM mass loss.}
\label{fig:A3}
\end{figure}

In the main body of this paper, specific dwarfs are highlighted in the figures. For completeness, the figures below include versions of \autoref{fig:3}, \autoref{fig:4}, and \autoref{fig:6} for all nine bright MW spheroidal dwarfs. For more detail regarding the data shown in these figures, see the captions provided for the figures in the main text. \autoref{fig:A2} has no corresponding figure in the main text; it shows the inferred inner slopes in the NIHAO and APOSTLE feedback emulators assuming the \RPSMHM{} SMHM relation and is referenced in \autoref{sec:3.2}. \autoref{fig:A3} also has no corresponding figure in the main text; it shows the inferred virial mass at the time of accretion on to the MW and the $z = 0$ stellar mass for all nine satellites in the NIHAO and APOSTLE feedback emulators assuming the \RPSMHM{} SMHM relation and is referenced in \autoref{sec:3.2}.
\onecolumn
\FloatBarrier
\begin{figure}
 \setlength\fboxrule{0.8pt}
	\centering
	\hspace{3.4em}\fbox{\parbox{55.5em}{{\color{NIHAO}NIHAO, RP17}\hfill{}{\color{APOSTLE}APOSTLE, RP17}\hfill{}{\color{SCALING}Simple scaling relations}\hfill{} \ERV{} cored \includegraphics{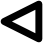}\; cuspy \includegraphics{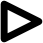}\hfill \KRV{} cored \includegraphics{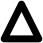}\; cuspy \includegraphics{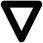}\hfill \ARV{} \includegraphics{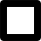}}}

	\includegraphics{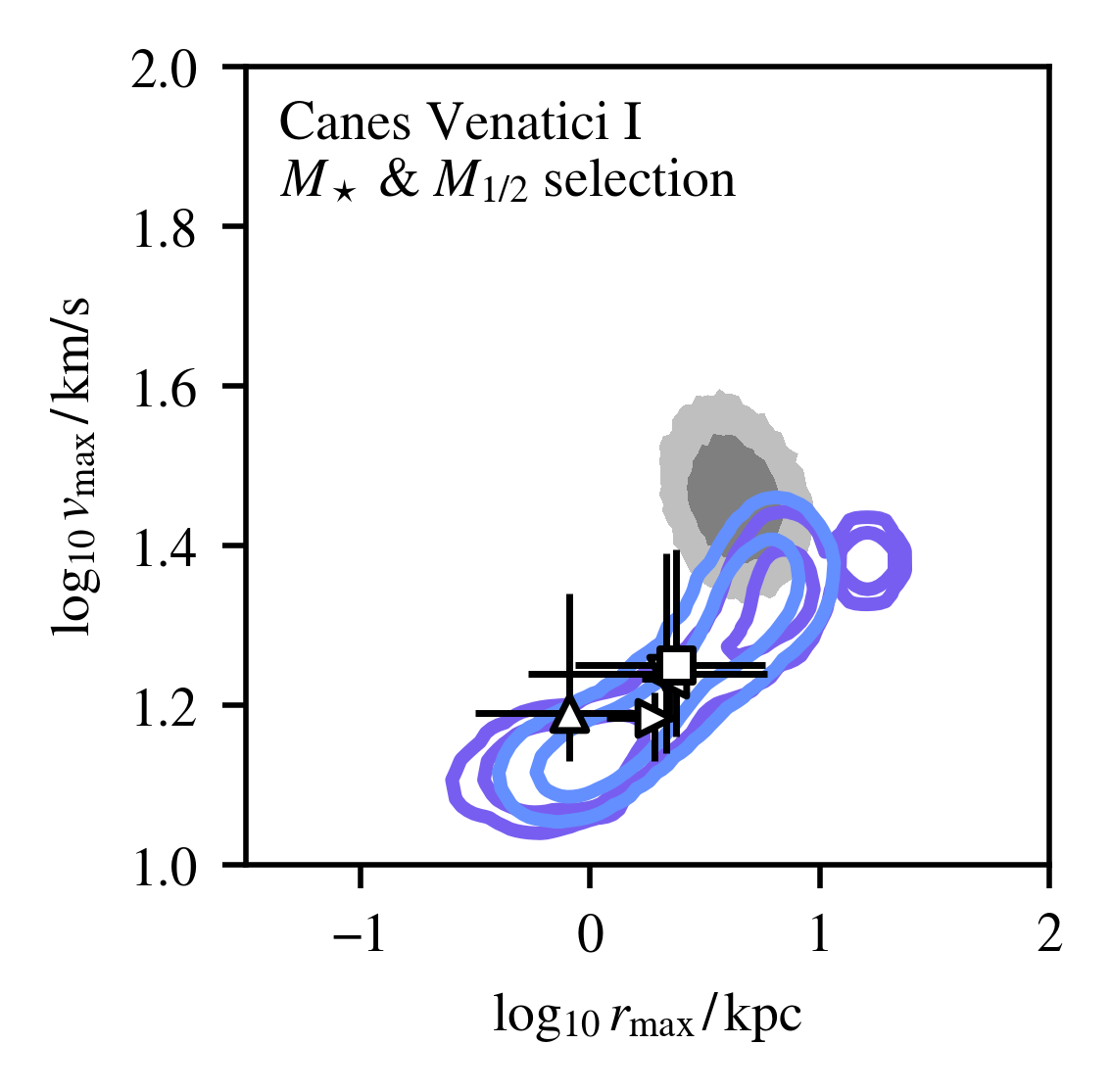}
	\includegraphics{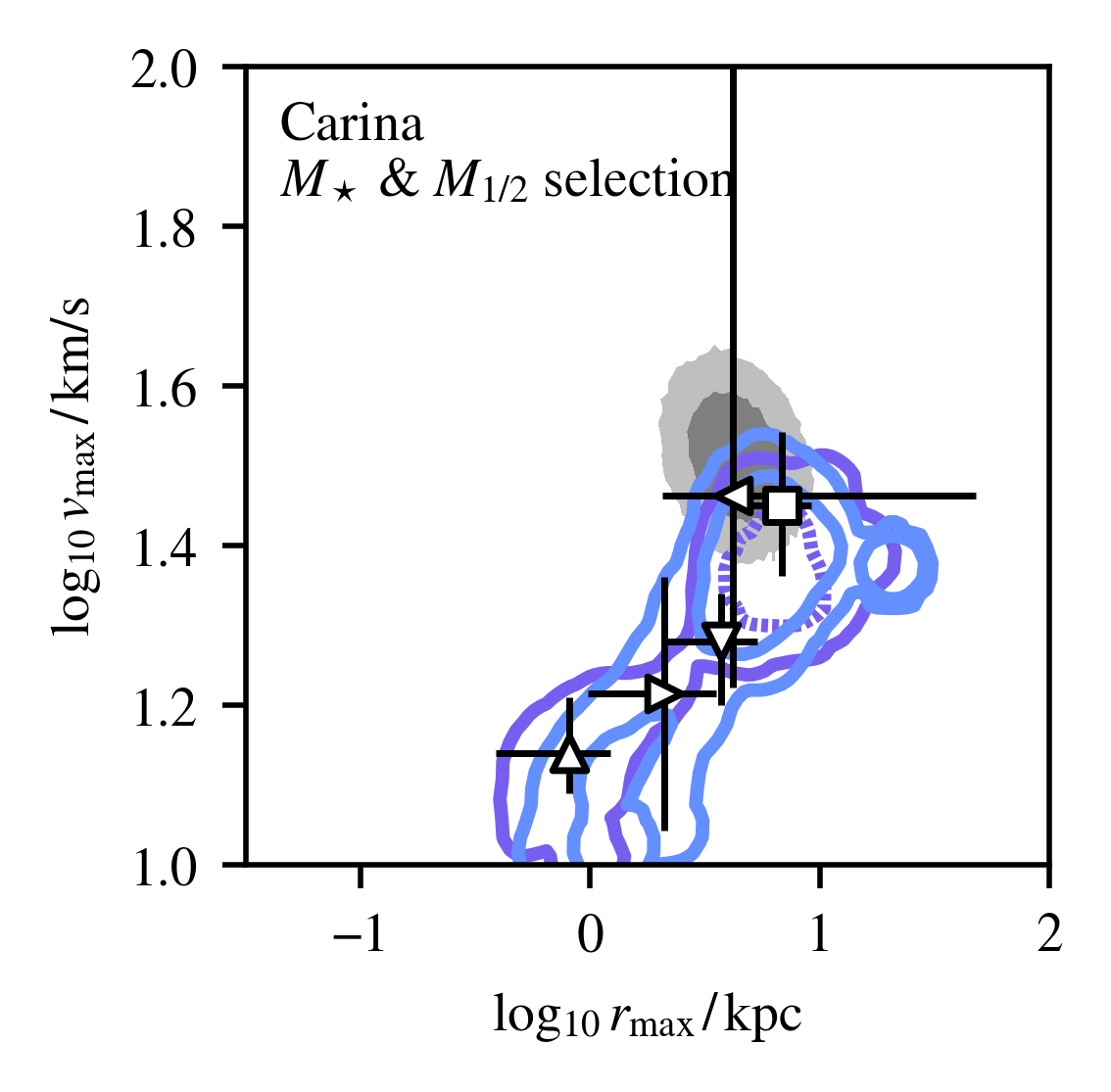}
	\includegraphics{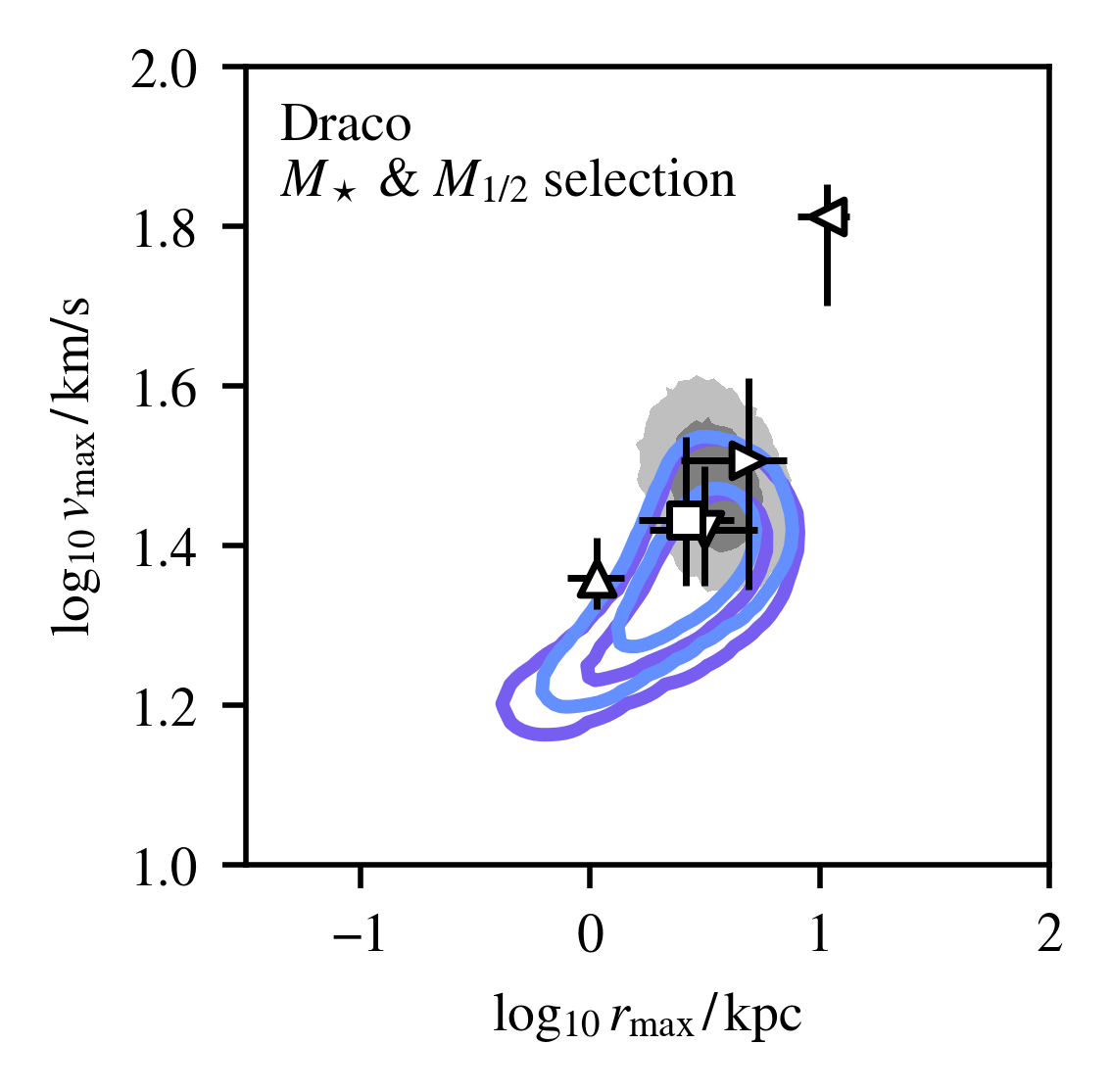}
	\includegraphics{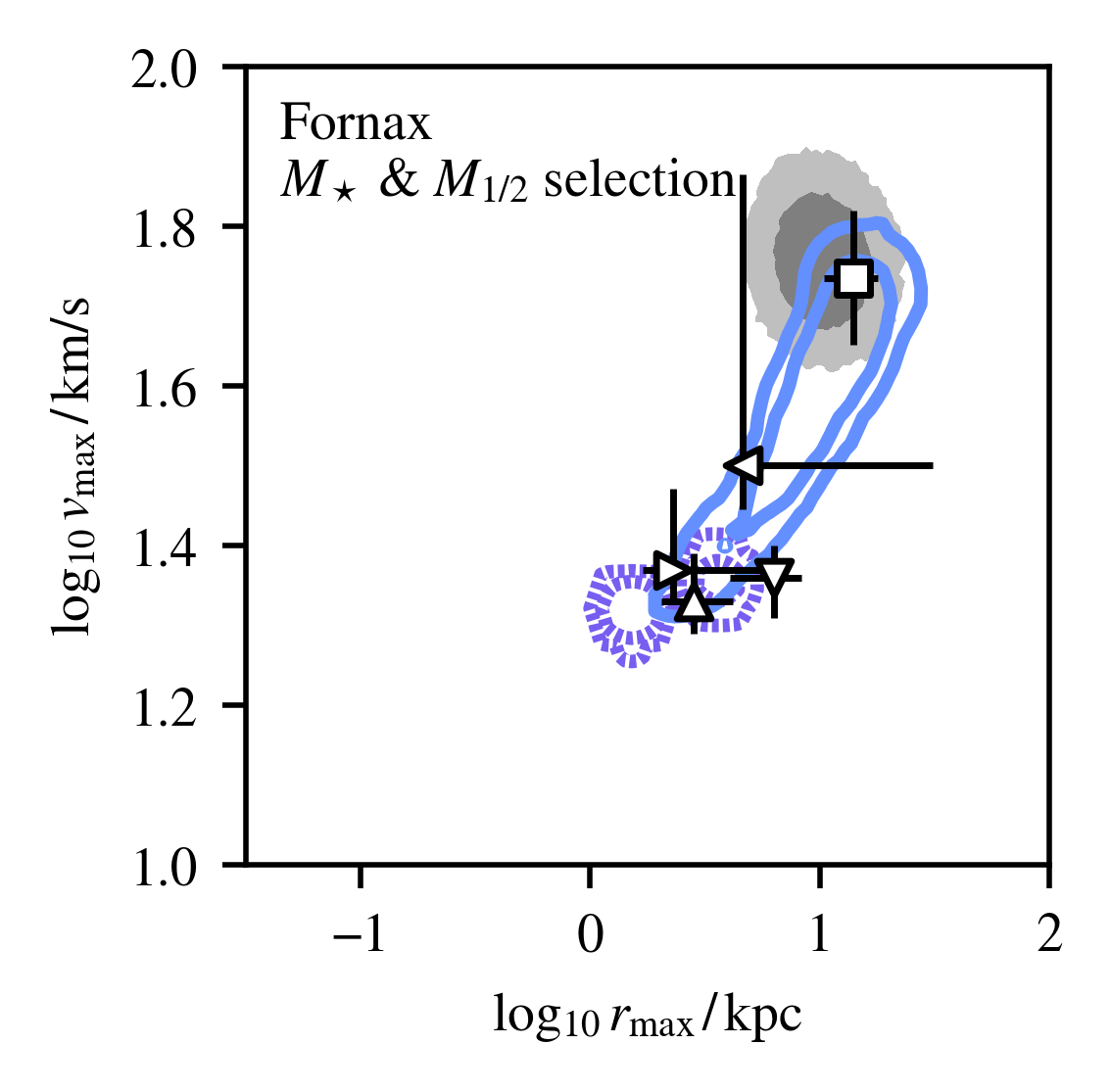}
	\includegraphics{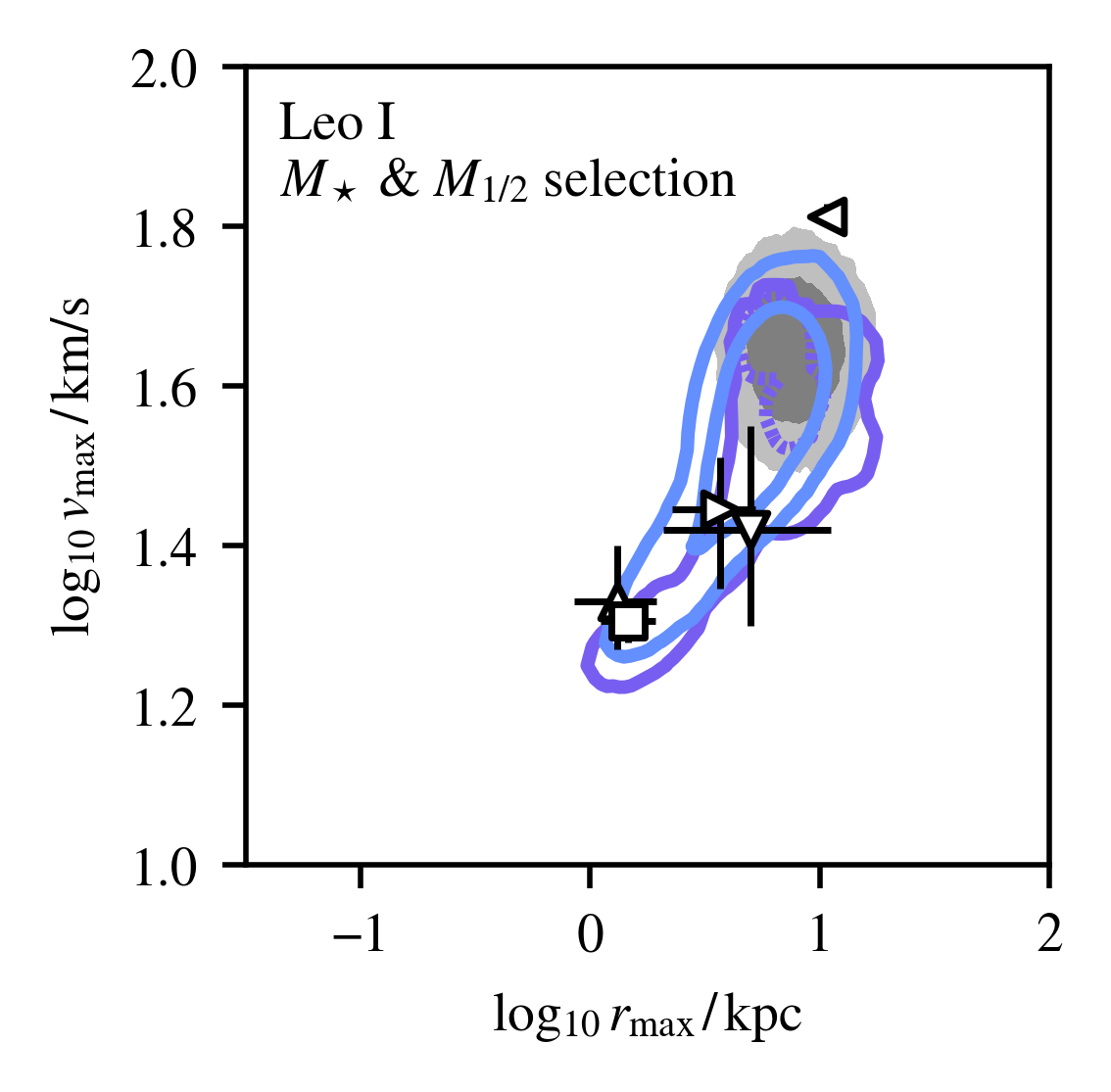}
	\includegraphics{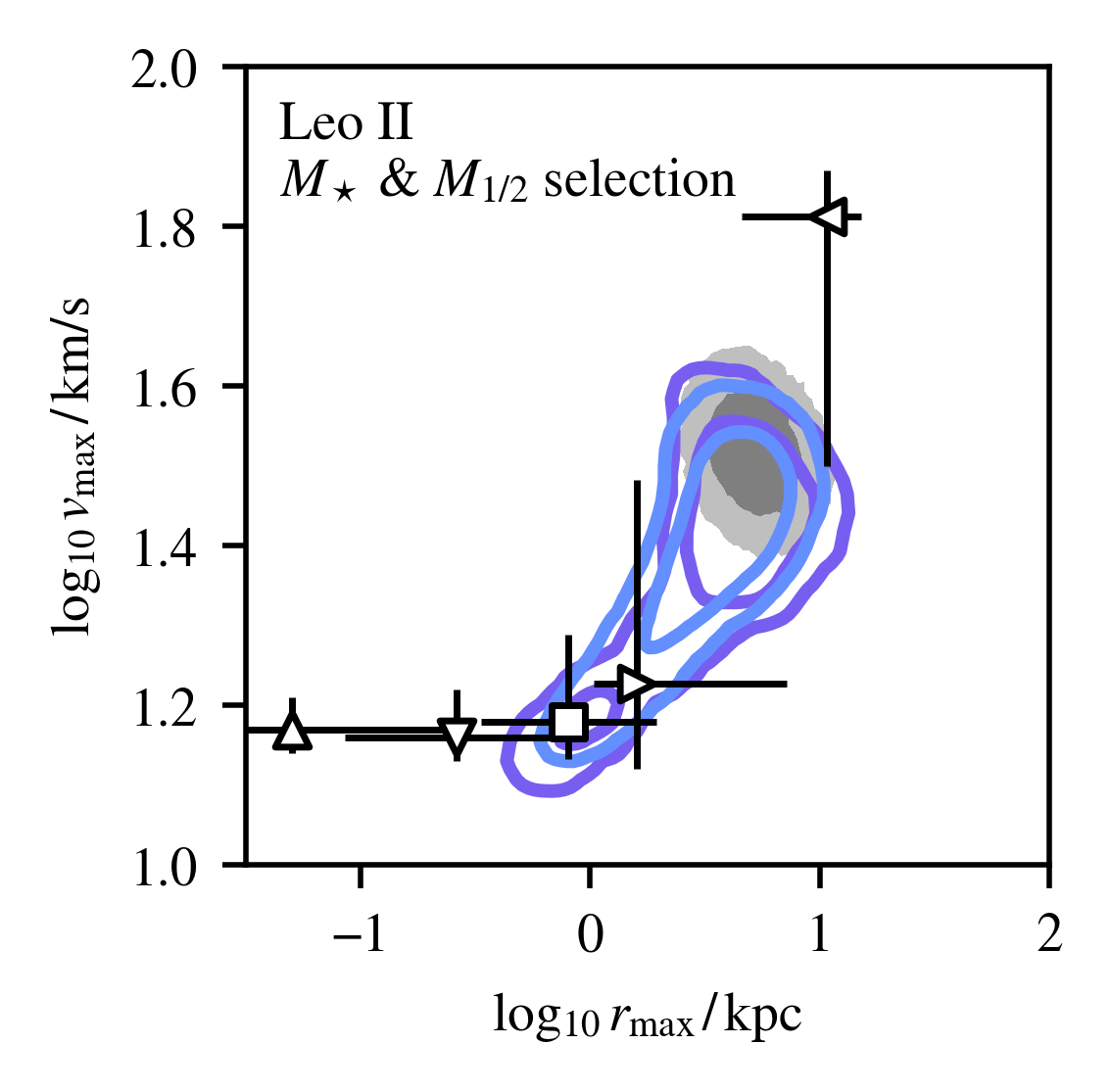}
	\includegraphics{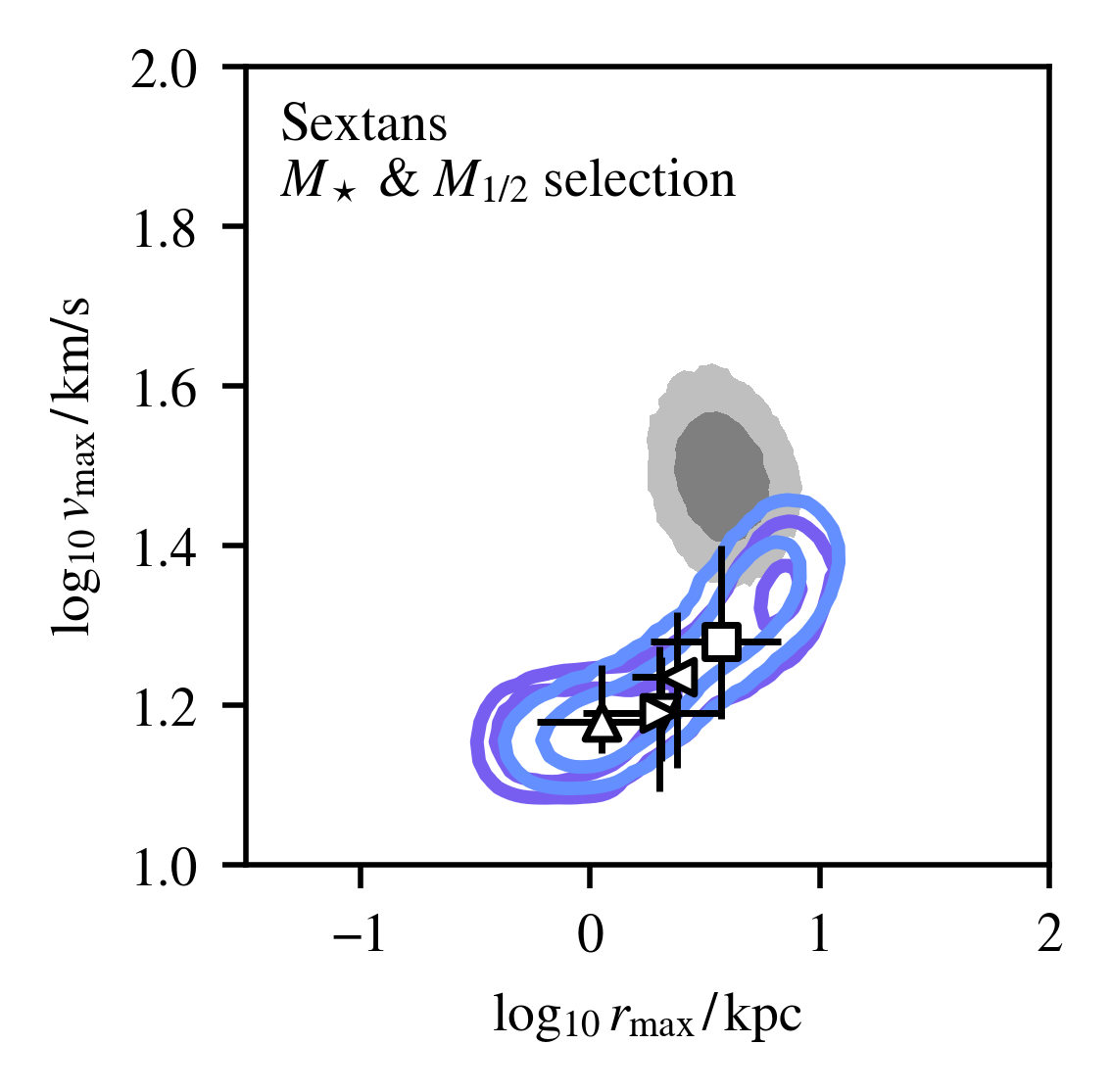}
	\includegraphics{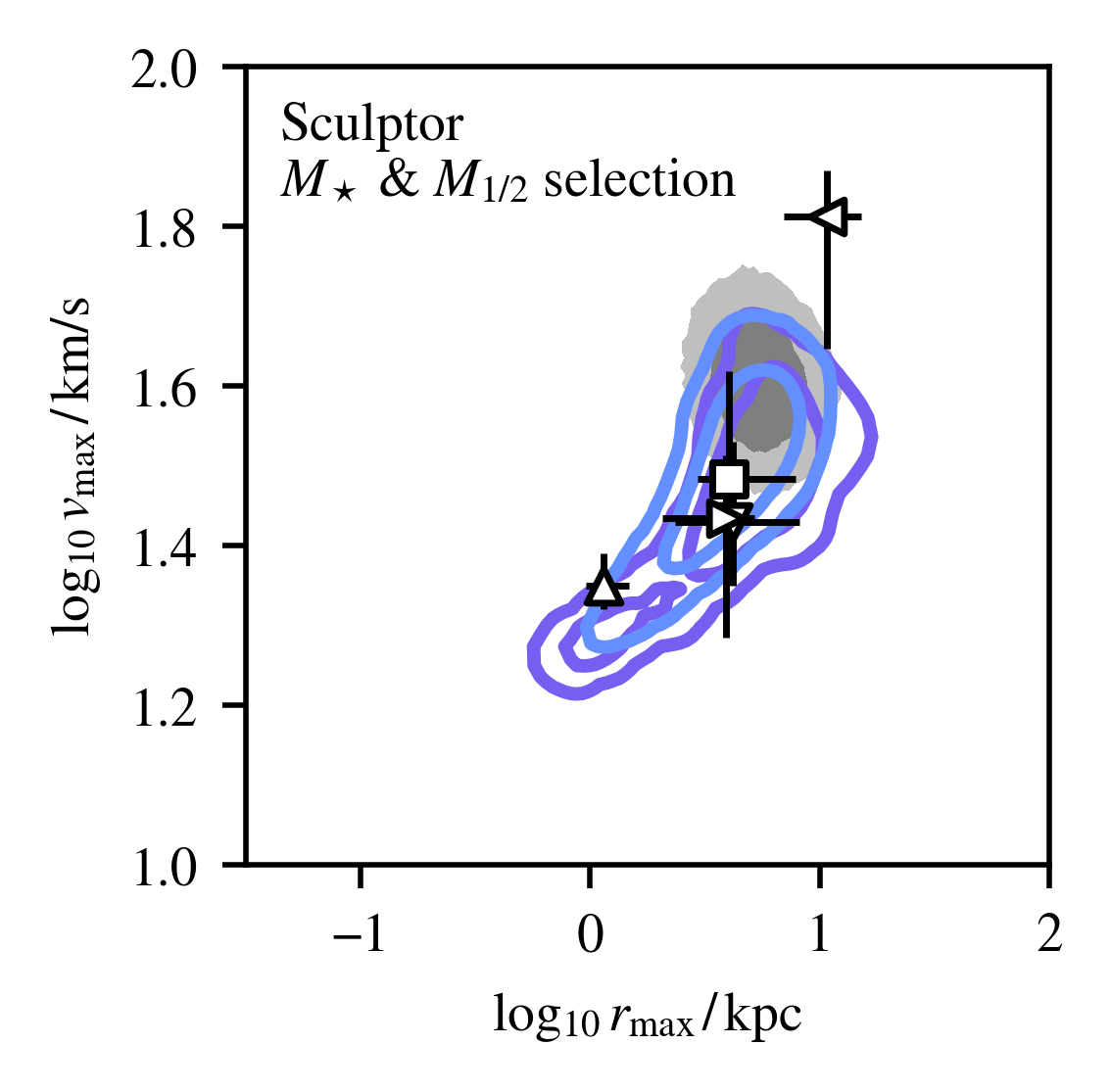}
	\includegraphics{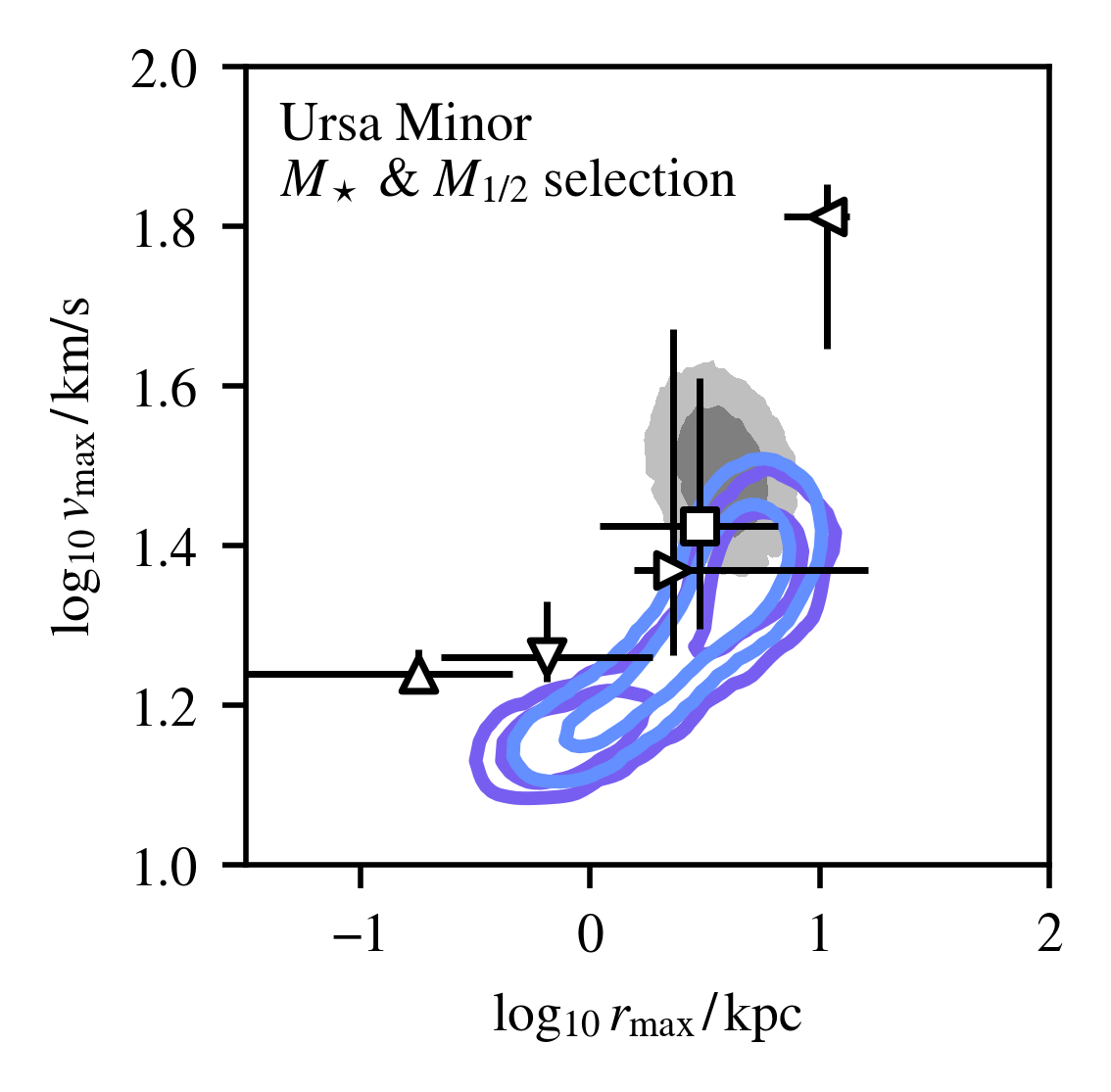}
	\caption{A version of \autoref{fig:3} for all classical satellites.}
 \label{fig:A1}
\end{figure}

\begin{figure}
	\centering
	\includegraphics{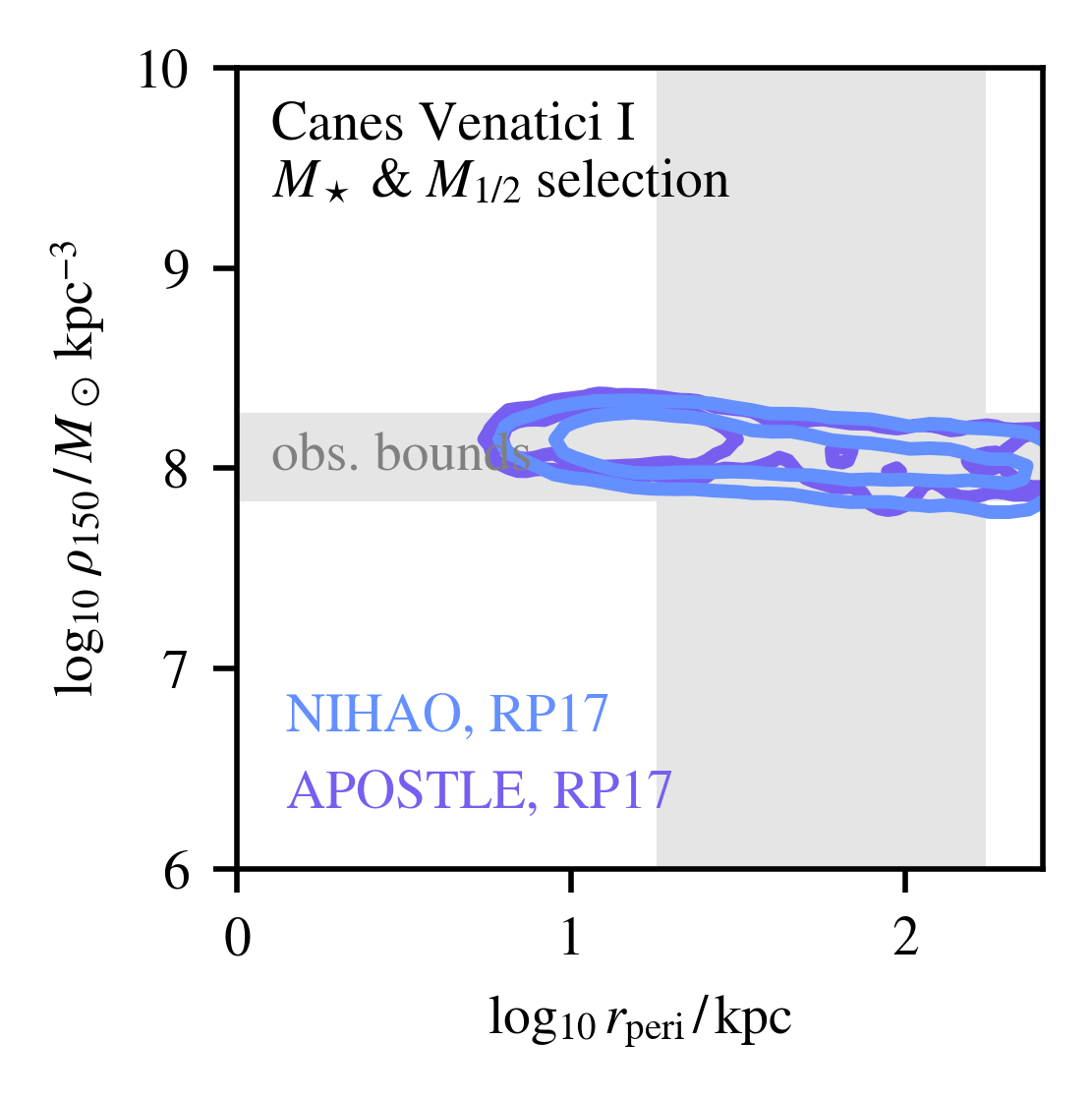}
	\includegraphics{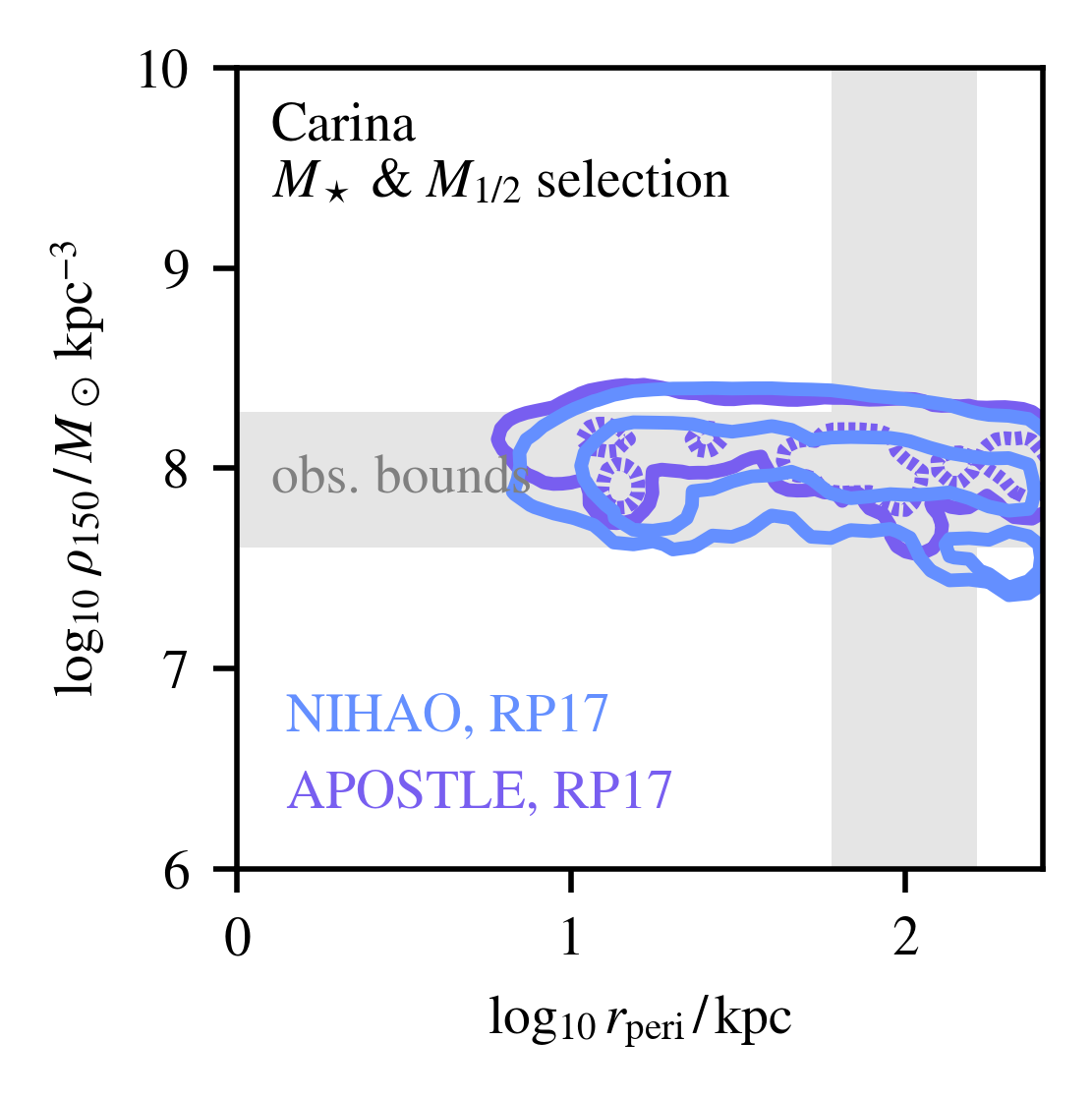}
	\includegraphics{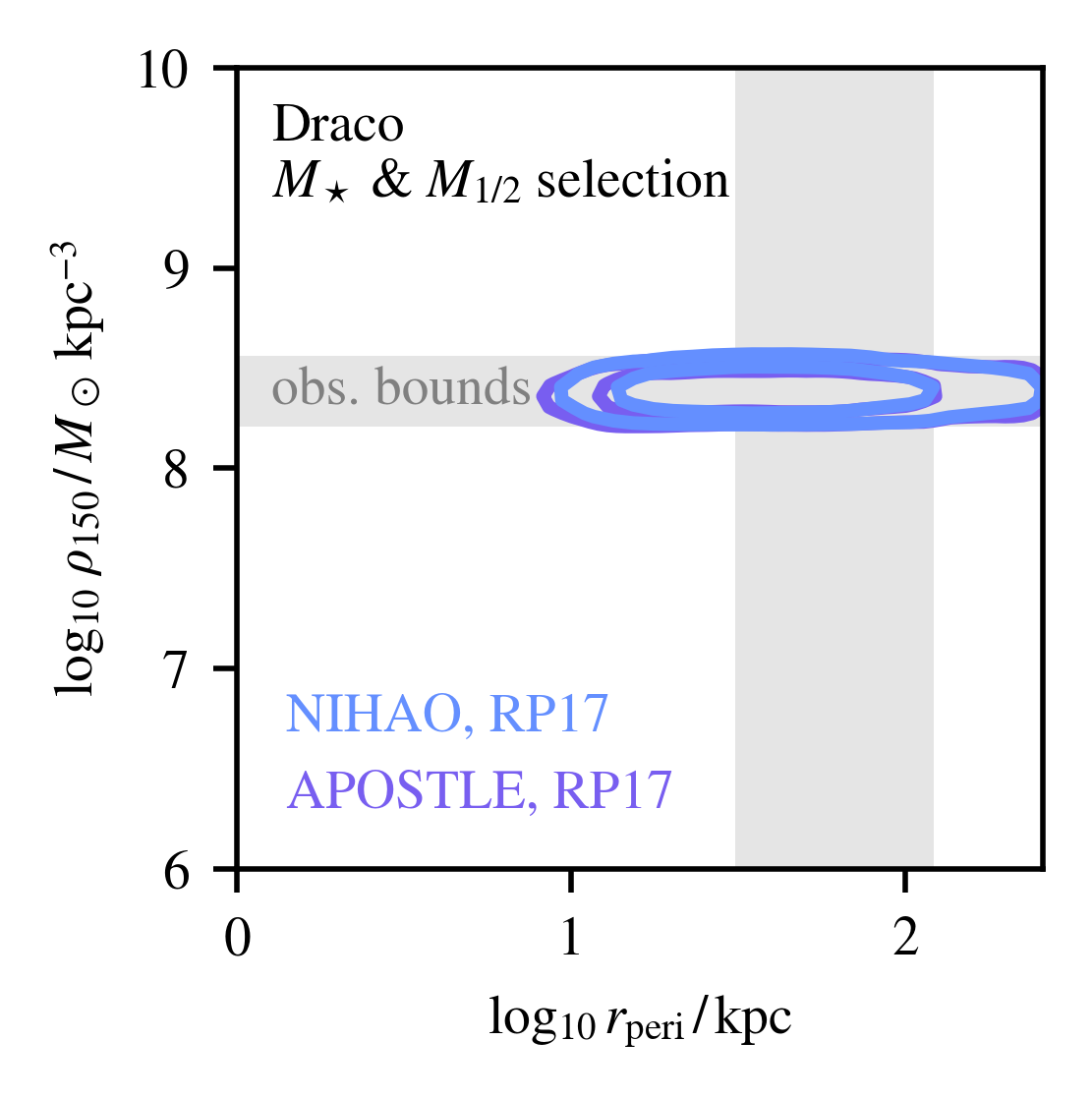}
	\includegraphics{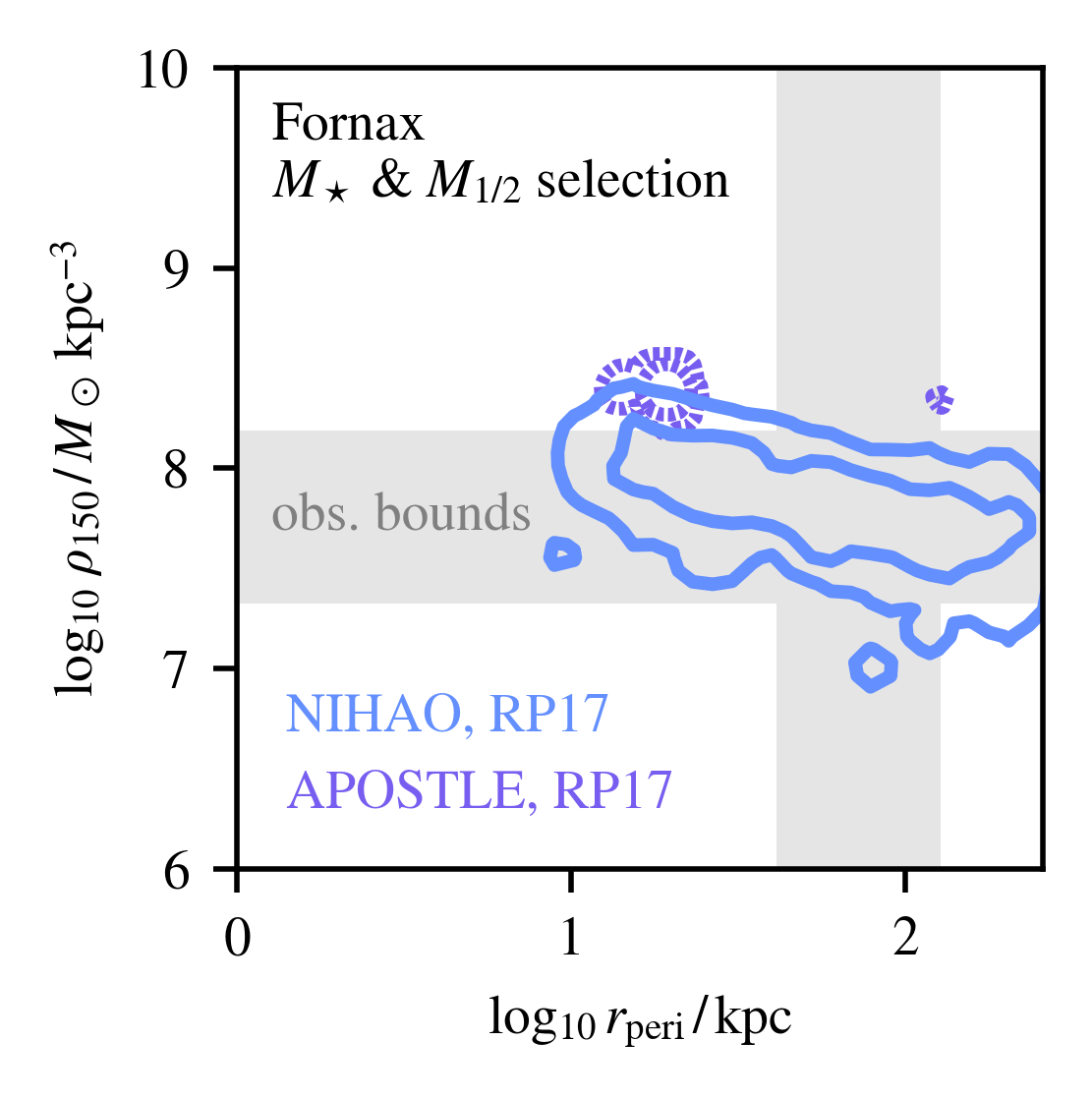}
	\includegraphics{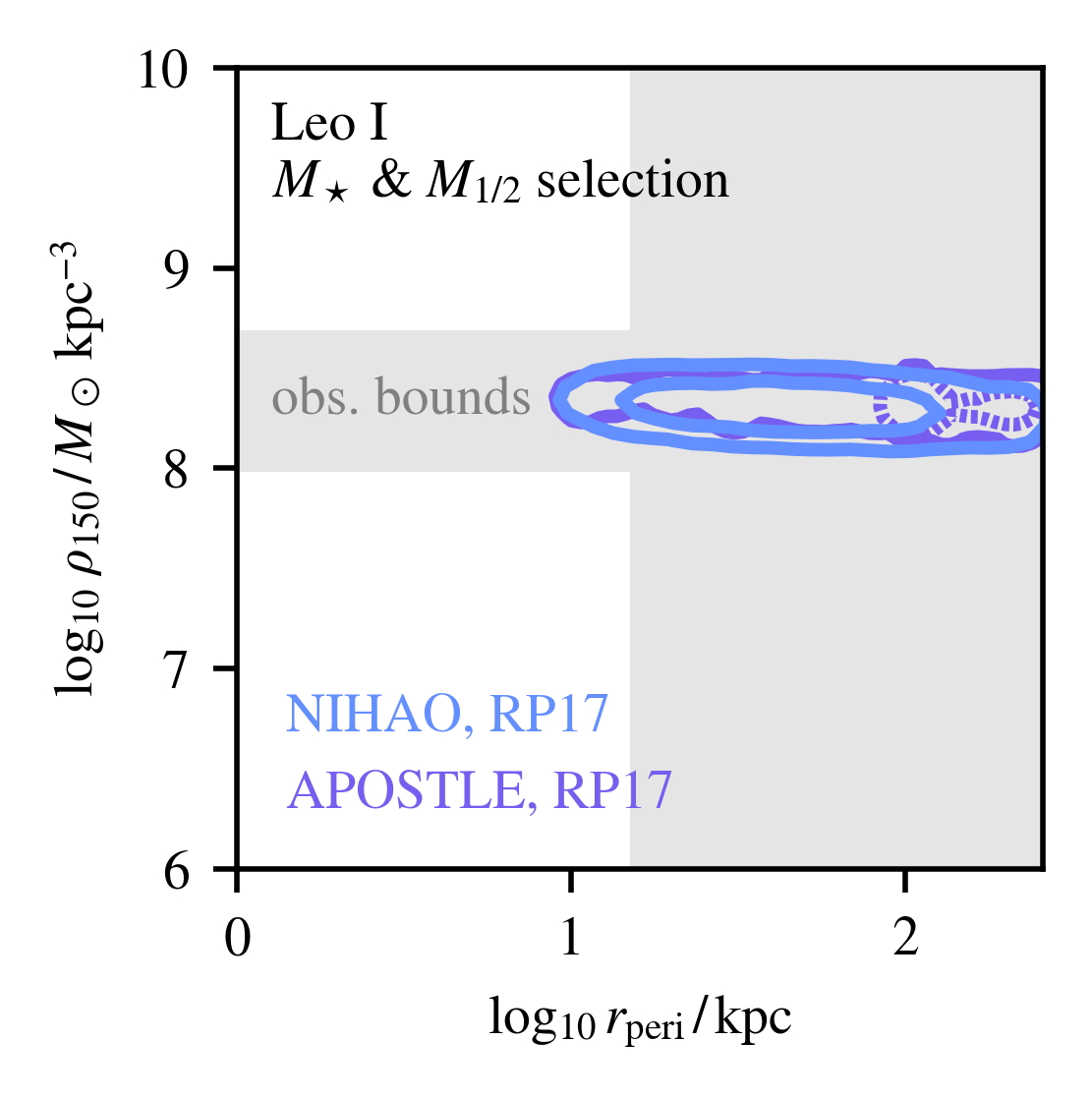}
	\includegraphics{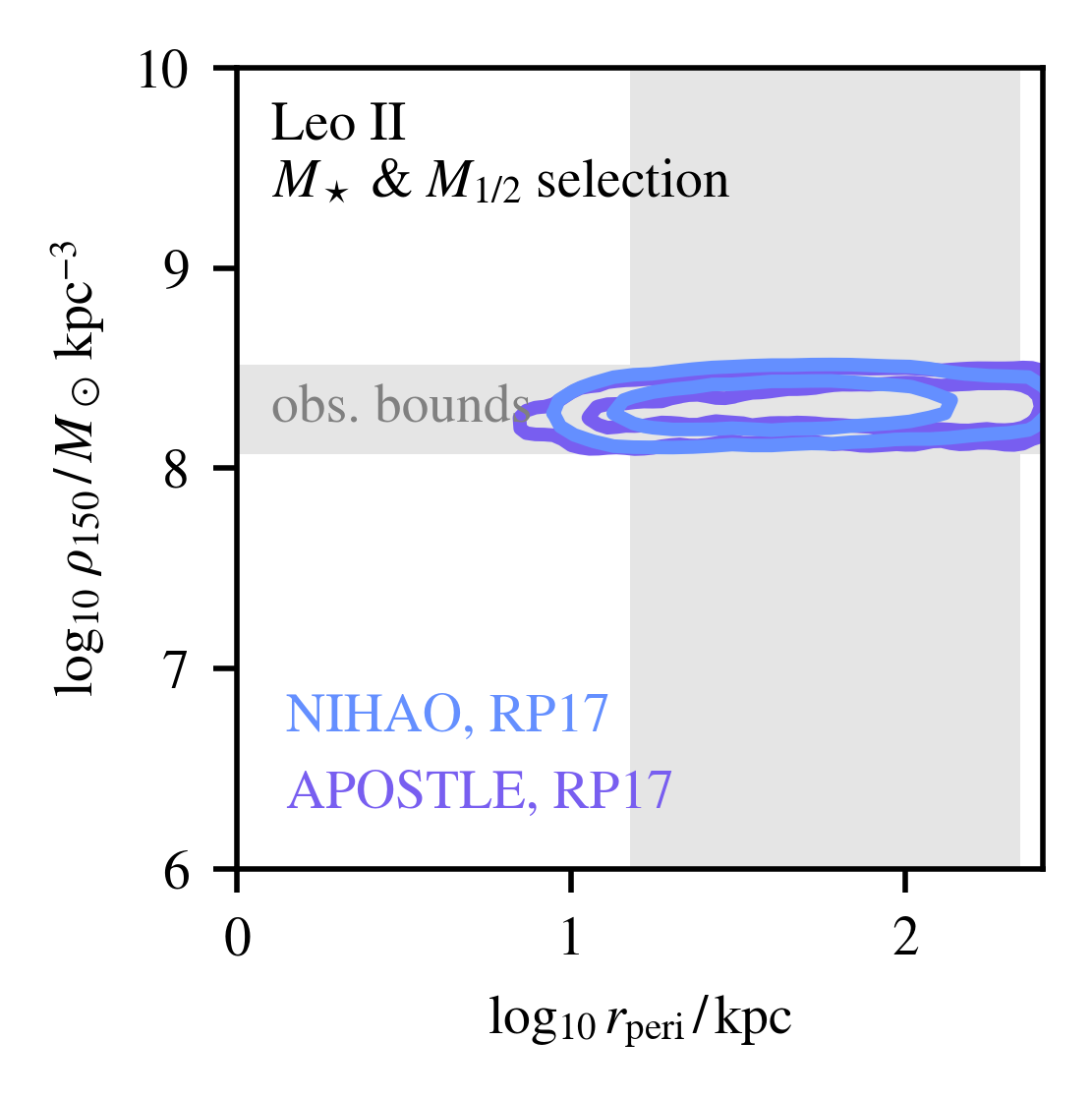}
	\includegraphics{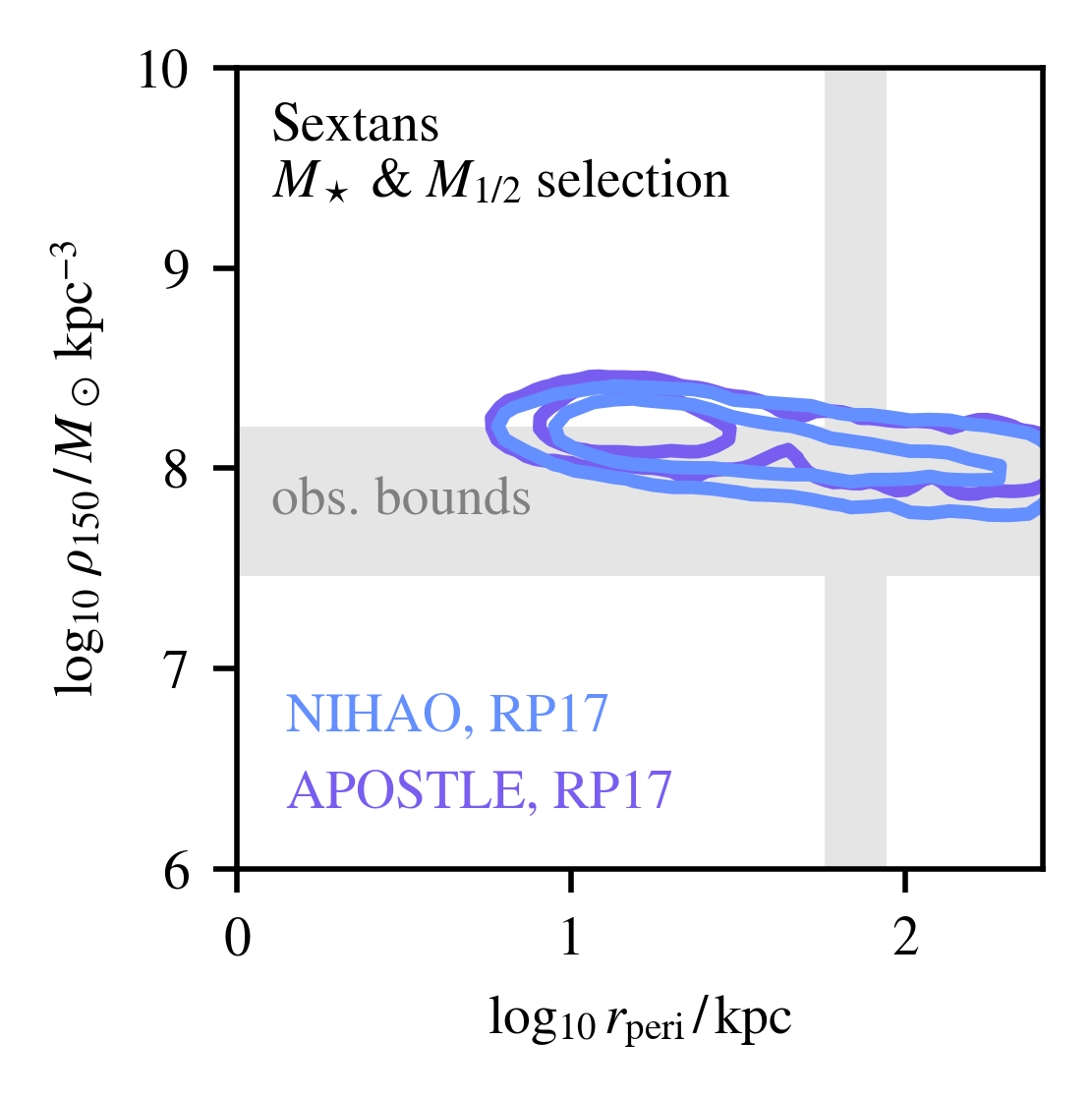}
	\includegraphics{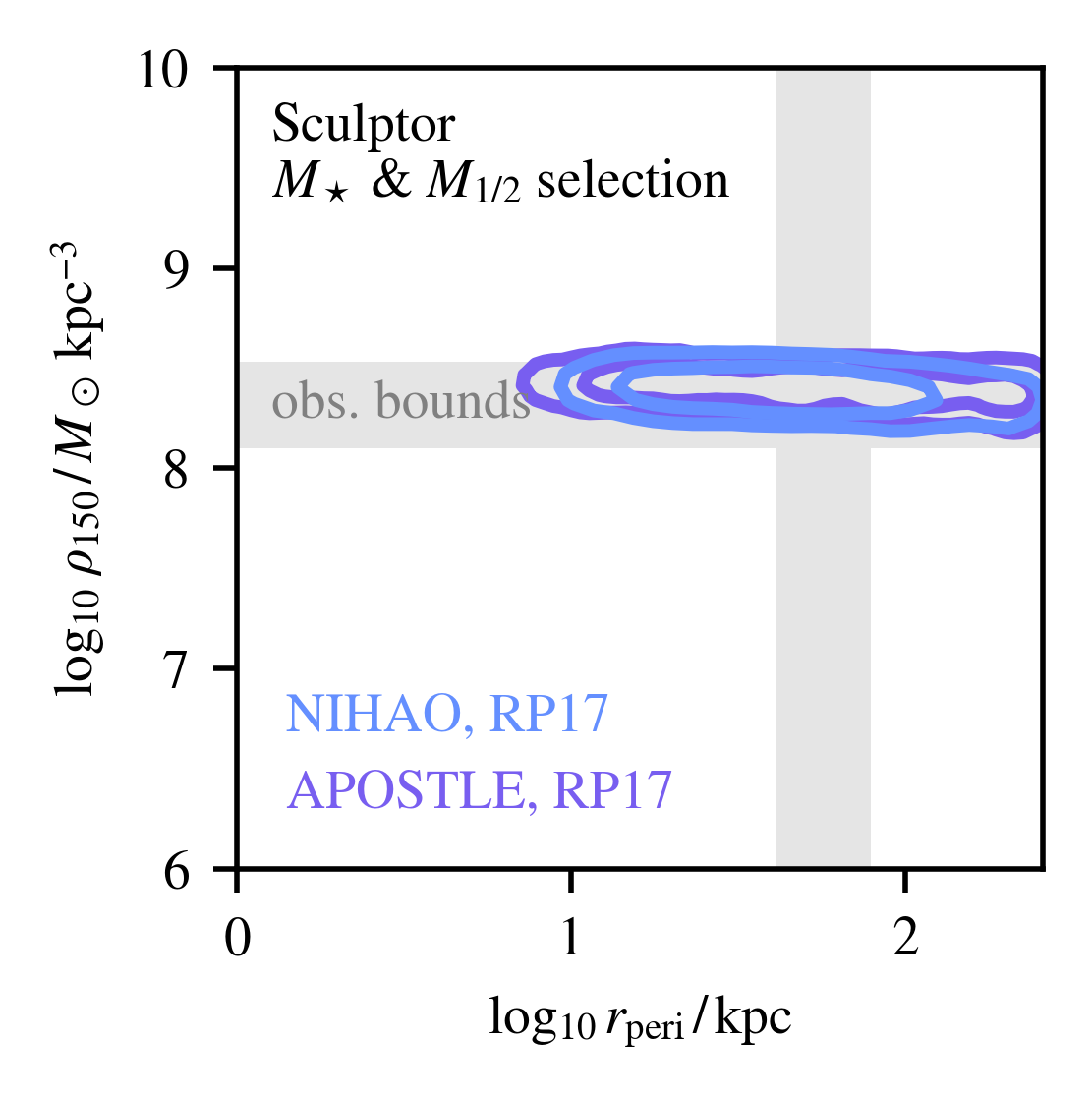}
	\includegraphics{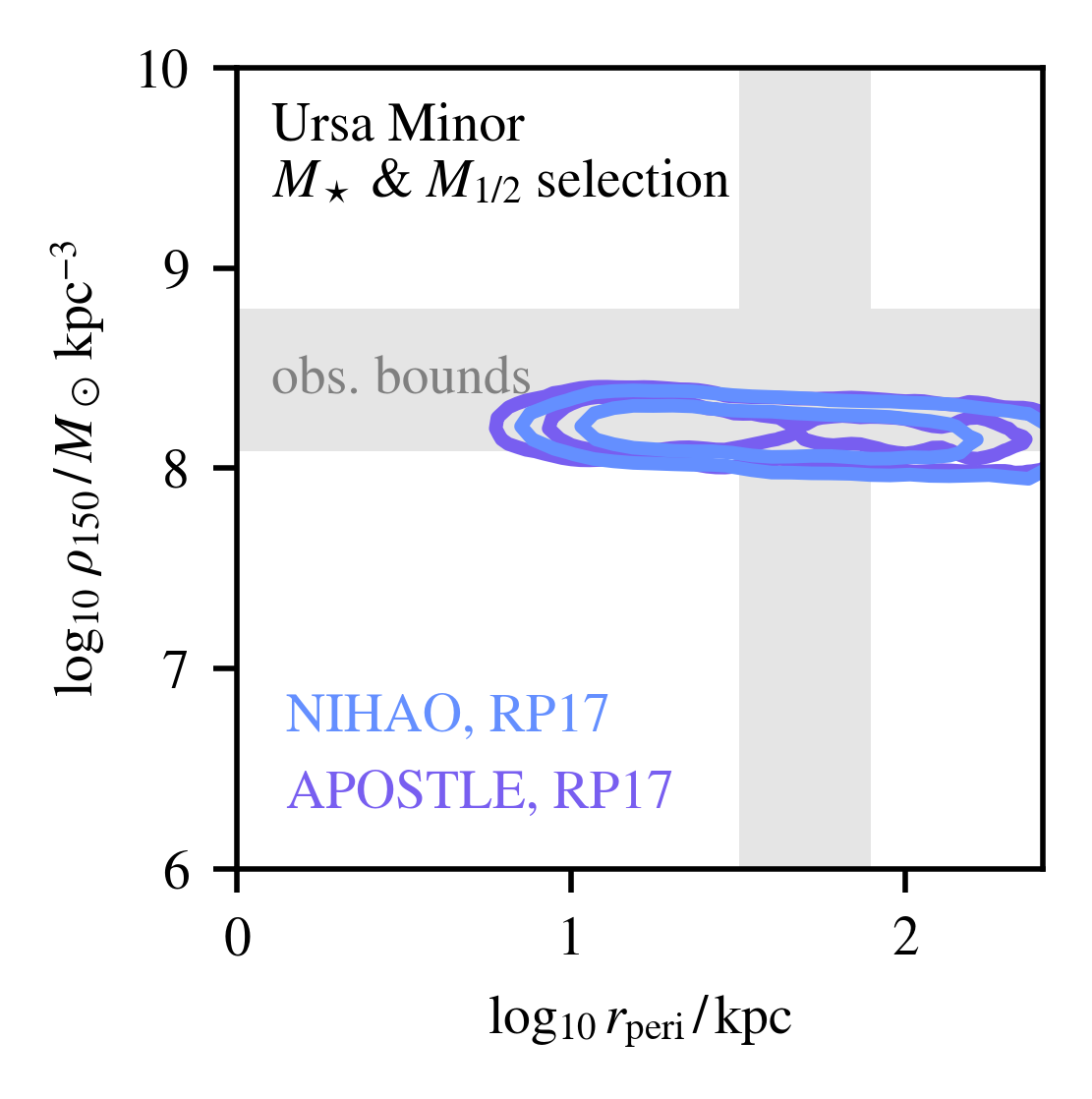}
	\caption{A version of \autoref{fig:4} for all classical satellites. Notably, the inference for $\rho_{150}$ is generally consistent with observation, though Sextans prefers the denser side of the observed region and Ursa Minor the less dense side.}
 \label{fig:A4}
\end{figure}
\begin{figure}
	\centering
	\includegraphics{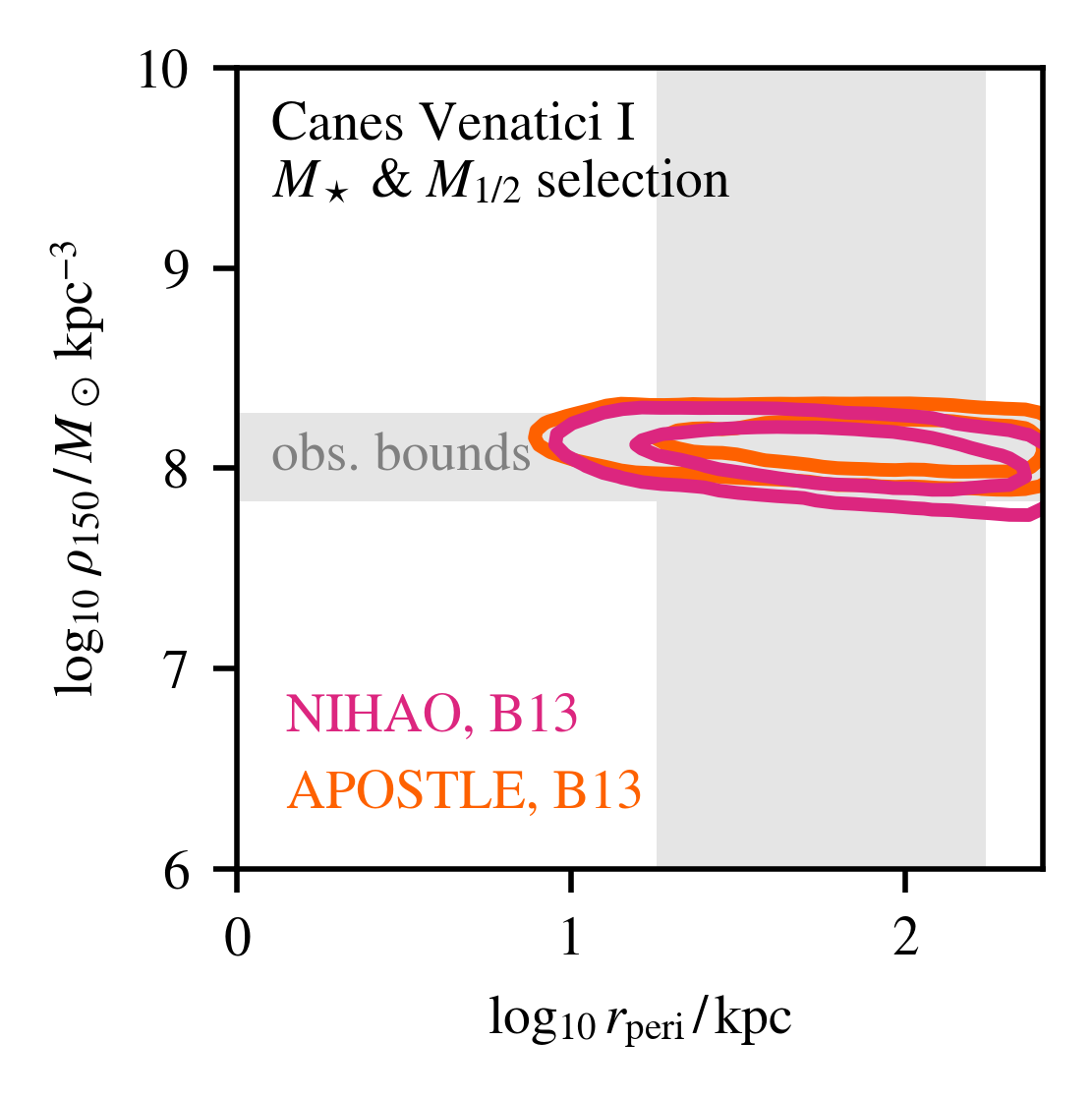}
	\includegraphics{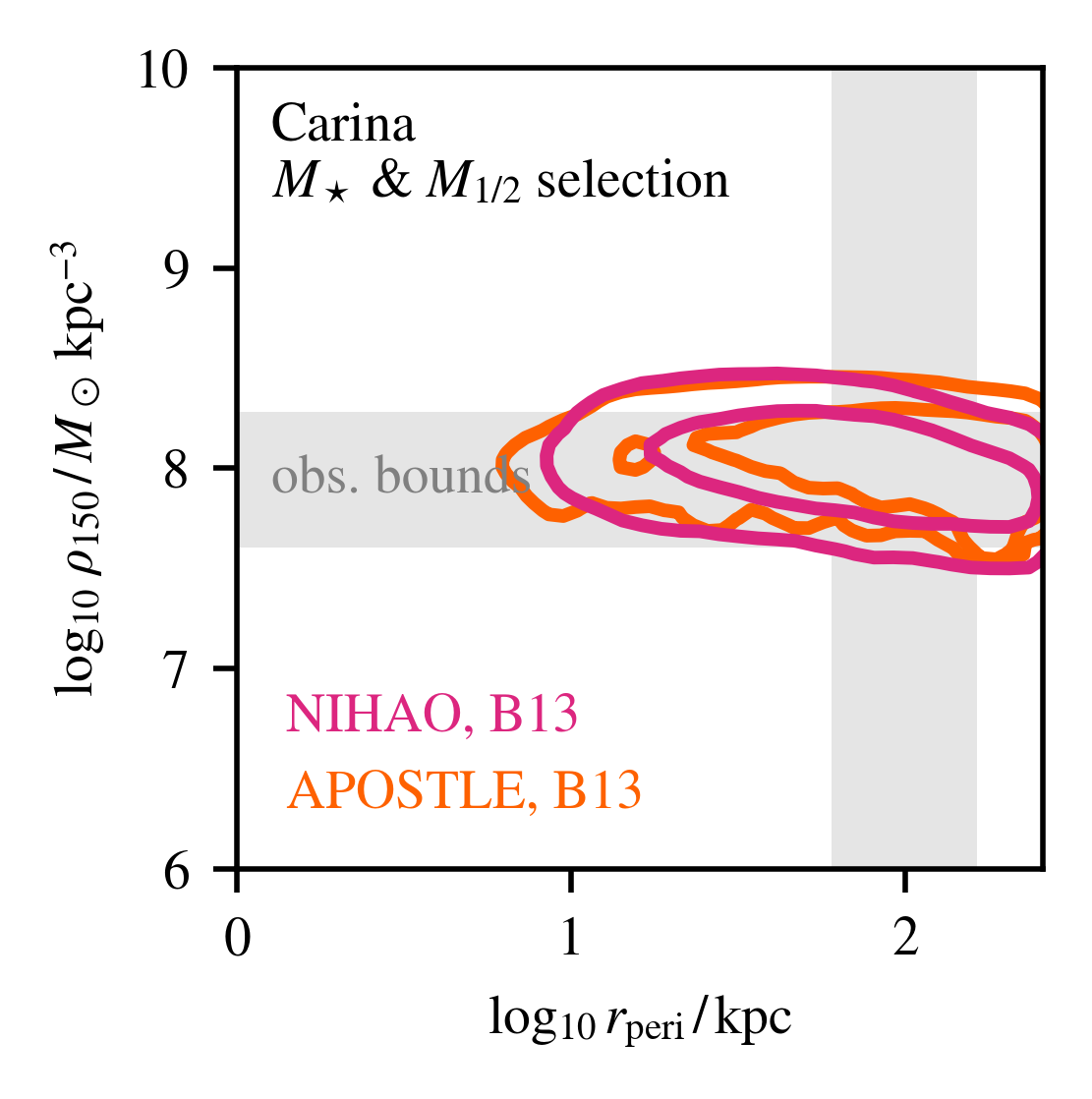}
	\includegraphics{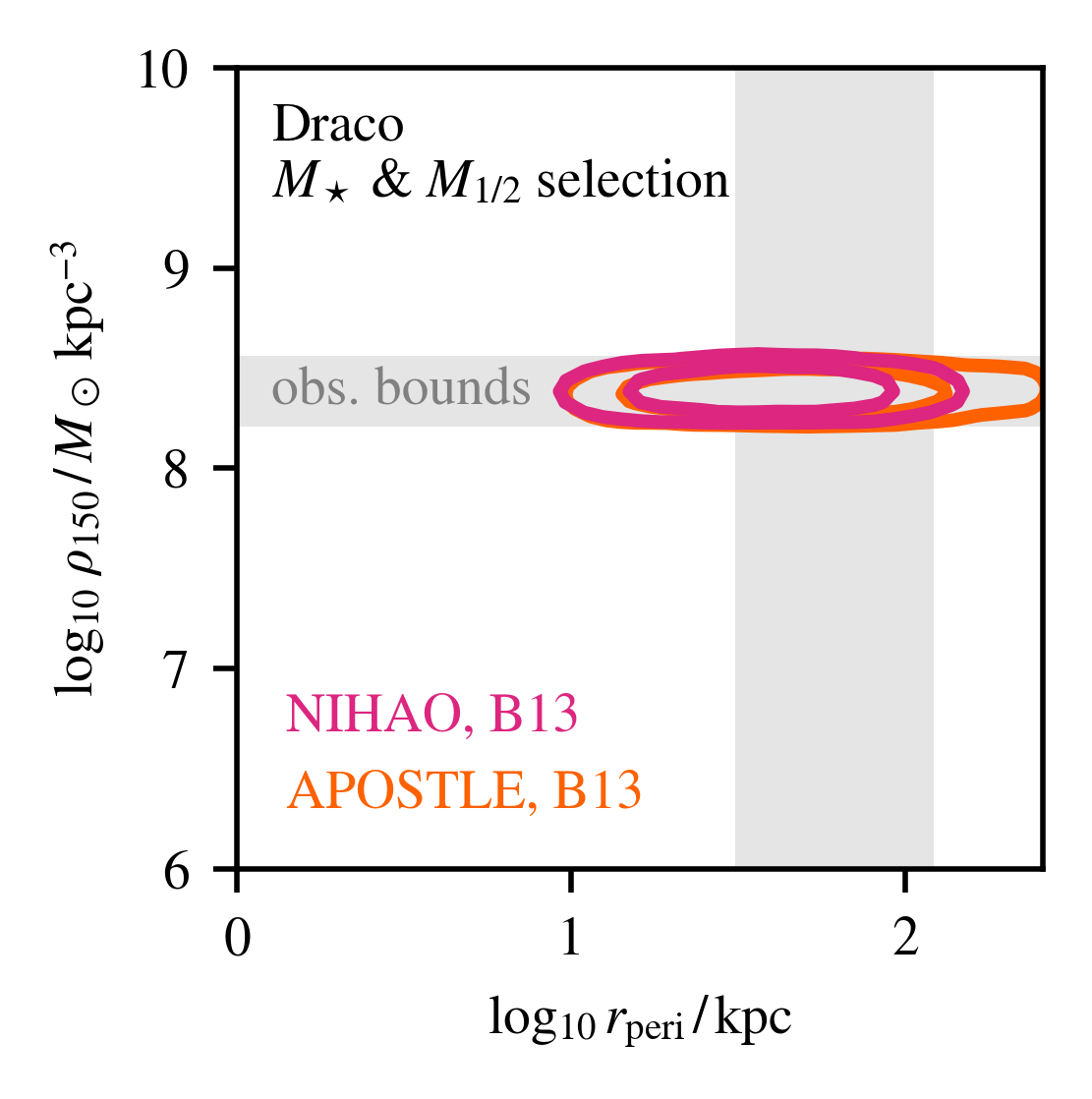}
	\includegraphics{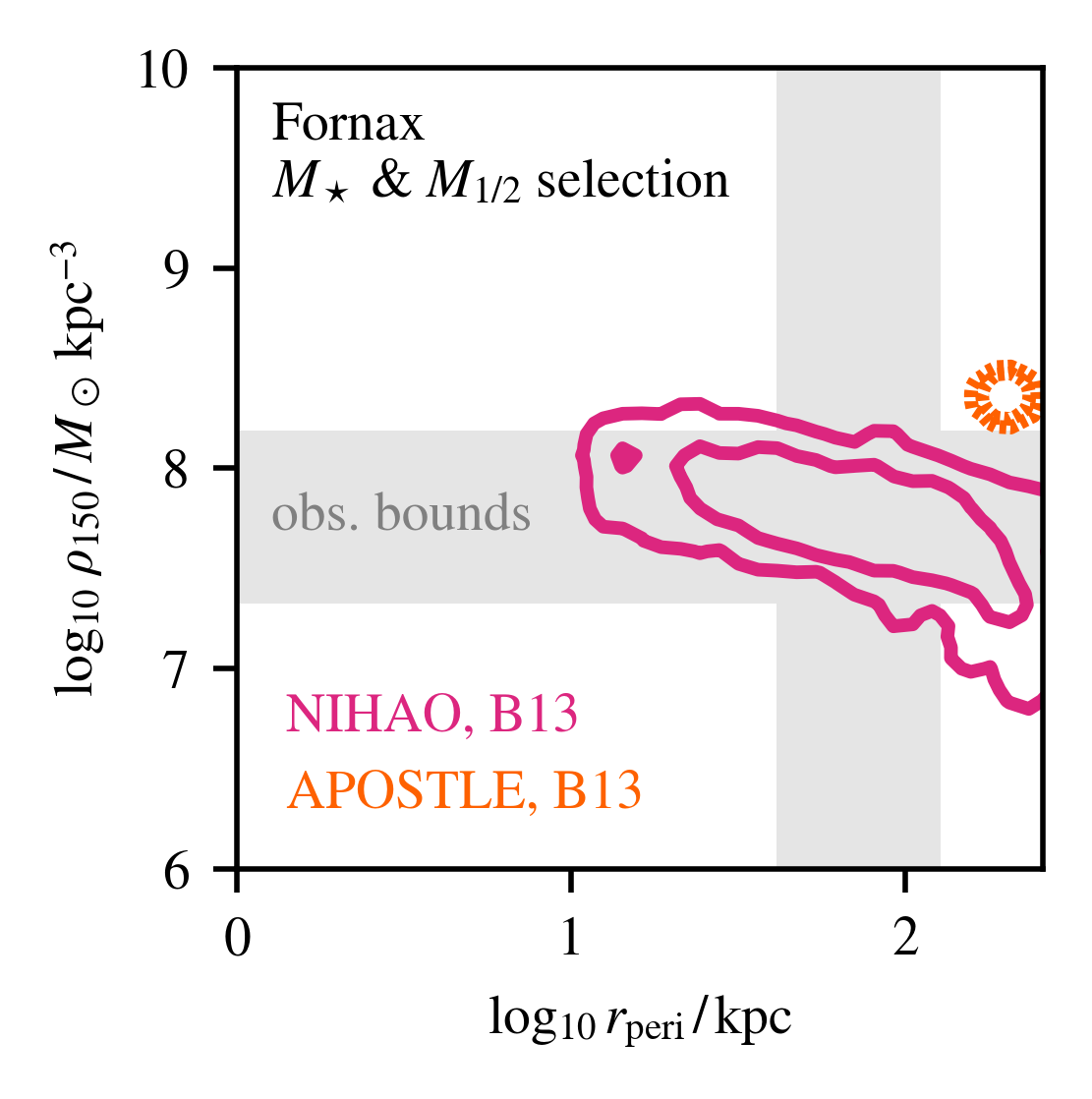}
	\includegraphics{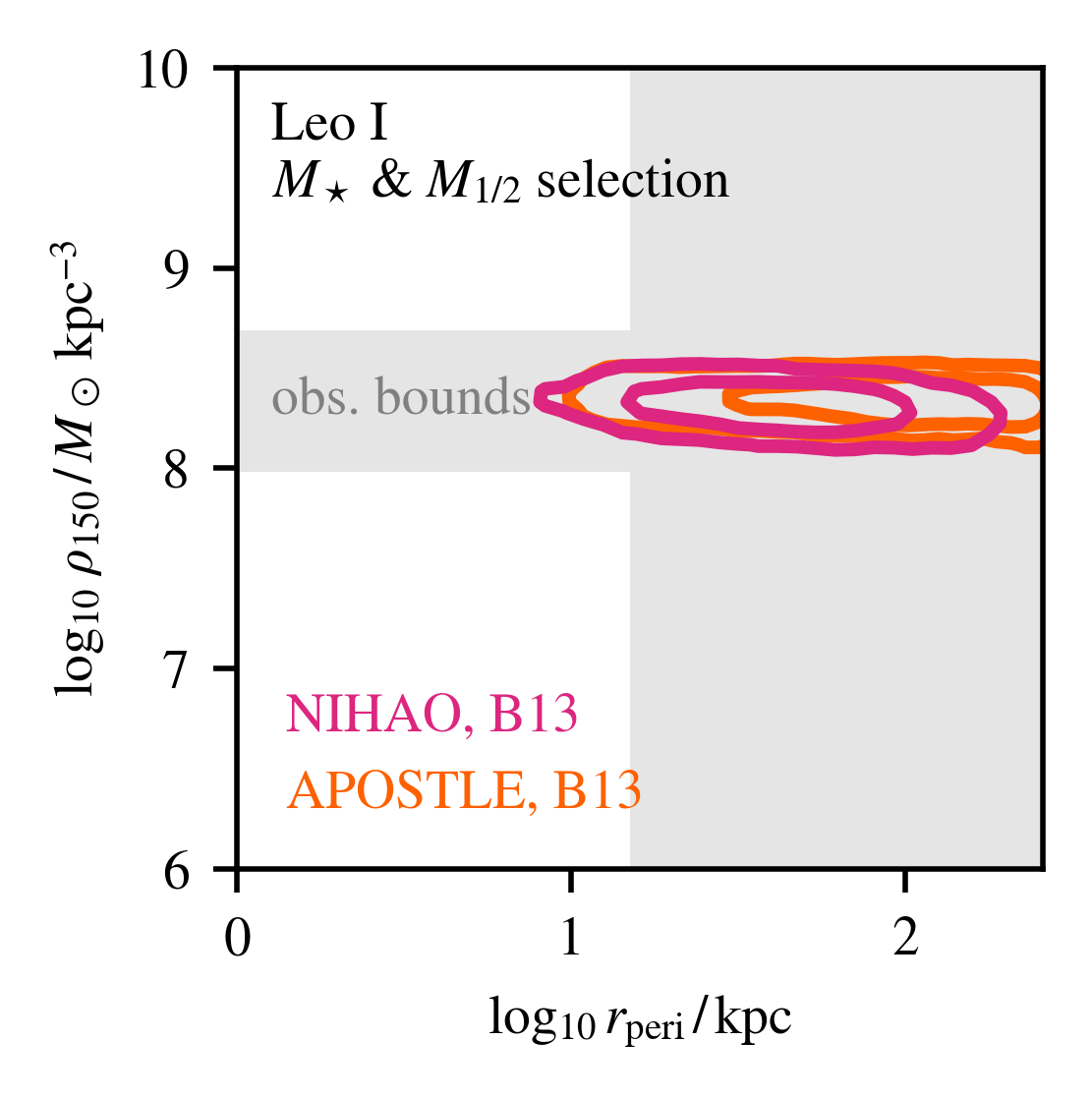}
	\includegraphics{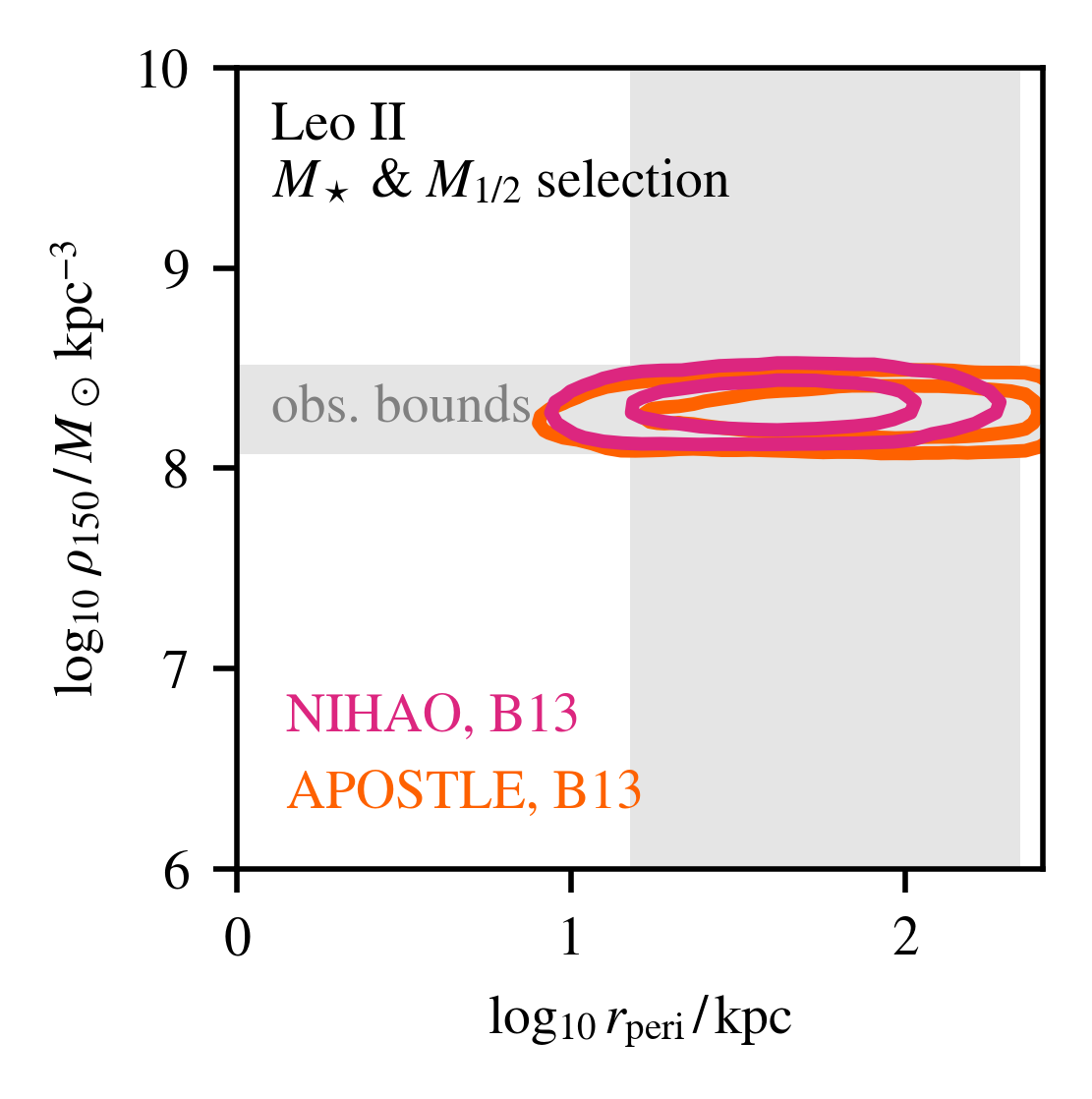}
	\includegraphics{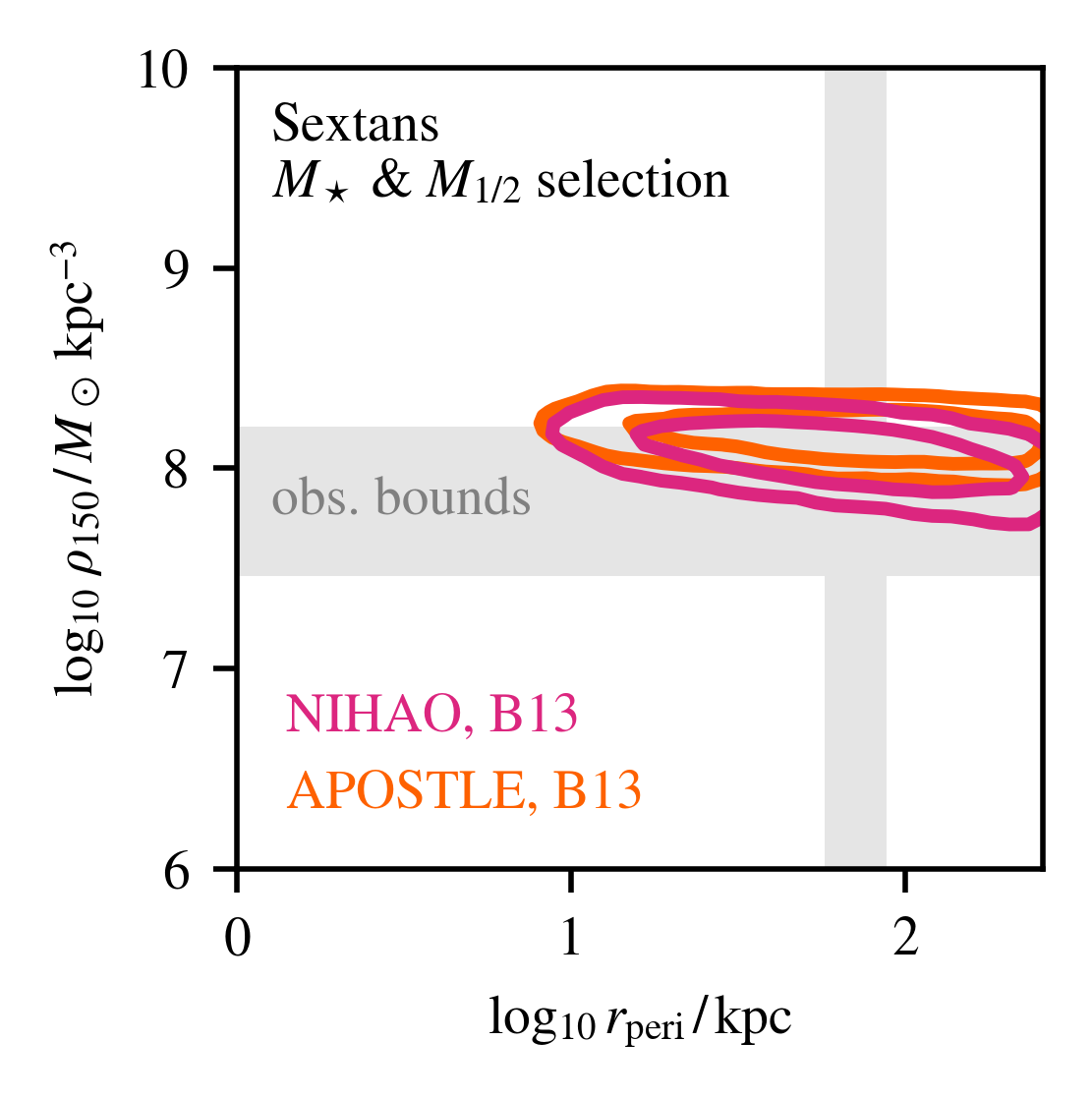}
	\includegraphics{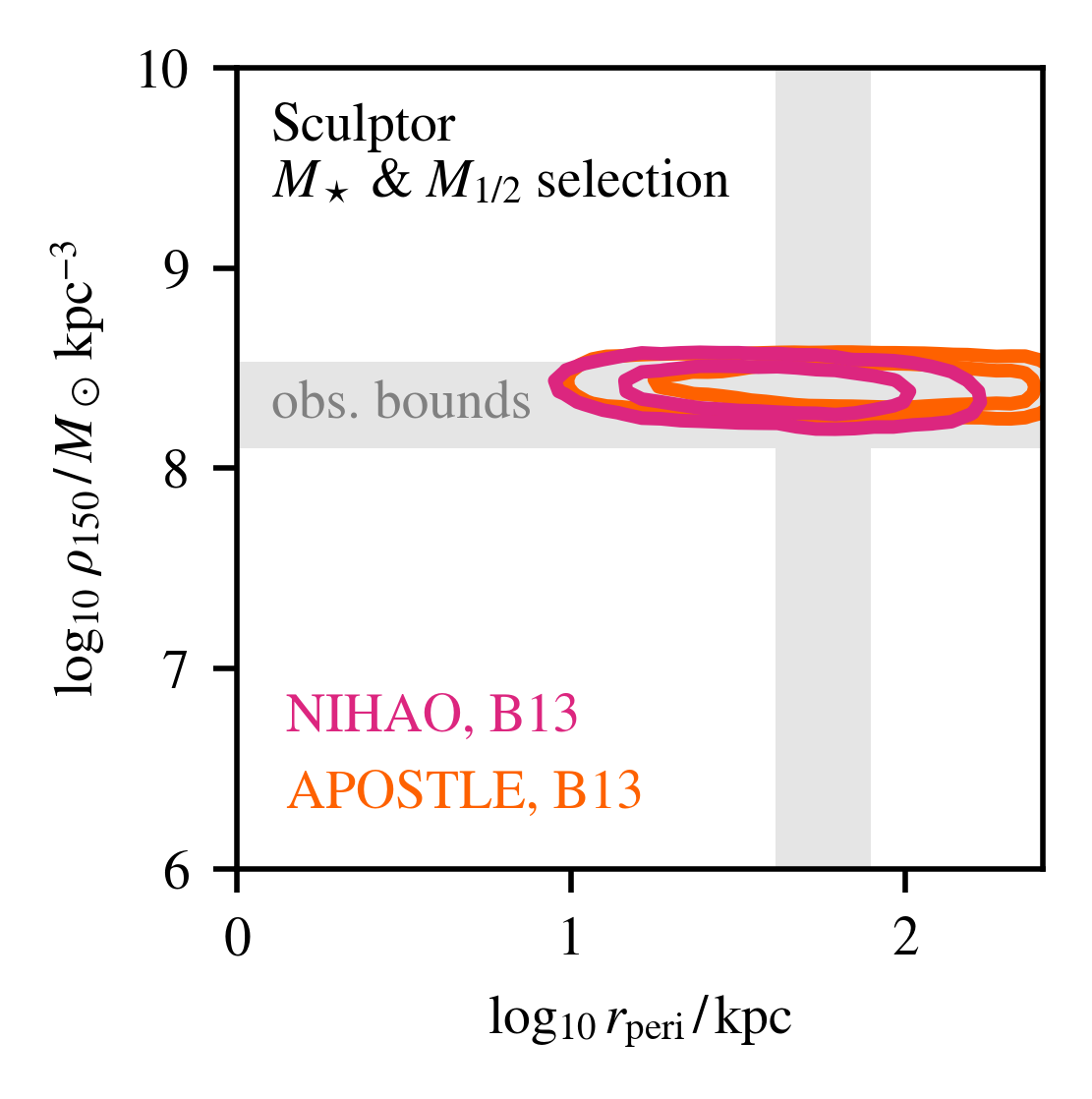}
	\includegraphics{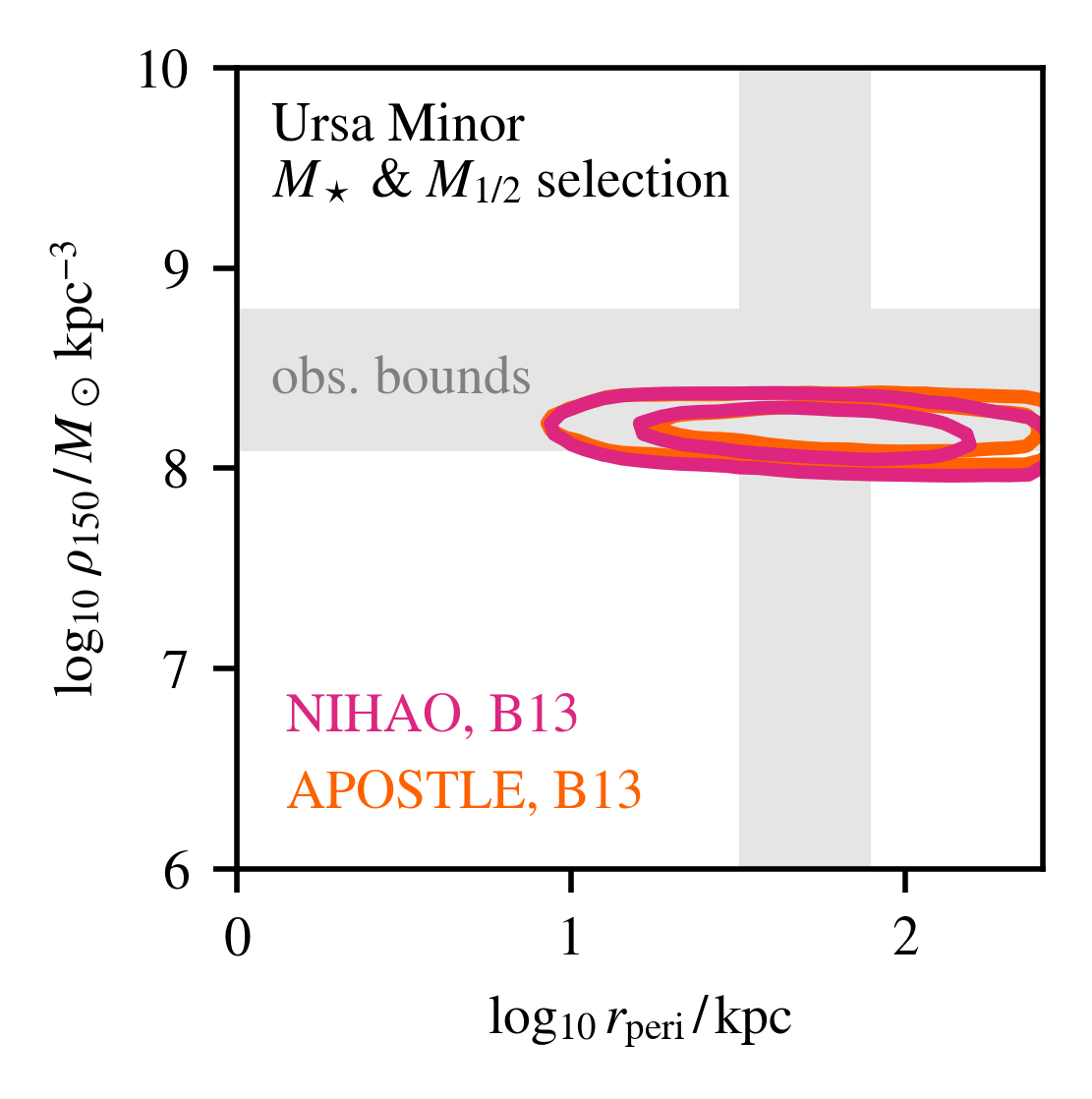}
	\caption{A version of \autoref{fig:4} for all classical satellites, using the \BSMHM{} SMHM relation. Despite placing larger galaxies into smaller haloes, the results are quite similar to the \RPSMHM{} model shown above in \autoref{fig:A4}.}
 \label{fig:A5}
\end{figure}

\begin{figure}
	\centering
	\includegraphics{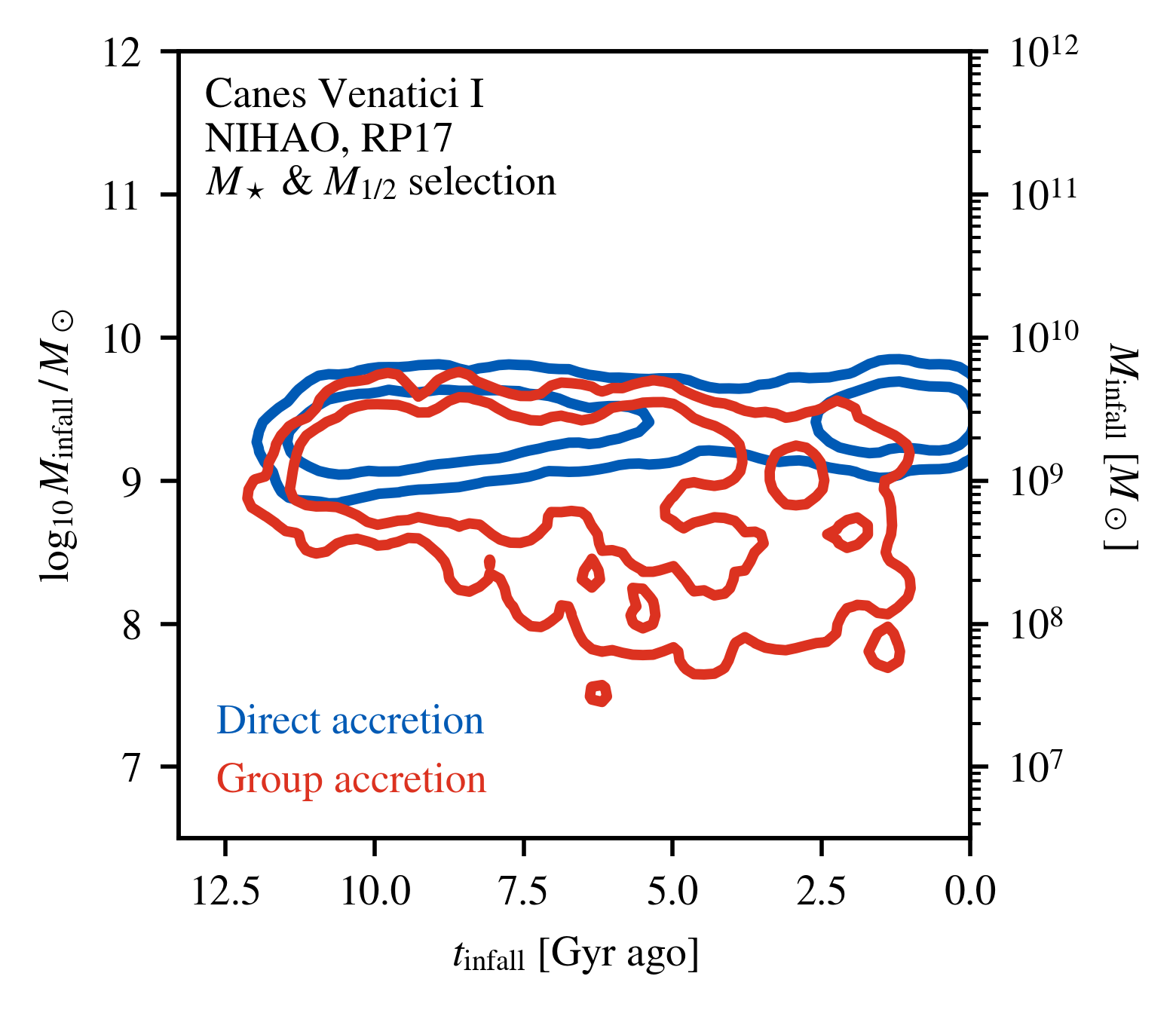}
	\includegraphics{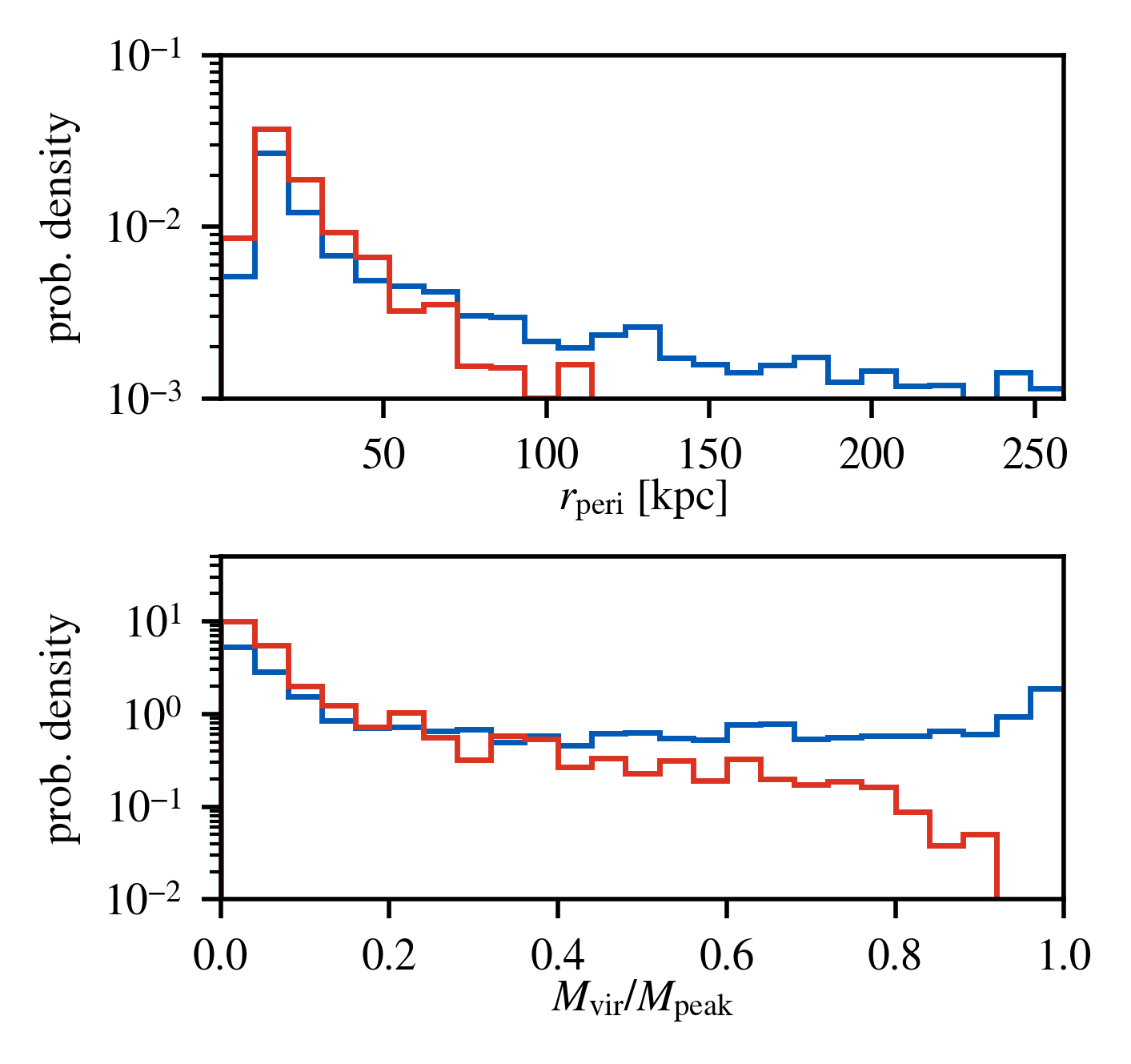}
	\includegraphics{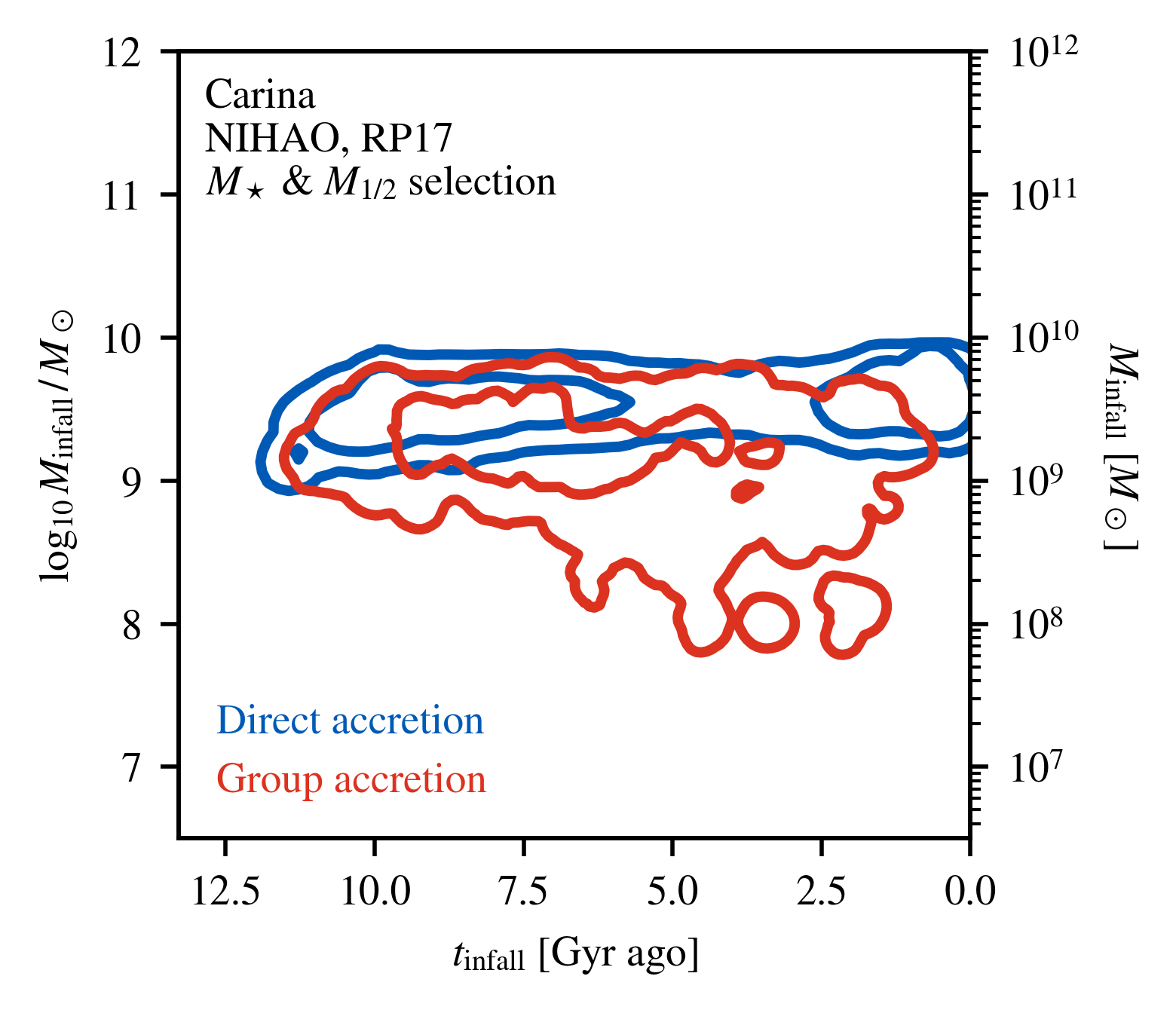}
	\includegraphics{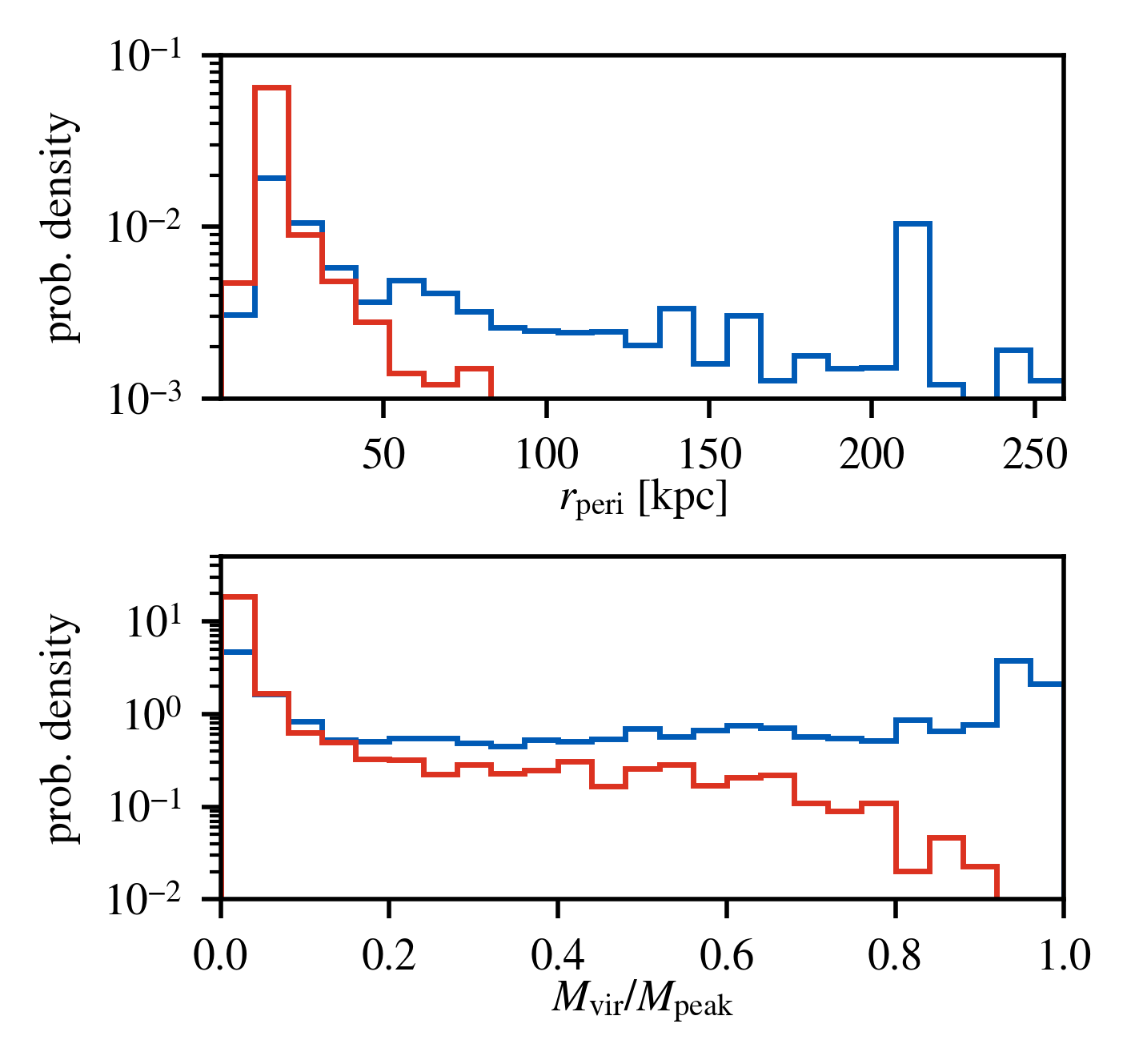}
	\includegraphics{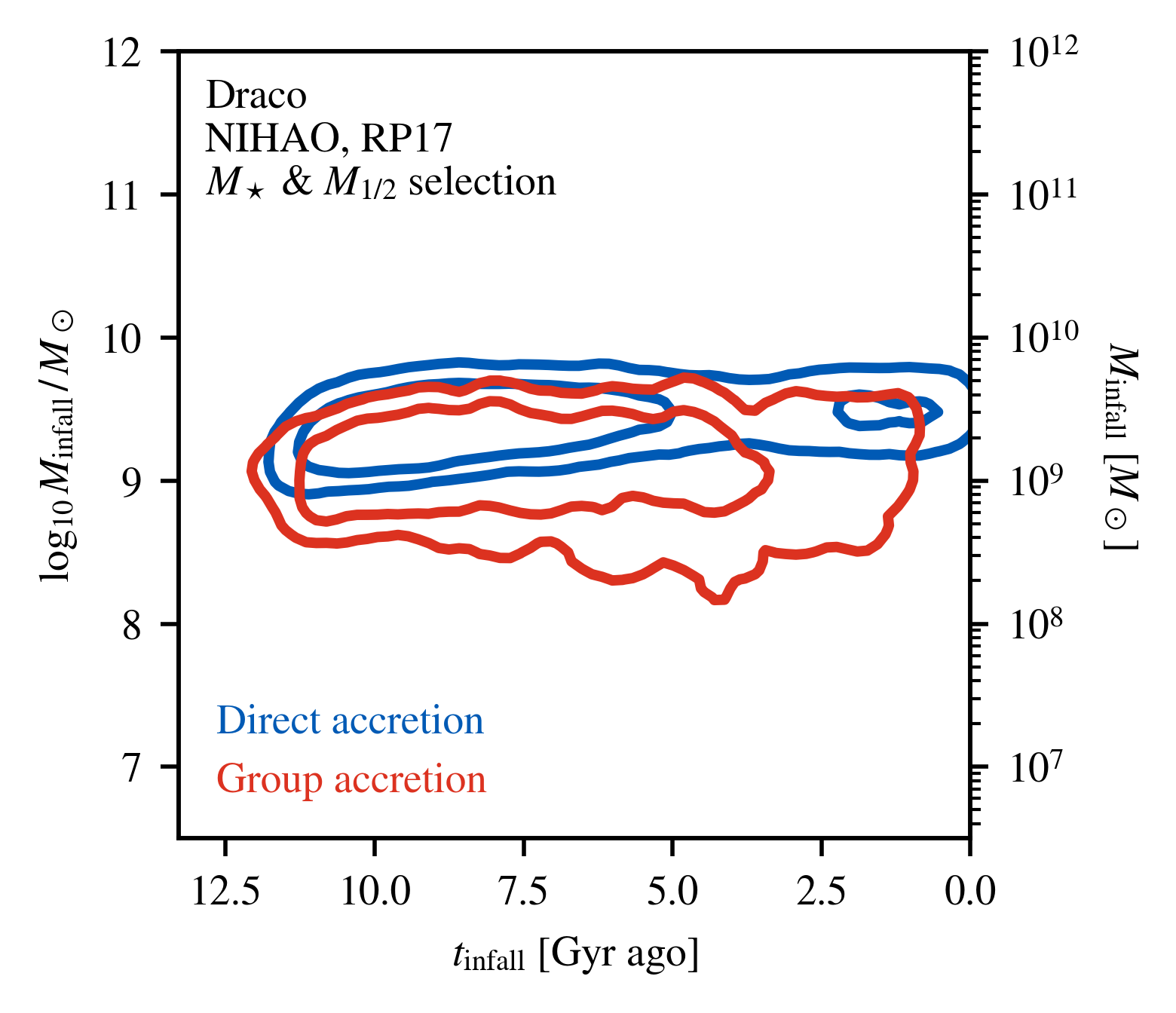}
	\includegraphics{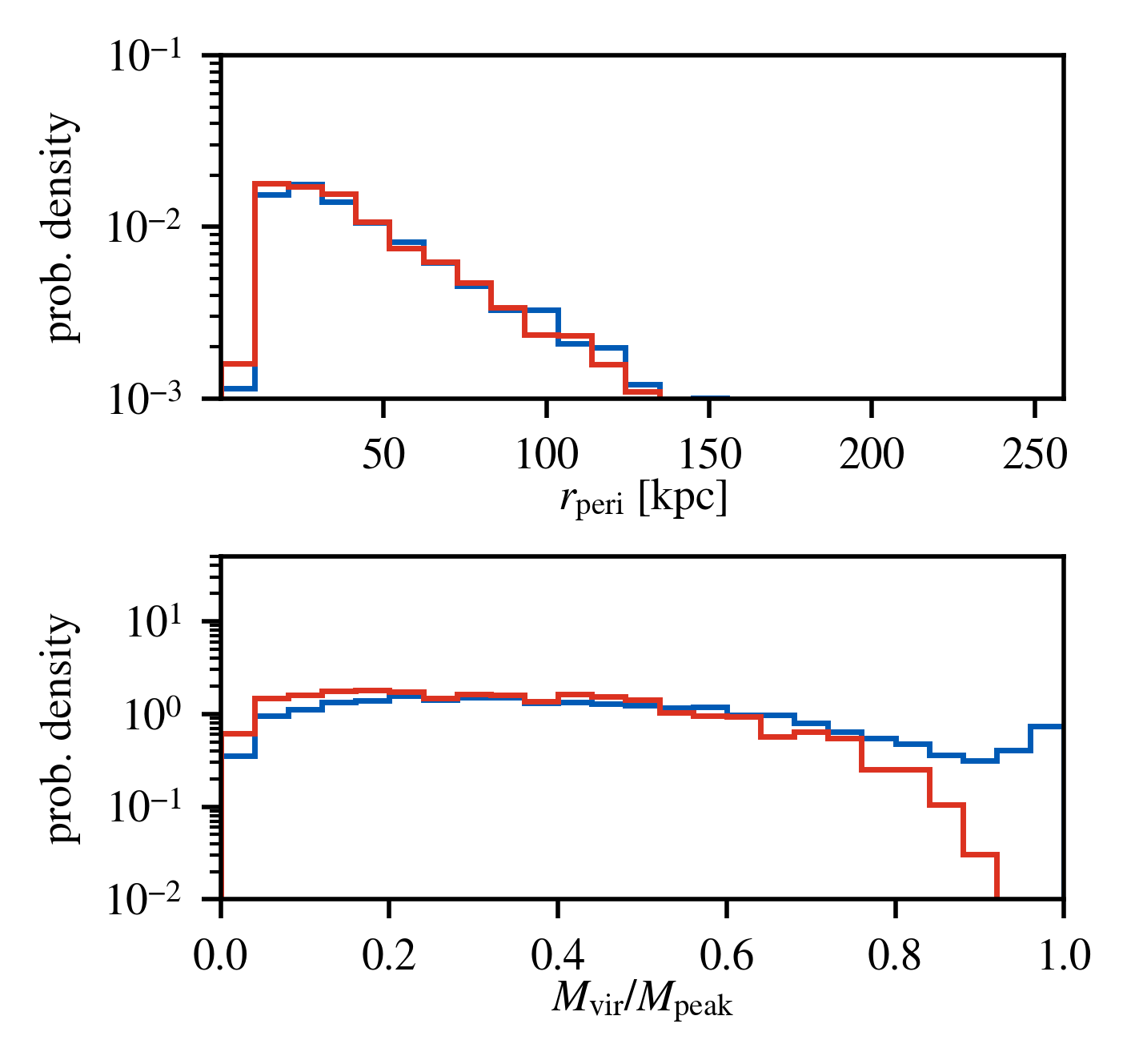}
	\caption{A version of \autoref{fig:6} for all classical satellites.}
	\label{fig:A6}
	\end{figure}
	\begin{figure}
	\addtocounter{figure}{-1}
	\centering
	\includegraphics{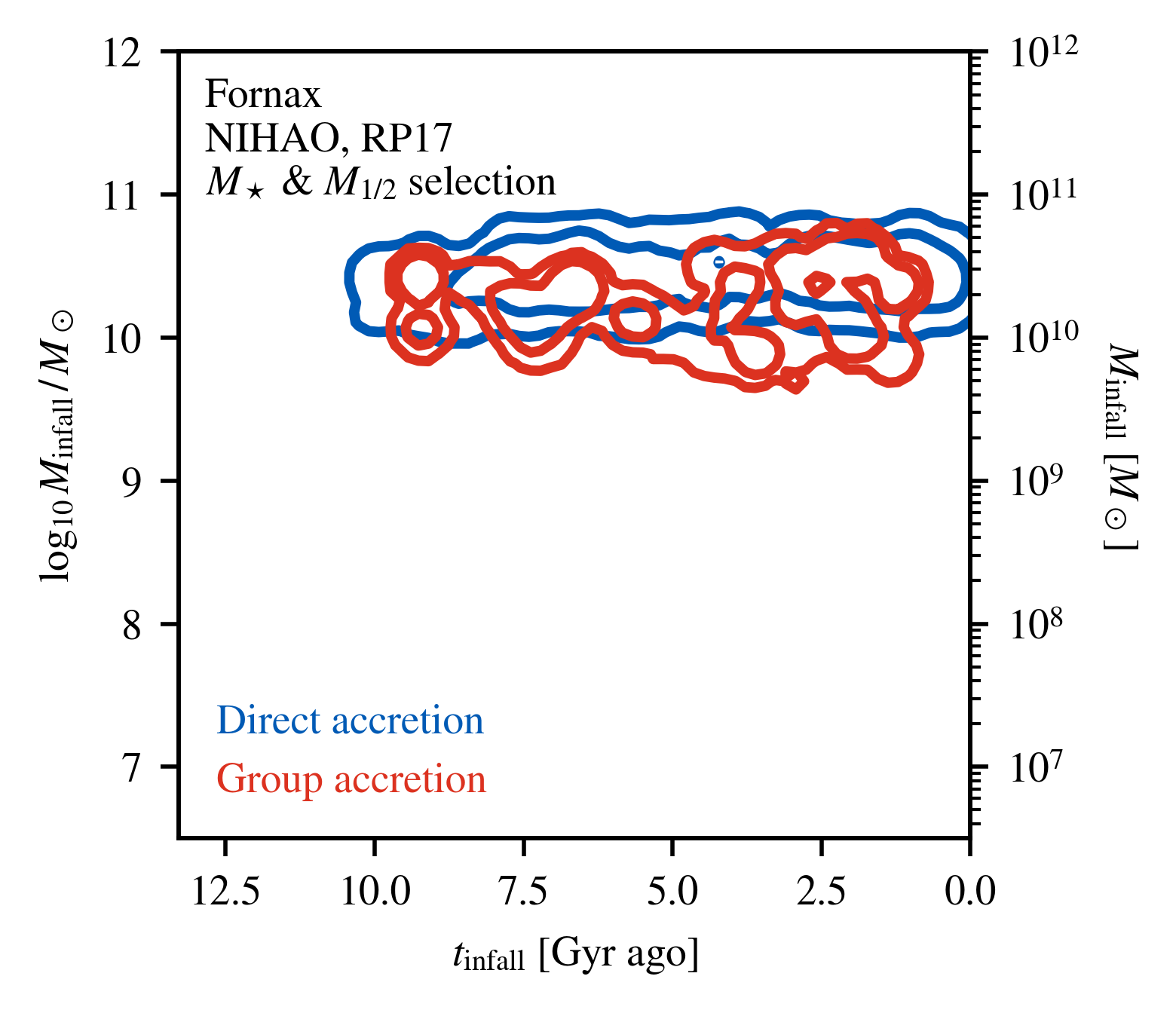}
	\includegraphics{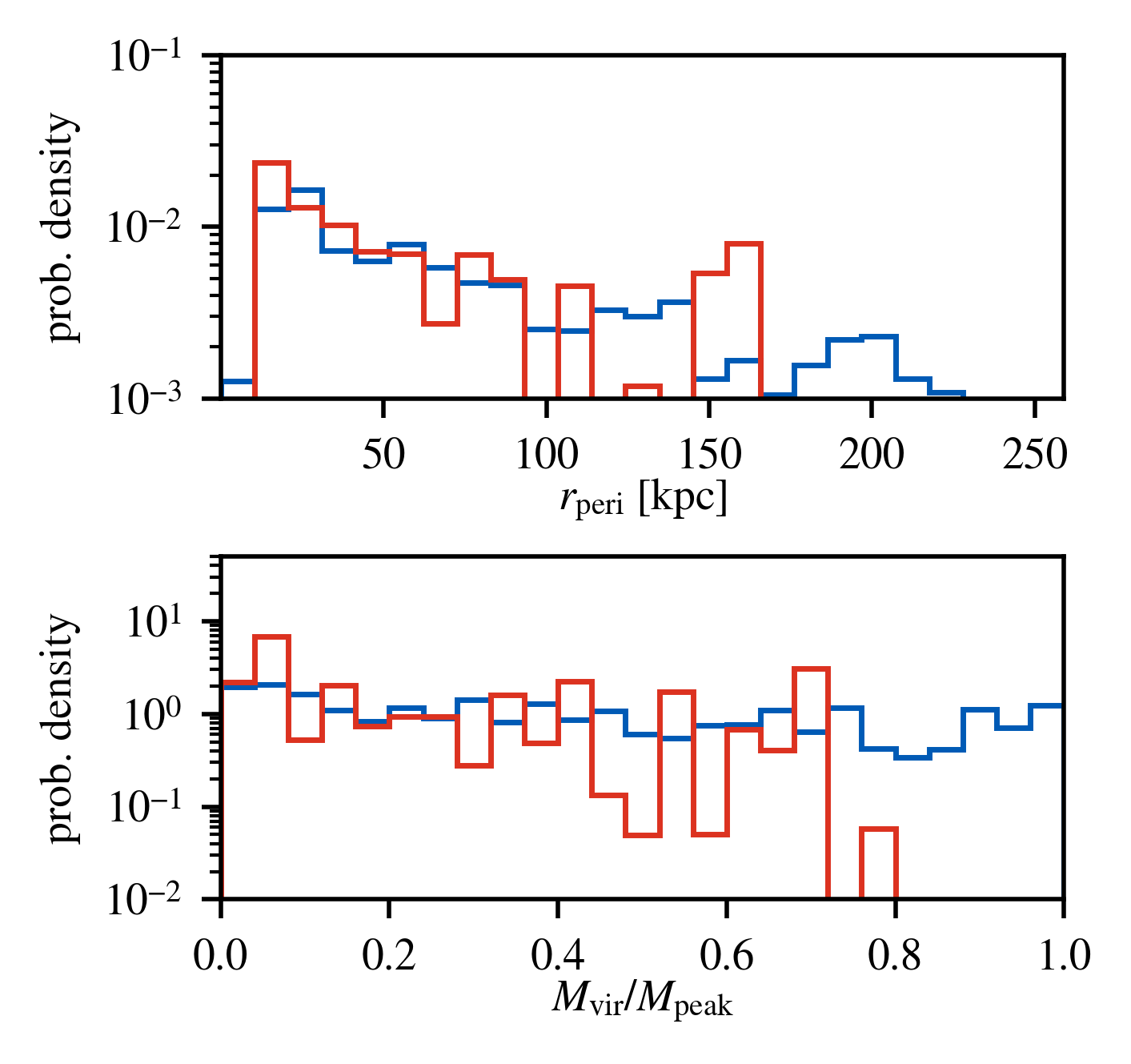}
	\includegraphics{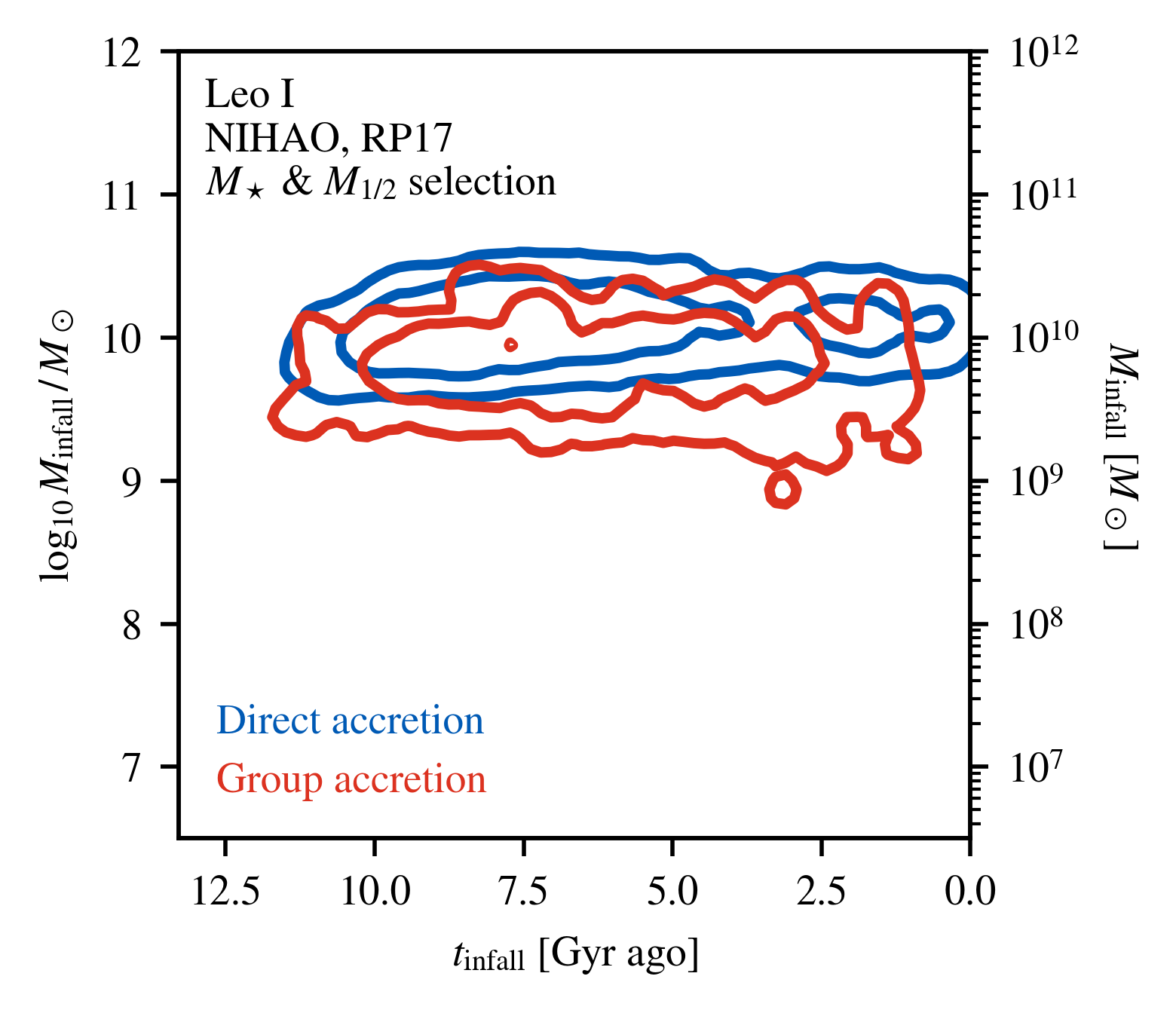}
	\includegraphics{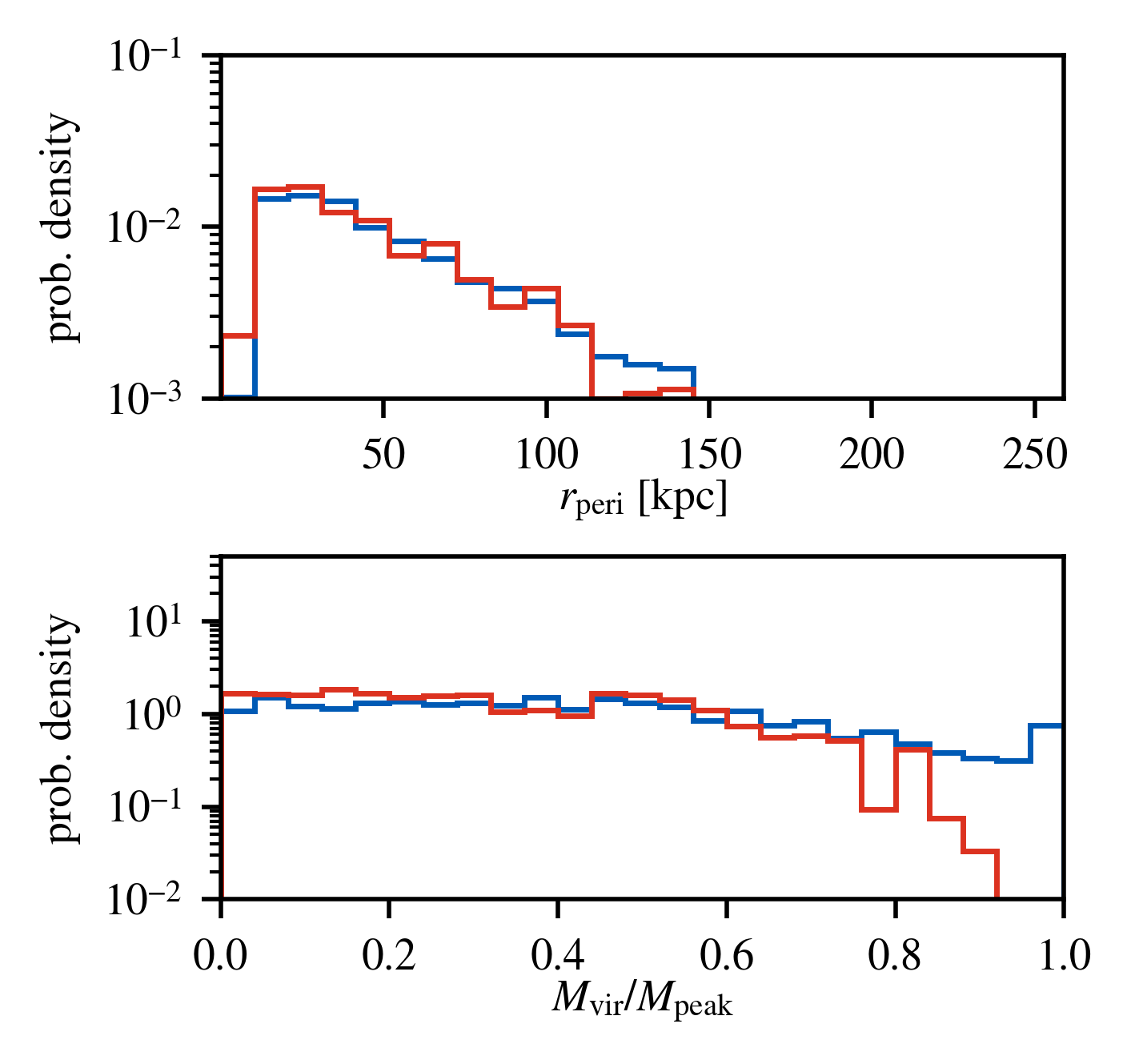}
	\includegraphics{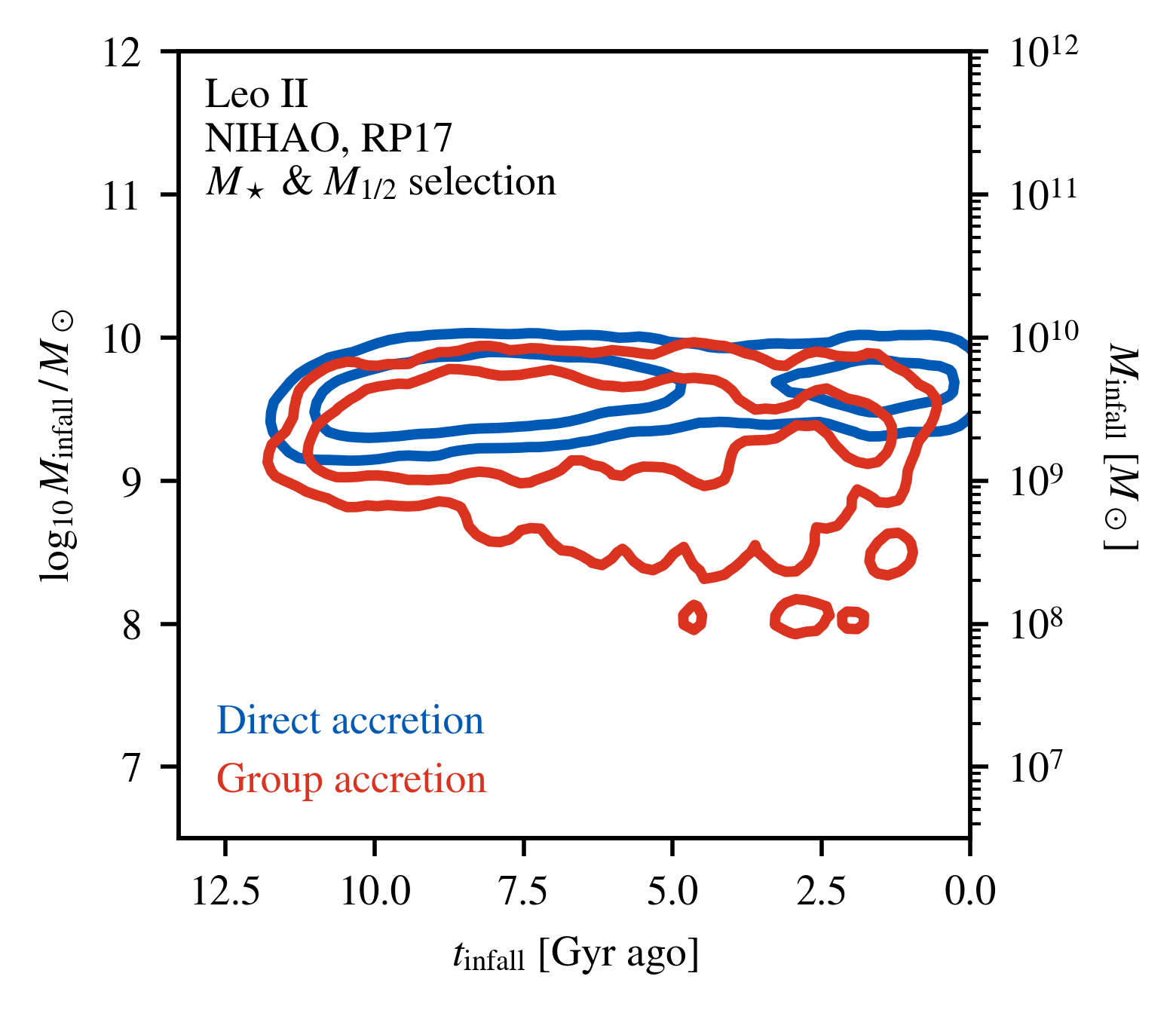}
	\includegraphics{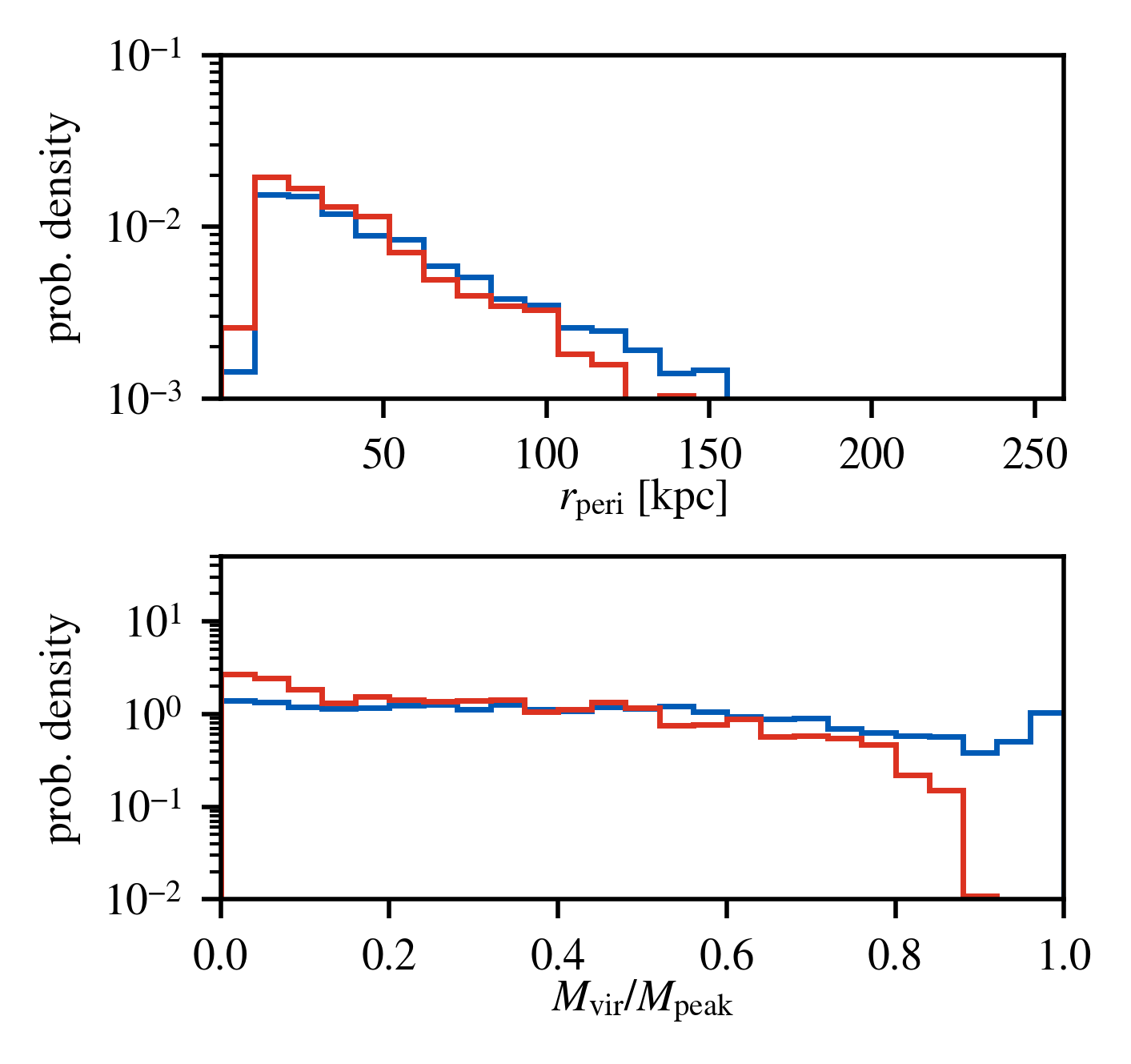}
	\caption{ -- \emph{continued}}
	\end{figure}
	\begin{figure}
	\addtocounter{figure}{-1}
	\centering
	\includegraphics{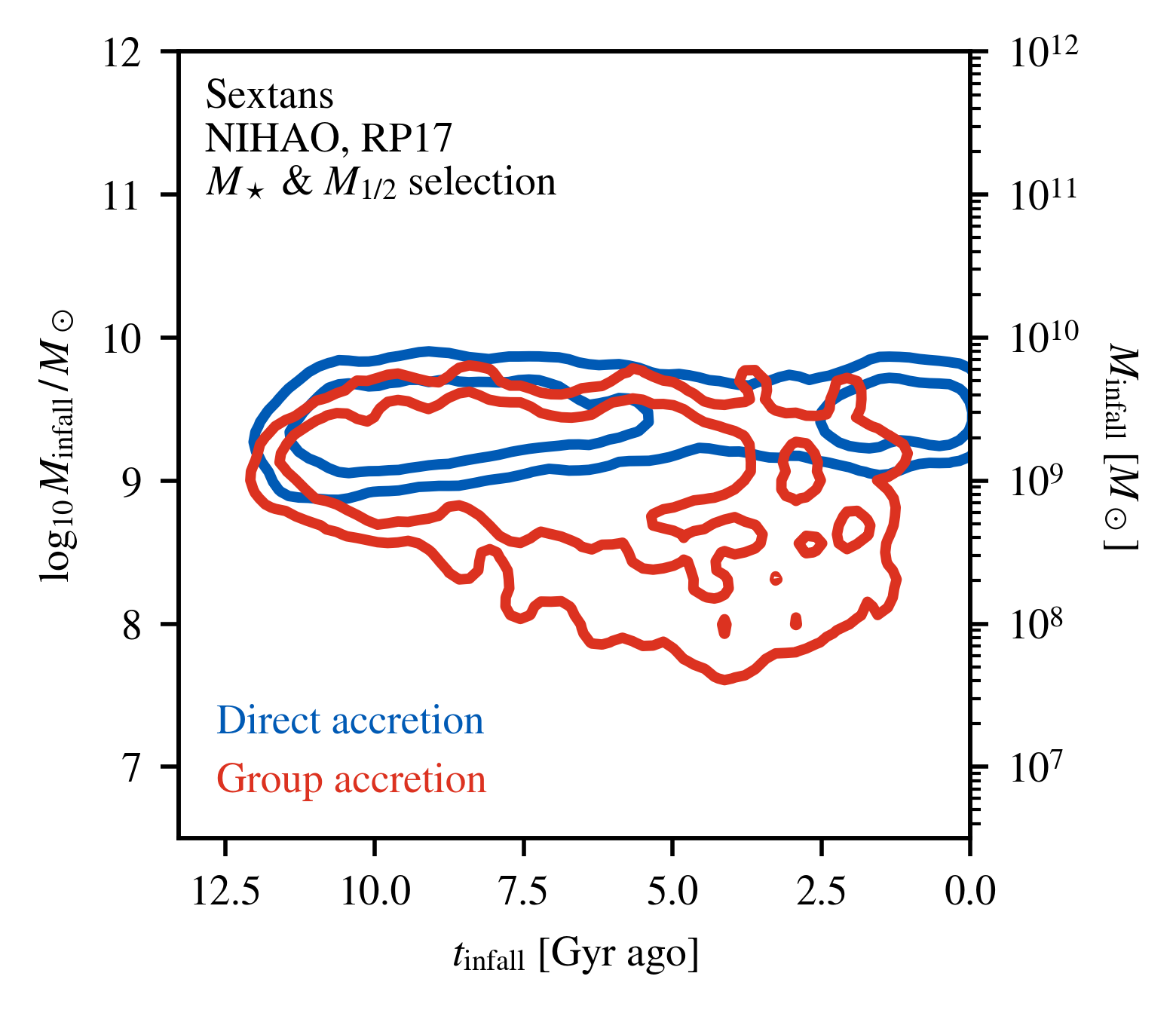}
	\includegraphics{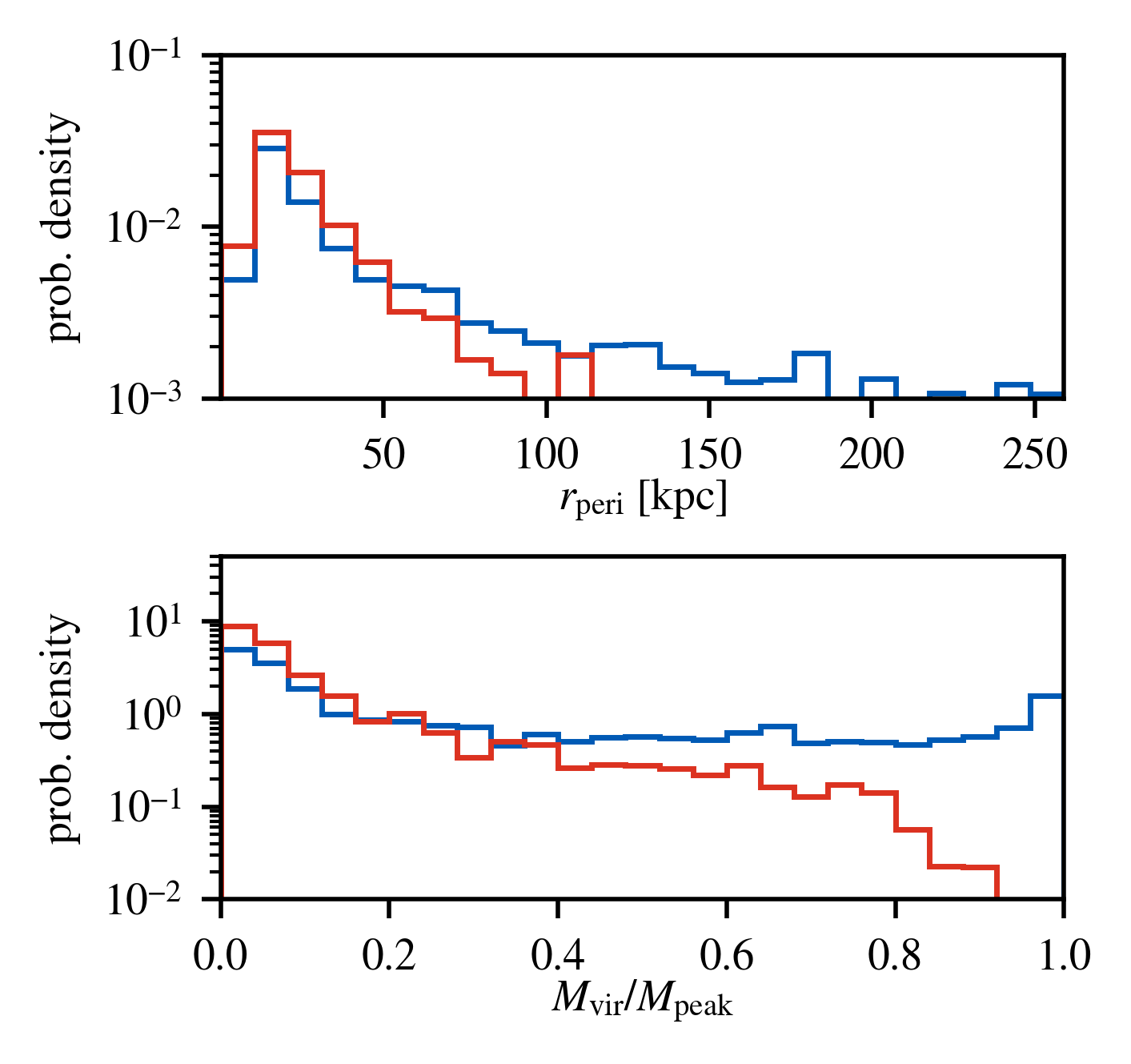}
	\includegraphics{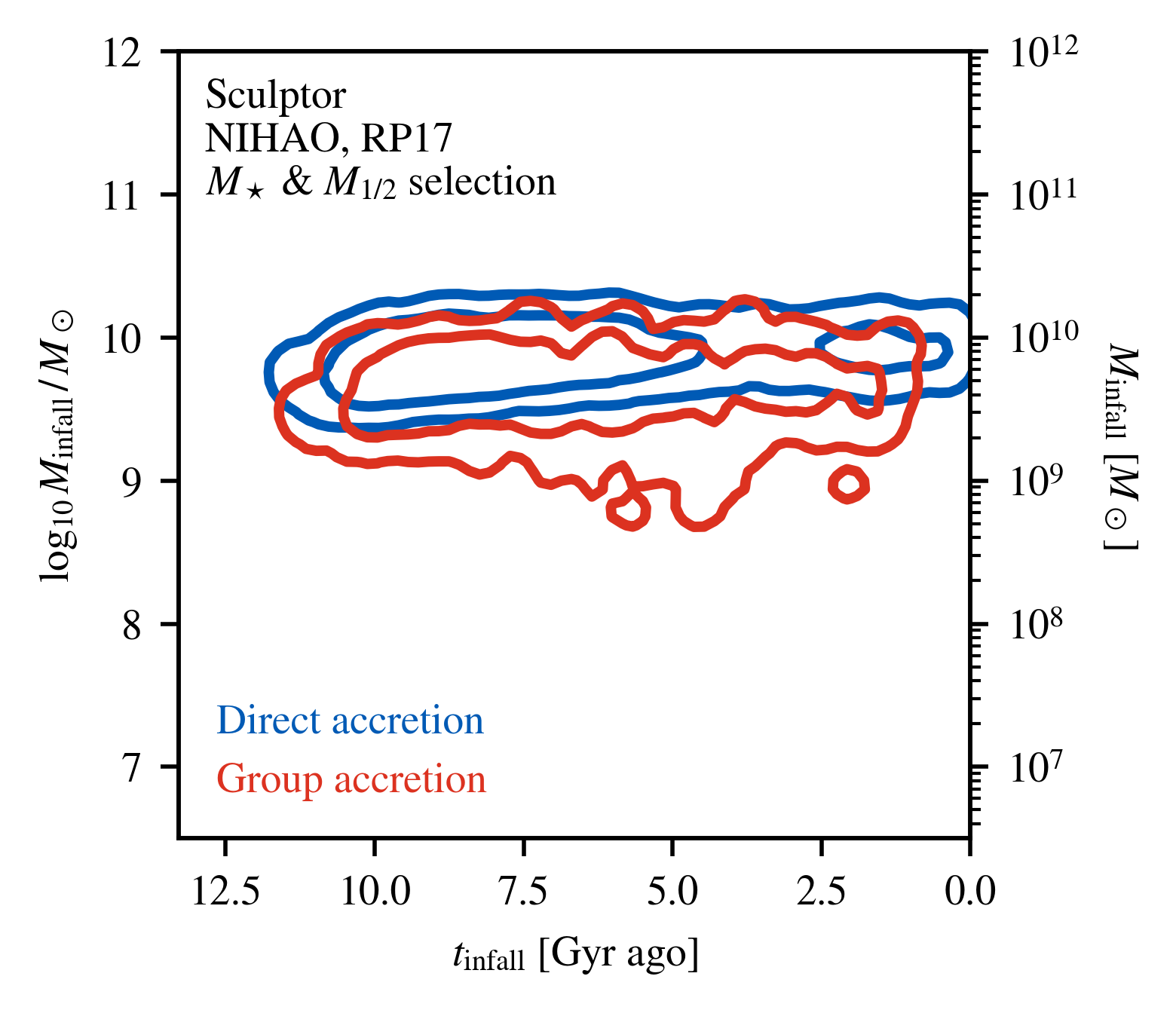}
	\includegraphics{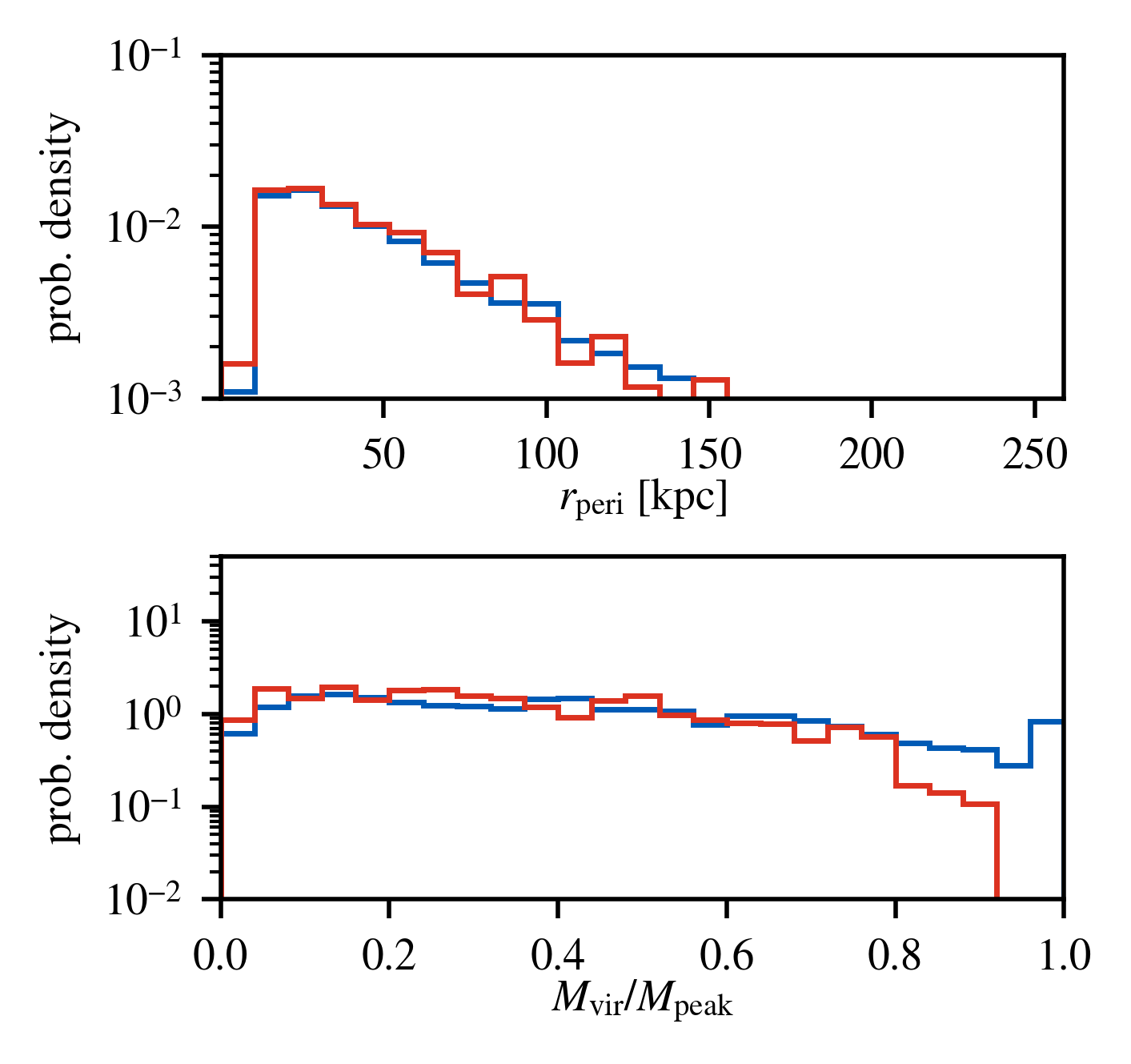}
	\includegraphics{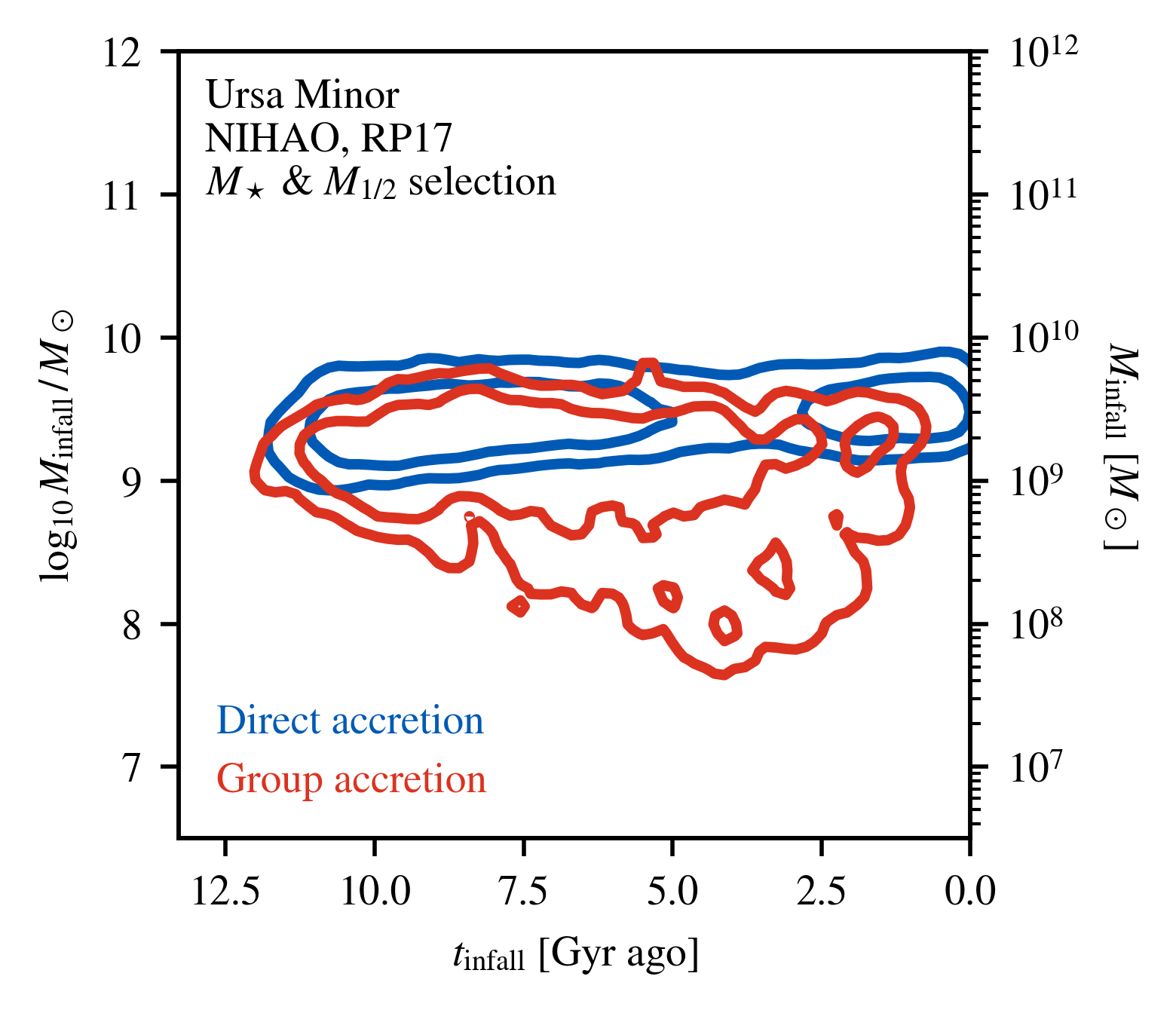}
	\includegraphics{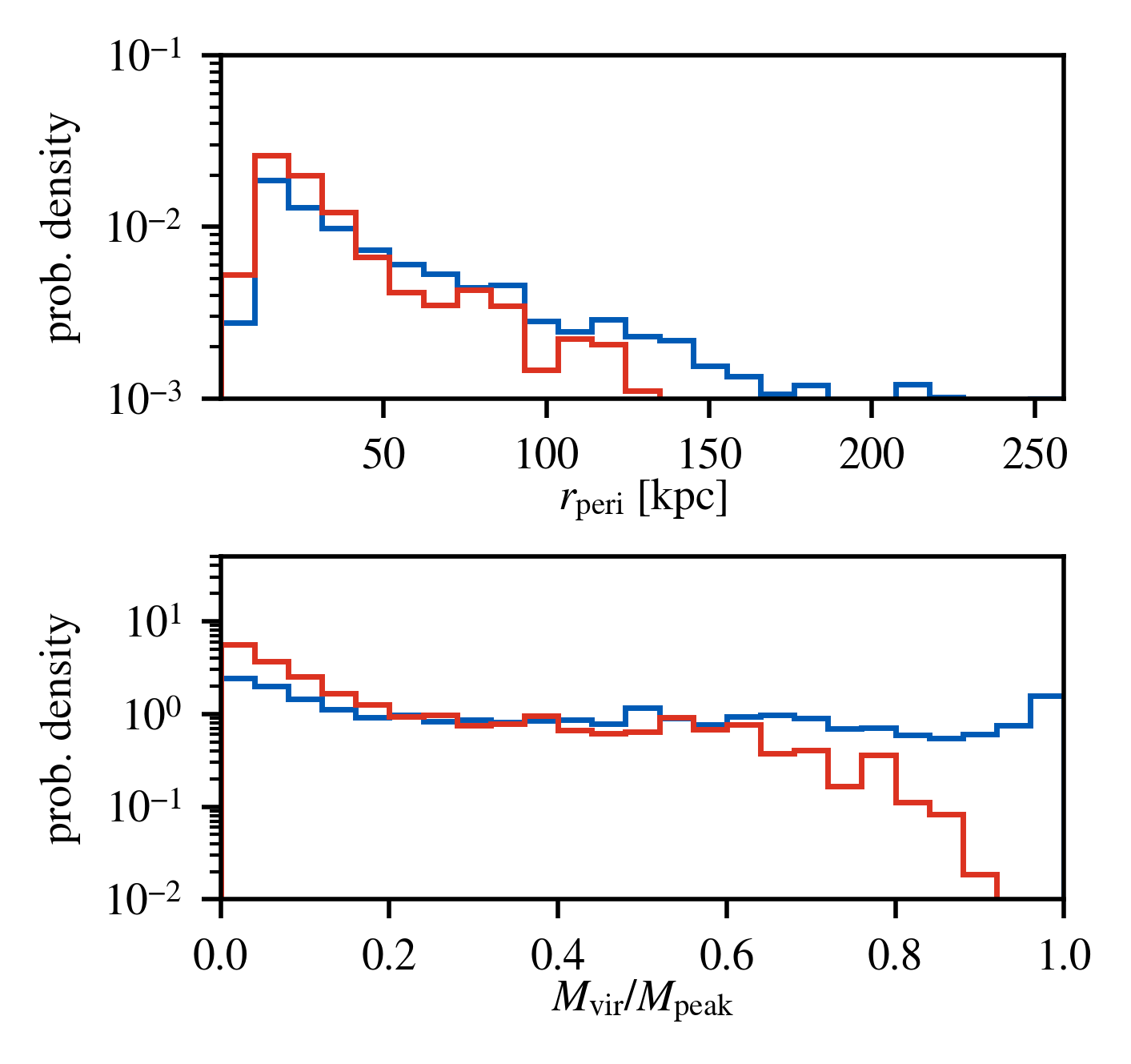}
	\caption{ -- \emph{continued}}
\end{figure}

\end{document}